\documentclass[%
,aps%
,pra%
,10pt%
,showpacs%
,showkeys%
,superscriptaddress%
,preprint%
,preprintnumbers%
,nofootinbib%
,amsmath%
,amssymb%
]{revtex4-2}

\usepackage{natbib}
\usepackage[utf8]{inputenc}
\usepackage[dvipsnames]{xcolor}
\usepackage{geometry}
\usepackage{graphicx}
\usepackage{dcolumn}

\usepackage{amsfonts}
\usepackage{amsmath}
\usepackage{amssymb}
\usepackage{bm}
\usepackage{mathrsfs}

\usepackage[caption=false]{subfig}

\begin{document}

\title{Reconstructing firm-level interactions:\\the Dutch input-output network}
\thanks{We thank ABN AMRO Bank N.V. and ING Bank N.V. for their support and active collaboration. A special thanks to the whole CADM team at ABN AMRO for their advice that helped shape this research.}

\author{Leonardo Niccol\`{o} Ialongo}
\email{leonardo.ialongo@sns.it}
\affiliation{Scuola Normale Superiore, Piazza dei Cavalieri 7, 56126 Pisa (Italy)}
\altaffiliation[Joint PhD with ]{IMT School for Advanced Studies Lucca (Lucca, Italy), Scuola Superiore Sant'Anna (Pisa, Italy), Università di Pisa (Pisa, Italy), Consiglio Nazionale delle Ricerce Pisa (Pisa, Italy)}
\affiliation{ABN AMRO Bank N.V. (The Netherlands)}

\author{Camille de Valk}
\affiliation{Lorentz Institute for Theoretical Physics, Leiden University, Niels Bohrweg 2, 2333 CA Leiden (The Netherlands)}
\affiliation{ING Bank N.V. (The Netherlands)}

\author{Emiliano Marchese}
\affiliation{IMT School for Advanced Studies Lucca, Piazza San Francesco 19, 55100 Lucca (Italy)}
\affiliation{ABN AMRO Bank N.V. (The Netherlands)}

\author{Fabian Jansen}
\affiliation{ING Bank N.V. (The Netherlands)}

\author{Hicham Zmarrou}
\affiliation{ABN AMRO Bank N.V. (The Netherlands)}

\author{Tiziano Squartini}
\affiliation{IMT School for Advanced Studies Lucca, Piazza San Francesco 19, 55100 Lucca (Italy)}
\affiliation{Institute for Advanced Study (IAS), University of Amsterdam, Oude Turfmarkt 145, 1012 GC Amsterdam (The Netherlands)}

\author{Diego Garlaschelli}
\affiliation{Lorentz Institute for Theoretical Physics, Leiden University, Niels Bohrweg 2, 2333 CA Leiden (The Netherlands)}
\affiliation{IMT School for Advanced Studies Lucca, Piazza San Francesco 19, 55100 Lucca (Italy)}

\date{\today}

\begin{abstract}
Recent crises have shown that the knowledge of the structure of input-output networks at the firm level is crucial when studying economic resilience from the microscopic point of view of firms that rewire their connections under supply and demand shocks. Unfortunately, empirical inter-firm network data are rarely accessible and protected by confidentiality. The available methods of network reconstruction from partial information, which have been devised for financial exposures, are inadequate for inter-firm relationships because they treat all pairs of nodes as potentially interacting, thereby overestimating the rewiring capabilities of the system. Here we use two big data sets of transactions in the Netherlands to represent a large portion of the Dutch inter-firm network and document the properties of one of the few analysed networks of this kind. We, then, introduce a generalized maximum-entropy reconstruction method that preserves the production function of each firm in the data, i.e. the input and output flows of each node for each product type. We confirm that the new method becomes increasingly more reliable as a finer product resolution is considered and can therefore be used as a generative model of inter-firm networks with fine production constraints. The likelihood of the model, being related to the entropy, proxies the rewiring capability of the system for a fixed input-output configuration.
\end{abstract}

\maketitle

\section{Introduction}

The structure of the input-output network formed by the superposition of multiple supply chains has important implications for the economy \cite{carvalho2019production}. Extreme weather events that disrupted supply chains in a given region have been shown to propagate economic distress over this network, indirectly affecting the whole economy \cite{carvalho2021jap, kashiwagi2021propagation}. The constraints that the input-output specificity of firms imposes on the economy can transform micro-economic shocks into fluctuations of aggregate demand and supply  \cite{acemoglu2012net}. This has been further highlighted by the recent COVID-19 pandemic, where the structure of the production network has driven the propagation of negative economic shocks, by determining the extent to which they can be absorbed or amplified, as a consequence of the lack of demand or limitations in the supply of intermediate products \cite{pichler2021shock}.

There are several models that attempt to understand how simple firm level interactions can lead to a complex behaviour of the aggregate: from agent-based models \cite{dawid2018agent}, to simple inventory dynamics \cite{bak1993sect}. In most cases, however, these studies are either conducted at the aggregate industry level or with unrealistic simulated network structures. This is mostly due to the lack of access to reliable data on firm level interactions: in fact, input-output relationships at firm level are difficult to observe due to the sensitive nature of the information and, for most countries, the absence of a dataset capturing the structure of these same relationships in a reliable way. As such, it is important to develop a statistical reconstruction framework able to accurately reproduce the structural properties of the real inter-firm network from limited information.

Maximum entropy graph ensembles, also known as Exponential Random Graphs (ERG), have successfully been applied to economic data in order to reconstruct features of real networks from partial information \cite{squartini2018reconstruction}. As it has been shown in several independent ``horse-races'' \cite{anand2018missing,mazzarisi2017methods, gandy2019adjustable,ramadiah2020reconstructing, lebacher2019search}, this analytical approach outperforms competing probabilistic recipes in providing an accurate prediction of a set of network features, ranging from microscopic (e.g. true positives, true negatives) to macroscopic ones (e.g. the trend of the average nearest neighbours degree, or the clustering coefficient, as a function of the nodes degree)\cite{squartini2018reconstruction}. In particular in the case of inter-bank networks a very efficient algorithm, the so called density-corrected Gravity Model (dcGM), has been proposed in order to obtain a similar performance to the Configuration Model in the absence of degree information \cite{cimini2015estimating}. This model entails a two step recipe that first draws links and then assigns them weights using only the strengths of each node and the total number of links of the graph \cite{cimini2015systemic}. In \cite{parisi2020horse} it has been shown that the dcGM, with exponentially distributed weights ($\text{CReM}_B$ in the paper, here simply dcGM) provides optimal performance while keeping computational complexity low. This model, although it preserves the nodes out- and in-strengths and the density of the graph, it does not constrain any network mesoscopic feature which may be known to play an important role - as the presence of sectoral interdependencies in the case of the inter-firm network. In fact, the dcGM has been developed in a context where the nature of the exchange is the same for every actor, hence where a link always represents a credit relationship. When firm-to-firm relationships are considered, instead, the nature of these connections is necessarily heterogeneous as the kind of goods and services exchanged will vary significantly between agents: naturally, this cannot be replicated by a model that does not constrain any property of the kind.

This contribution is devoted to extending the dcGM to incorporate the available information on sector linkages. From an economic point of view, these quantities reflect constraints `affecting' the production of firms. To the best of our knowledge, this is the first reconstruction model that explicitly embodies this kind of product constraints.

\section{Methodology}

Let us represent a directed, weighted network via an asymmetric adjacency matrix $W$ whose generic entry $w_{ij}$ is the weight of the connection from node $i$ to node $j$. In turn, $W$ induces the binary adjacency matrix $A$ where $a_{ij}=1$ if a connection from node $i$ to node $j$ is present and $a_{ij}=0$ otherwise. In what follows, we use the $*$ symbol to mark the original graph, as in $A^*$ and $W^*$, and the empirical value of the objective quantities. An important observation concerns self-loops that, as usual, are not allowed: their removal introduces a small distortion on the expected value of the strengths (see \cite{parisi2020horse} for a detailed discussion) that, however, does not affect the overall performance of the method.

\subsection{The density-corrected Gravity Model (dcGM)}
Let us recall the definition of the density-corrected Gravity Model. It is a two-step probabilistic recipe prescribing to connect any two nodes $i$ and $j$ with a probability reading
\begin{equation}
p_{ij}^\text{dcGM}=\frac{zs_i^{out}s_j^{in}}{1+zs_i^{out}s_j^{in}},\quad\forall\:i\neq j
\end{equation}
where $s_i^{out} = \sum_{i\neq j} w_{ij}$ is the out-strength of node $i$ and $s_j^{in} = \sum_{j\neq i} w_{ij}$ is the in-strength of node $j$. $z$ is the only parameter of the model that can be determined upon fixing the expected link density of the network by solving the equation
\begin{equation}
\langle L \rangle = \sum_{i\neq j}p_{ij}^\text{dcGM}=\sum_{i\neq j}a^*_{ij}=L^*.
\end{equation}

The second step of the model embodies a prescription to determine weights. In its simplest version, the second step of the dcGM is defined by the Bernoulli recipe $w_{ij}\sim\text{Ber}(p_{ij}^\text{dcGM})$ prescribing to draw the value $w_{ij}=0$ with probability $1-p_{ij}^\text{dcGM}$ and the value $w_{ij}=s_i^{out}s_j^{in}/(w^{tot}\cdot p_{ij}^\text{dcGM})$ with probability $p_{ij}^\text{dcGM}$, where $w^{\text{tot}}=\sum_is_i^{out}=\sum_is_i^{in}$ is the total weight of the network. Such a prescription can be further refined by letting the weights be sampled from either a geometric distribution $w_{ij}\sim\text{Geo}(\beta_{ij})$ in the discrete case, or an exponential distribution $w_{ij}\sim\text{Exp}(\beta_{ij})$ in the continuous one. Notice that while the Bernoulli prescription ensures that
\begin{equation}
\langle w_{ij}\rangle=\frac{s_i^{out}s_j^{in}}{w^{tot}},\quad\forall\:i\neq j
\end{equation}
and hence we recover the analytic expression of the expected weight between nodes $i$ and $j$ defining the Gravity Model, in the other two cases the parameters of the distribution must be chosen to ensure that the condition above holds. Specifically, $\beta_{ij}=(s_i^{out}s_j^{in}-w^{tot})/s_i^{out}s_j^{in}$ for the geometric distribution and $\beta_{ij}=w^{tot}\cdot p_{ij}^\text{dcGM}/(s_i^{out}s_j^{in})$ for the exponential distribution.

\subsection{The limitations of the dcGM}
One of the main limitations of the dcGM concerns the inaccurate reconstruction of the mesoscopic features of a network. As an example, let us focus on an inter-firm network and consider the weight between two groups\footnote{We will use the following notation to denote groups of node (i.e. sectors): $g_i$ is the group to which node $i$ belongs such that $i \in g_i\:\forall i$. Hence, $g_i$ and $g_j$ can be the same group even if $i\neq j$ but $i$ and $j$ belong to the same sector. We assume that each node belongs to only one group.} of firms, say $g_i$ and $g_j$, i.e.
\begin{equation}
s_{g_i\rightarrow g_j}=\sum_{i\in g_i}\sum_{j\in g_j}w_{ij};
\end{equation}
its expected value under the dcGM reads
\begin{align}
\langle s_{g_i\rightarrow g_j}\rangle &= \sum_{i\in g_i}\sum_{j\in g_j}\langle w_{ij}\rangle\\
&=\sum_{i\in g_i}\sum_{j\in g_j}\frac{s_i^{out}s_j^{in}}{w^{tot}}=\frac{s_{g_i}^{out}s_{g_j}^{in}}{w^{tot}},\quad\forall\:g_i\neq g_j
\end{align}
which, in general, will differ from the empirical value $s_{g_i\rightarrow g_j}$. The reason lies in the fact that the dcGM predicts any two firms to be linked with a probability that is unaffected by the size of the inter-sector flows, but only depends on the individual node strengths. We know empirically that this cannot be the case.

\subsection{The stripe-corrected Gravity Model} \label{sub:scGM}
Let us now introduce a model constraining a peculiar kind of mesoscale structures, hereby called `stripes'. We do so by defining a new quantity, i.e. the in-strength by sector:
\begin{equation}
s_{g_i\rightarrow j}=\sum_{i\in g_i}w_{ij},\quad\forall\:j.
\end{equation}

This quantity is important because it represents the list of quantities by industry of origin ($g_i$), which is a good proxy for product kind, each node $j$ uses to produce $s_j^{out}$ units of output. Therefore, it has a clear meaning in terms economic input-output analysis and Leontief production functions. Hence, we may define the \emph{stripe-corrected Gravity Model} (scGM) as the same two step procedure of the dcGM with the parameter $\beta_{ij}$ chosen such that
\begin{equation}\label{eq:exp_sc_weight}
\langle w_{ij} | a_{ij} = 1 \rangle= \frac{s_i^{out}s_{g_i\rightarrow j}}{w_{g_i}^{tot}\cdot p_{ij}^\text{scGM}},\quad\forall\:i\neq j
\end{equation}
with
\begin{equation}
p_{ij}^\text{scGM}=\frac{zs_i^{out}s_{g_i\rightarrow j}}{1+zs_i^{out}s_{g_i\rightarrow j}},\quad\forall\:i\neq j
\label{eq:prob-block-variant-I}
\end{equation}
and where we have normalized with the total outgoing strength of sector $g_i$, i.e.
\begin{equation}\label{norm}
\sum_{i\in g_i}s_i^{out}=\sum_{i\in g_i}\sum_j w_{ij}=\sum_js_{g_i\rightarrow j}\equiv w^{tot}_{g_i},\quad\forall\:g_i;
\end{equation}
notice that the first and the last sum simply account for the same addenda, just sorting them differently. While the relationship $\sum_is_i^{out}=\sum_js_j^{in}=w^{tot}$ ensures the dcGM to be normalized in such a way to reproduce in-strengths and out-strengths, eq. (\ref{norm}) now ensures the scGM to be normalized in such a way to reproduce out-strengths and sectorial in-strengths. In fact,
\begin{align}
\langle s_i^{out}\rangle &=\sum_j\langle w_{ij}\rangle=\sum_j\frac{s_i^{out}s_{g_i\rightarrow j}}{w_{g_i}^{tot}}=s_i^{out}\frac{\sum_js_{g_i\rightarrow j}}{w_{g_i}^{tot}}=s_i^{out}\quad\forall\:i,\label{eqs_out}\\
\langle s_{g_i\rightarrow j}\rangle &=\sum_{i\in g_i}\langle w_{ij}\rangle=\sum_{i\in g_i}\frac{s_i^{out}s_{g_i\rightarrow j}}{w_{g_i}^{tot}}=s_{g_i\rightarrow j}\frac{\sum_{i\in g_i}s_i^{out}}{w_{g_i}^{tot}}= s_{g_i\rightarrow j}\quad\forall\:g_i,\:j;
\label{eq:sl_in}
\end{align}
removing self contributions, instead, leads to the slightly distorted relationships $\langle s_i^{out}\rangle=s_i^{out}-\frac{s_i^{out} s_{g_i\rightarrow i}}{w_{g_i}^{tot}}$, $\forall\:i$ and $\langle s_{g_j\rightarrow j}\rangle=s_{g_j\rightarrow j}-\frac{s_j^{out}s_{g_j\rightarrow j}}{w^{tot}_{g_j}}$, $\forall\:j\in g_j$. 

Furthermore,
\begin{align}
\langle s_j^{in}\rangle&=\sum_i\langle w_{ij}\rangle=\sum_i\frac{s_i^{out}s_{g_i\rightarrow j}}{w_{g_i}^{tot}}=\sum_{g_i}\sum_{i\in g_i}\frac{s_i^{out}s_{g_i\rightarrow j}}{w_{g_i}^{tot}}=\sum_{g_i}s_{g_i\rightarrow j}=s_j^{in},\quad\forall\:j,\\
\langle s_{g_i}^{out}\rangle&=\sum_{i\in g_i}\sum_j\langle w_{ij}\rangle=\sum_{i\in g_i}\sum_j\frac{s_i^{out}s_{g_i\rightarrow j}}{w_{g_i}^{tot}}=\sum_{i\in g_i}s_i^{out}\frac{\sum_js_{g_i\rightarrow j}}{w_{g_i}^{tot}}=\sum_{i\in g_i}s_i^{out}=s_{g_i}^{out},\quad\forall\:g_i;
\end{align}
where again removing self contributions introduces the slightly distorted relationships $\langle s_j^{in}\rangle=s_j^{in}-\frac{s_j^{out}s_{g_j\rightarrow j}}{w^{tot}_{g_j}}$, $\forall\:j$ and $\langle s_{g_i}^{out}\rangle=s_{g_i}^{out}-\sum_{i\in g_i} \frac{s_i^{out} s_{g_i\rightarrow i}}{w_{g_i}^{tot}}$, $\forall\:g_i$. Notice that, in this model, the diagonal elements $w_{ii}$ can be zero also `naturally' such that the expected values of all the strengths above match exactly the empirical values. This happens if $s_{g_i\rightarrow i}=0$, $\forall\:i$, that is, if there are no links between nodes in the same sector. We note that in the ideal case where we could distinguish companies exactly by the product they produce this would likely be the case. On the other hand it must be noted that, if the above is not true, for some nodes the weight to redistribute might be more than in the dcGM as the denominator is now the total strength of the sector and not of the whole graph. 

The only parameter $z$ can be found by solving the equation
\begin{equation}\label{eq:scGMdensfit}
\langle L\rangle=\sum_{j\neq i}\frac{zs_i^{out}s_{g_i\rightarrow j}}{1+zs_i^{out}s_{g_i\rightarrow j}}=L^*.
\end{equation}

One issue that may arise is related to the number of outgoing links from each sector $g_i$. Let us define the number of outgoing links per sector as
\begin{equation}
\sum_{i\in g_i}\sum_{j} a_{ij} \equiv L_{g_i},\quad\forall\:g_i;
\end{equation}
clearly, in equation \eqref{eq:scGMdensfit} we are constraining $\sum_{g_i} L_{g_i}$ but we have no guarantees over the single $L_{g_i}$. Therefore, should we have access to the coefficients $L_{g_i}$, $\forall\:g_i$ we could  account for the different densities in the sectors by defining a multi-z variant of the model. The only difference with the basic definition of the model provided above will be given by the probabilities $p_{ij}^\text{scGM}$ that would now be defined as
\begin{equation}
p_{ij}^\text{scGM}=\frac{z_{g_i}s_i^{out}s_{g_i\rightarrow j}}{1+z_{g_i}s_i^{out}s_{g_i\rightarrow j}},\quad\forall\:i\neq j.
\end{equation}

The same quantities are preserved in the multi-z variant and the parameters $z_{g_i}$, $\forall\:g_i$ can be found by solving the equations
\begin{equation}\label{eq:multi_z}
\langle L_{g_i}\rangle=\sum_{i\in g_i}\sum_j\frac{z_{g_i}s_i^{out} s_{g_i\rightarrow j}}{1+z_{g_i}s_i^{out}s_{g_i\rightarrow j}}=L_{g_i}^*,\quad\forall\:g_i.
\end{equation}

It is interesting to notice that one of the main advantages of the stripe-corrected model is that it introduces a strong restriction on the number of available links. In fact, the probability of observing a link between $i$ and $j$ according to the dcGM cannot be zero unless either $s_i^{out}$ or $s_j^{in}$ are zero. In case a sufficiently-detailed sector classification is provided, instead, we expect to observe a significant sparsity in the sectors; such an evidence embodies the restriction that if a company does not buy any product from a given sector, no connection can exist between that company and any other in the said sector. Such a possibility, that in mathematical terms translates into the fact that many $s_{g_i\rightarrow j}$ terms will be zero, is successfully handled by the scGM and ensures it to perform better than the dcGM in reproducing the true zeros of the adjacency matrix, by construction.

\section{Datasets}

We have validated our methodology on the properties of two similar datasets, made available for this research by ABN AMRO Bank N.V. (ABN) and ING Bank N.V. (ING). The information was not shared by the two banks but rather the analysis was conducted in parallel such that no sensitive information was transferred between the two institutions. Only the anonymous results have been made available for the purpose of writing the present paper.

The raw data consists of a table of money flows to and from the accounts of the commercial clients of the banks from 2018 (January for ABN and October for ING) to the present date. Our analysis will therefore focus on the year 2019, for which the data is fully available and during which there have not been any extreme events that might affect the result of our analysis. This data is primarily constituted of SEPA transactions which are by far the most common type of cashless payment for these accounts. Given that we see on the clients accounts movements coming and going to non-client accounts, it would be possible for us to include non-clients as nodes in our network. However, the available information is severely limited and we were often unable to distinguish between accounts belonging to a firm or a private individual; hence, in order to avoid possible biases introduced by this discrepancy, we have limited ourselves to build our networks by just considering client-to-client transactions. 

We define the networks as directed, the direction of the flow being opposite to the direction of the money transfer. We further assign each node to a sector using its industrial classification code (see Supplementary Material). These codes are hierarchical in nature allowing us to define five hierarchical levels of sector definitions with level five being the most specific (see table \ref{tab:sector_levels} for a summary of the various hierarchical levels in both datasets). Hence in our networks, an edge from firm $i$ in Sector A to firm $j$ in Sector B with weight 100, will represent the sale of a product of type A by firm $i$ to firm $j$ for the value of 100 Euros. After this construction, the two networks, one for ING and one for ABN clients, have (order) of $10^5$ nodes and $10^6$ links each (see the Supplementary Material section \ref{sub:trx_filters} for further discussion on the data).

\begin{table}[t!]
\caption{\label{tab:sector_levels}%
Number of distinct sectors by hierarchical level used in classification codes.}
\begin{ruledtabular}
\begin{tabular}{lllrr}
\textrm{Level}&
\textrm{SBI rule}&
\textrm{NAICS rule}&
\textrm{ABN}&
\textrm{ING}\\
\colrule
1 & Area        & 2 digits          & 20    & 23\\
2 & 2 digits    & 3 digits          & 82    & 89\\
3 & 3 digits    & 4 digits          & 252   & 303\\
4 & 4 digits    & 5 digits          & 566   & 642\\
5 & All digits  & All (6) digits    & 886   & 953\\
\end{tabular}
\end{ruledtabular}
\end{table}

\section{Results}

\subsection{Fitness ansatz}
\begin{figure}[p]
    \centering
    \subfloat[All sectors \label{fig:ansatz_0}]{%
    \includegraphics[width=0.9\textwidth]{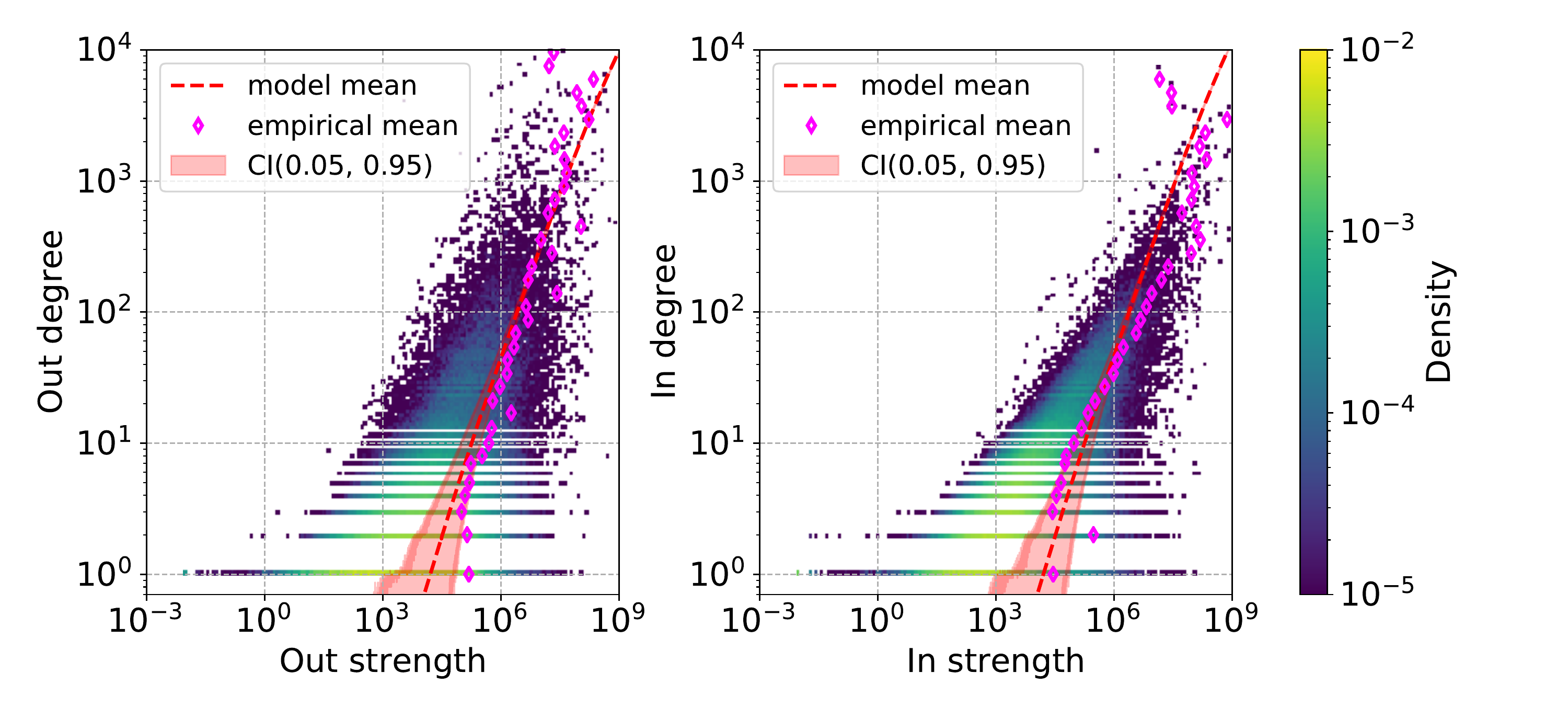}
    }\vfil
    \subfloat[C - Manufacturing\label{fig:ansatz_1}]{%
    \includegraphics[width=0.9\textwidth]{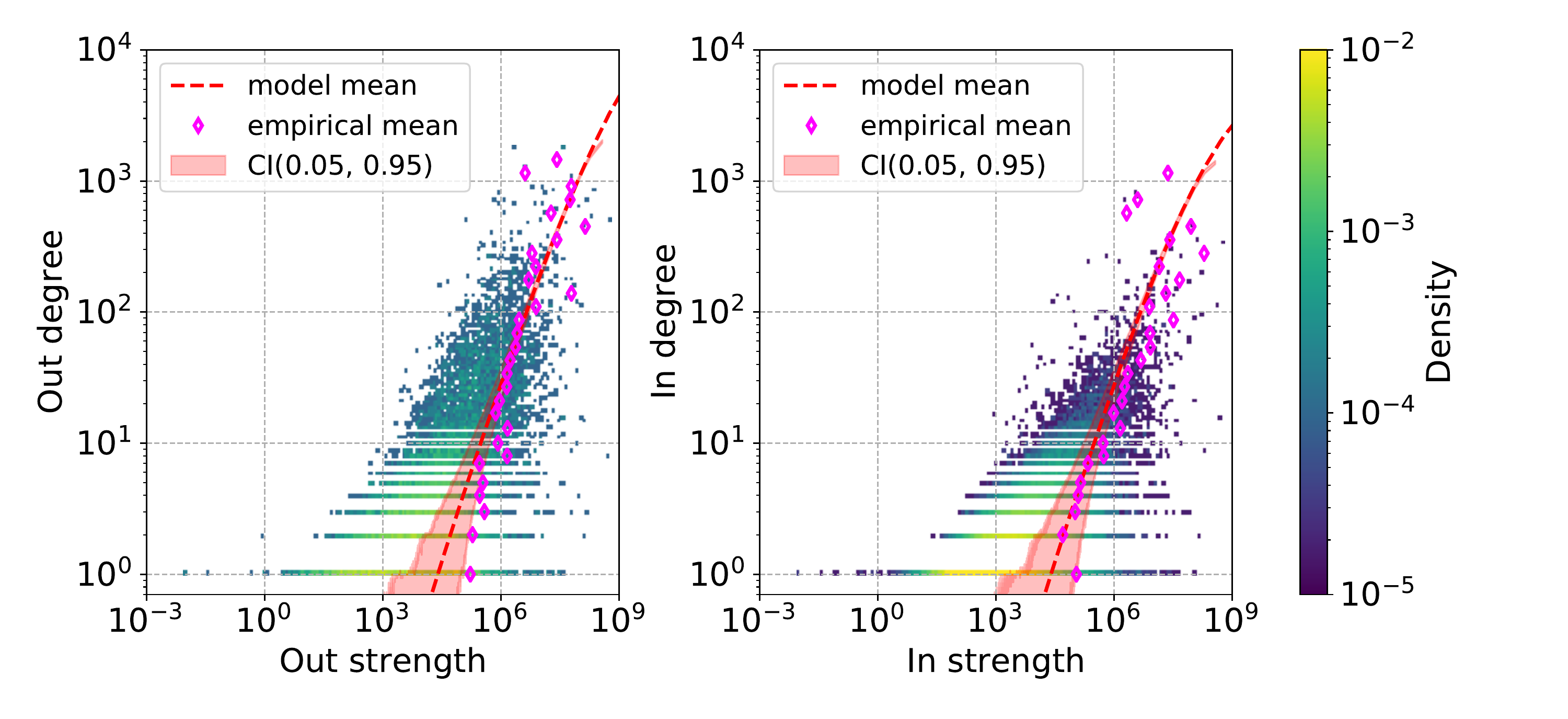}%
    }
    \caption{Degree as a function of strength for the specified sector. Both in the left and right plots we represent degree and strength by sector only for those firms that have non-zero values. This means firms in the sector for the left and firms with a relation to the sector for the right. Coloured area represents the density of points in the empirical network of Institution 1. In red is the fitted model mean and the confidence interval computed from one hundred samples taken from the ensemble and smoothed using a moving average.}\label{fig:ansatzA}
\end{figure} 

\begin{figure}[p]
    \centering
    \subfloat[33 - Repair and installation of machinery and equipment\label{fig:ansatz_2}]{%
        \includegraphics[width=0.9\textwidth]{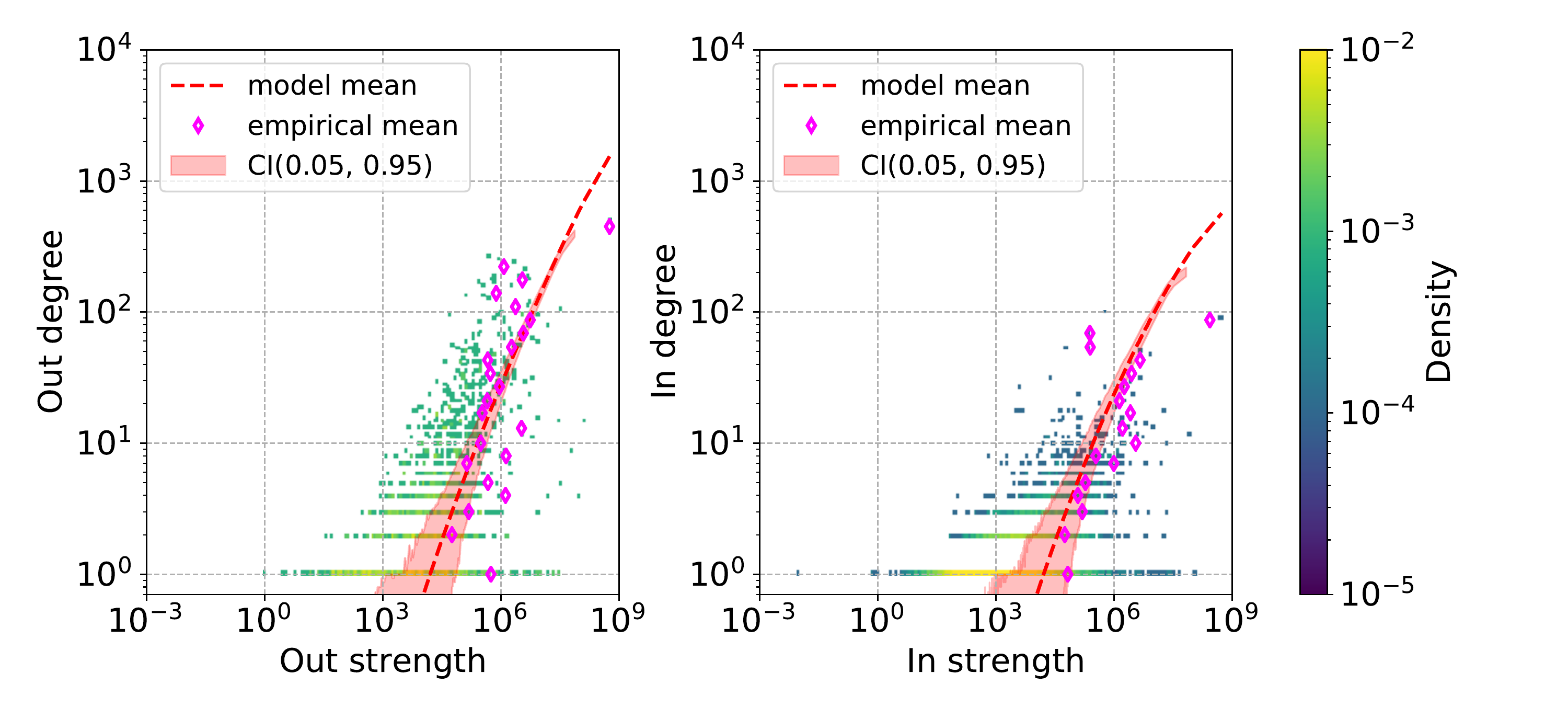}%
    }\vfil
    \subfloat[33121 - Repair and maintenance of machinery for general use and machine parts \label{fig:ansatz_5}]{%
        \includegraphics[width=0.9\textwidth]{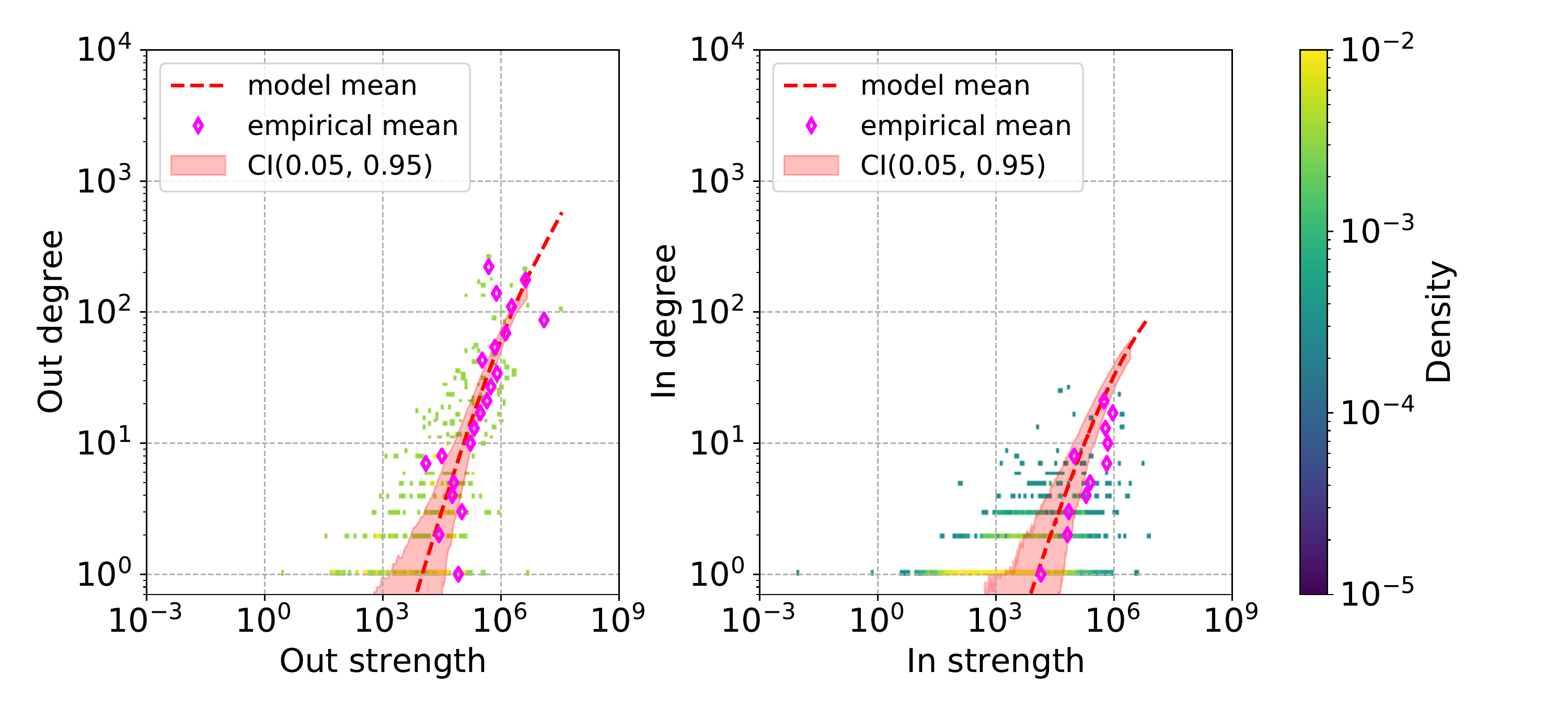}%
    }
    \caption{Degree as a function of strength for the specified sector. Both in the left and right plots we represent degree and strength by sector only for those firms that have non-zero values. This means firms in the sector for the left and firms with a relation to the sector for the right. Coloured area represents the density of points in the empirical network of Institution 1. In red is the fitted model mean and the confidence interval computed from one hundred samples taken from the ensemble and smoothed using a moving average.}\label{fig:ansatzB}
\end{figure} 

Both the dcGM and the scGM rely on the simple hypothesis that strengths and degrees are positively correlated: if this were not the case for our datasets, then both models would perform poorly. We qualitatively assess the veracity of this hypothesis by checking how well in the empirical data this correlation holds at the various levels of sector definitions. In figures \ref{fig:ansatzA} and \ref{fig:ansatzB} we plotted the relationship between out-strengths and out-degrees as well as between in-strengths and in-degrees by sector, for an increasingly specific sector definition. We find that the fitness ansatz is well supported by the data in terms of trend and average, with both models being able to correctly capture the trend at both the aggregate and the sector-specific level. Notice, however, that the spread of data points around the mean is not reproduced by either model - the predicted variance of the degree being less than, or of the same order of magnitude of, its expected value; this discrepancy is again due to the fitness ansatz which leads our models to predict degrees that are, by construction, proportional to the strengths - hence the expected monotonic trend.


\subsection{Sector information}\label{sub:sect_info}
\begin{figure}[t!]
    \centering
    \subfloat[Institution 1\label{fig:aic_params_1}]{%
        \includegraphics[width=0.49\textwidth]{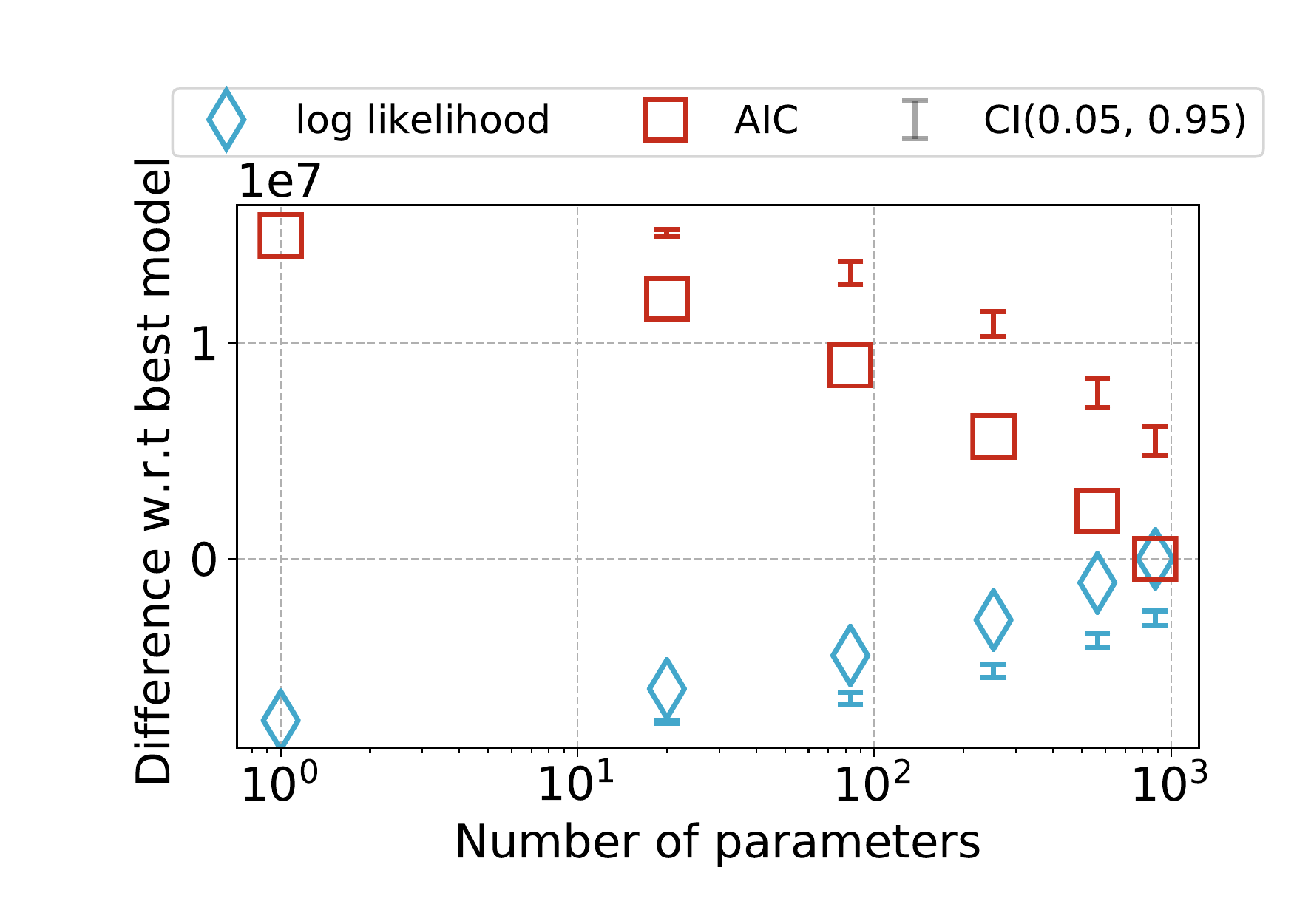}%
    }\hfil
    \subfloat[Institution 2\label{fig:aic_params_2}]{%
      \includegraphics[width=0.49\textwidth]{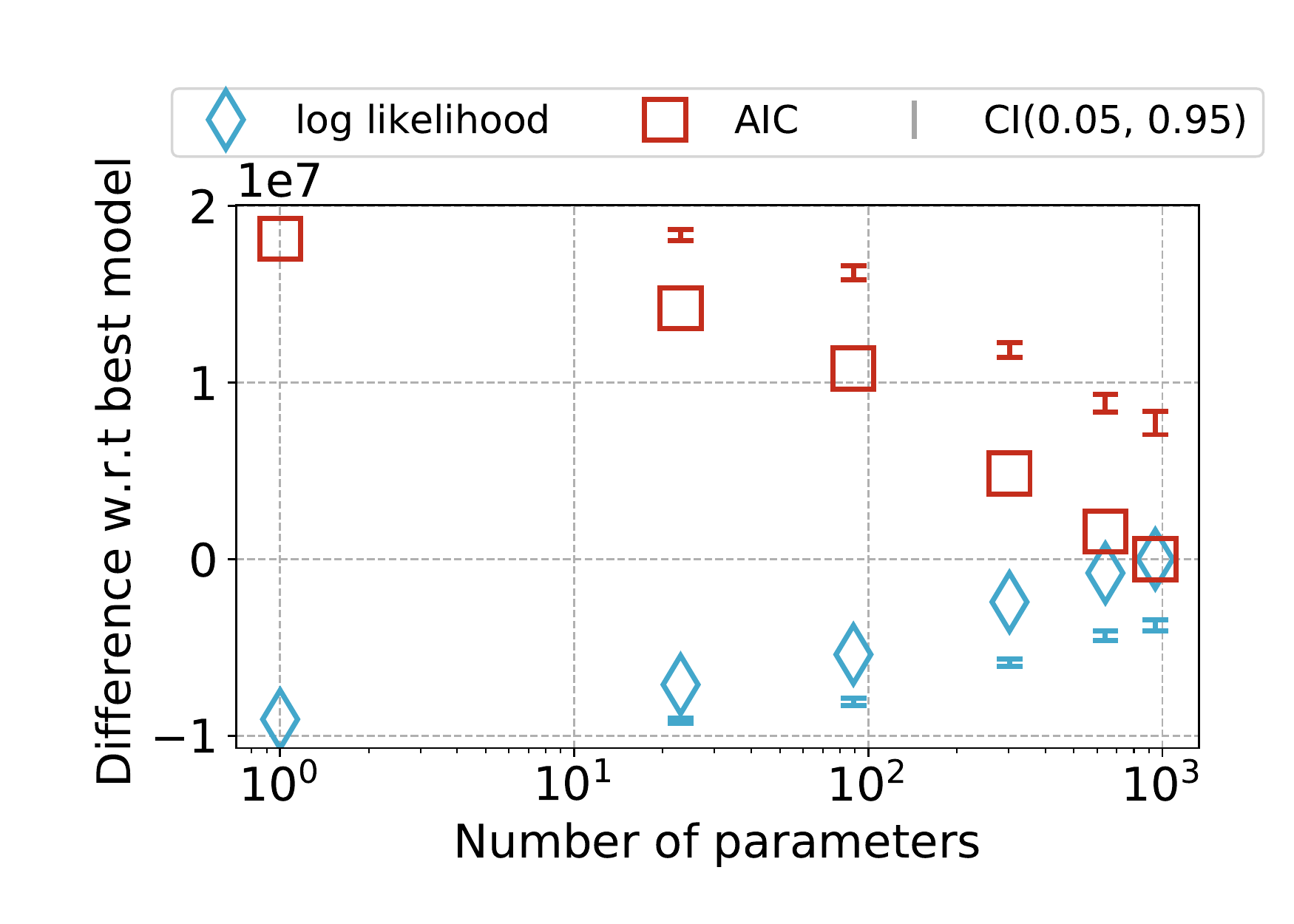}%
    }
    \caption{Difference in log likelihood and AIC of the fitted models for an increasing number of sectors. The model has one parameter per sector and the results are given with respect to the best performing model. The two plots correspond to the results computed on the datasets of the two collaborating institutions, while the confidence interval corresponds to the fitted models where the sector labels have been randomized.}\label{fig:aic_params}
\end{figure}

The main advantage in the formulation of the scGM is the introduction of information about the sectors. In order to quantitatively test the added value of introducing this structure, let us compare the likelihood of observing the binary component of the empirical graphs given the fitted ensembles defined at the various hierarchical sector levels. By considering the formulation of the scGM provided in equation \eqref{eq:multi_z} - notice that, in this case, the number of free parameters in the model is equal to the number of distinct sector labels. As figure \ref{fig:aic_params} shows, the scGM outperforms the dcGM for both networks, at all hierarchical levels: the increase in the log-likelihood, in fact, more than offsets the increase in number of parameters, as the Akaike Information Criterion (AIC) reveals.

Although the AIC allows us to rule out the possibility of over-fitting by increasing the number of free parameters, we also want to exclude the possibility that the introduction of any stripe-like meso-structure could lead to the same bias. To this end, we have compared the log-likelihood and the AIC of our scGM with a variant of it defined by randomly permuting all sector labels, while keeping the same number of sectors and sector densities. We have performed one hundred permutations and fitted the scGM to the graphs with permuted labels at the various hierarchical levels. The results are shown in figure \ref{fig:aic_params} as confidence interval and clearly demonstrate that the real sector dependencies carry much more information than random labels. 

To understand why this may be the case, one can think about the relationship that exists between the in-strength by sector and the entropy of the scGM. Every entry in the in-strength by sector sequence that is zero restricts the number of configurations of the system by disallowing any connection between the node and any firm in the sector: it is, then, clear that the entropy of this model is maximized when no entry of the in-strength by sector is zero. In the real data, the relationship between firms and sectors tends to be very sparse, with firms having, on average, a connection with less than ten sectors at the highest grouping resolution. Permuting the sector labels increases the probability that firms with the same label now belong to different groups: this, in turn, increases the number of non-zero $s_{g_i\rightarrow j}$ terms, hence increasing the entropy of the ensemble. 

Given that the likelihood in ERG models is equal to the negative entropy of the ensemble (see Supplementary Materials section \ref{sub:entropy}) we can also interpret the increase in likelihood in figure \ref{fig:aic_params} as a reduction in the entropy of the system, i.e. in the number of alternative network configurations that preserve the same numbers of in- and out-neighbours (along with their input and output flows) as estimated by the model. The reduction in entropy indicates that the model with a single sector or good leads to an over-estimation of the number of alternative configurations and therefore of the rewiring capability of the system. Finer product resolution improves our estimate of the entropy by disallowing connection between firms and the sectors that do not belong to their input. This is certainly true with respect to the dcGM, that permits interaction between firms in all sectors, however the permutation experiment confirms that the true industry classification minimizes, in a sense, the entropy of the ensemble. This is important as it provides ground for the scGM to be used as a generative model of inter-firm networks subject to production constraints and, by estimating the rewiring opportunities for firms, as an estimator of the resilience of the system to shocks and node removals.

\subsection{Structural properties}

\begin{figure}[ptb]
    \centering
    \subfloat[Institution 1\label{fig:dout_cdf_1}]{%
        \includegraphics[width=0.49\textwidth]{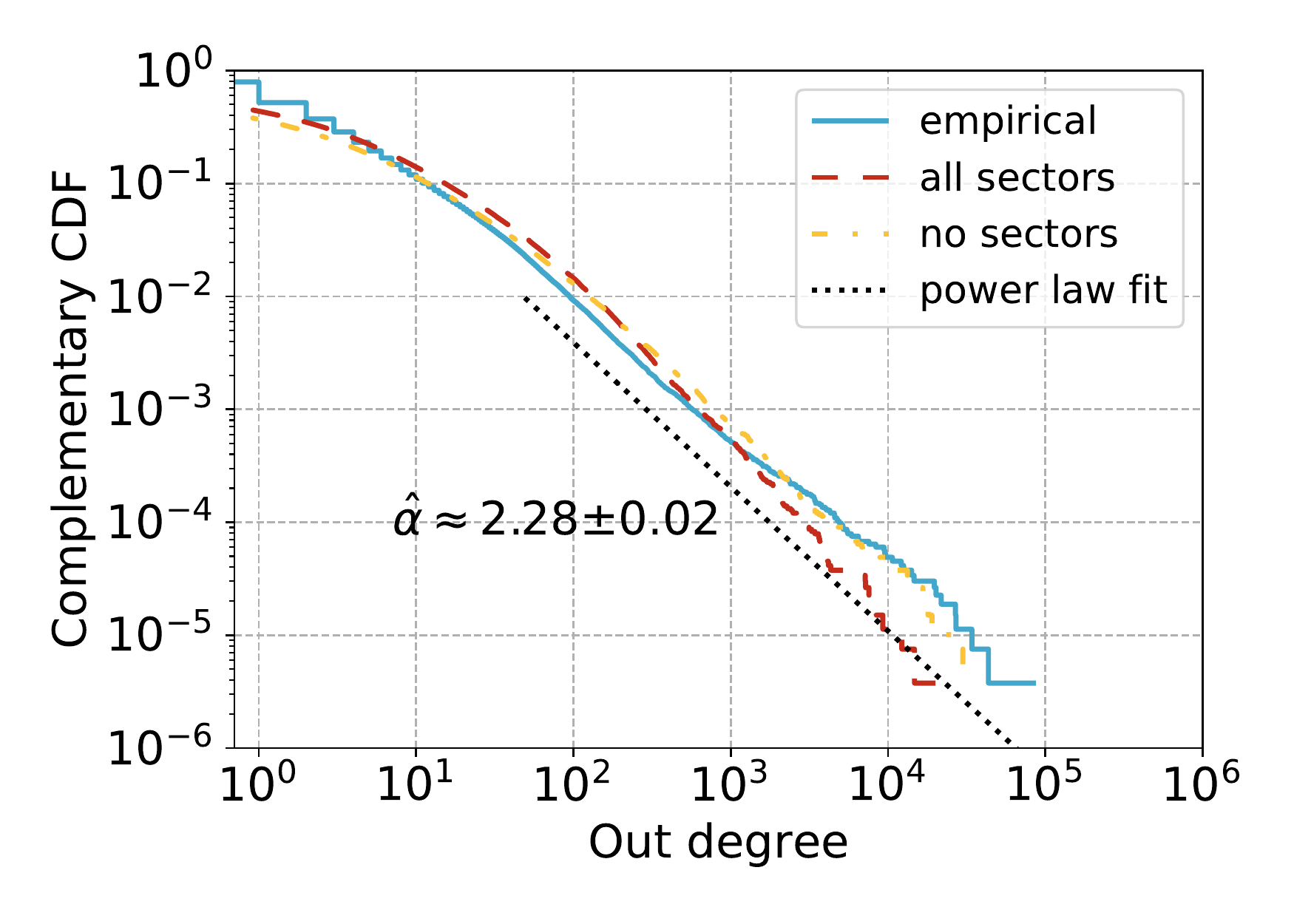}%
    }\hfil
    \subfloat[Institution 2\label{fig:dout_cdf_2}]{%
      \includegraphics[width=0.49\textwidth]{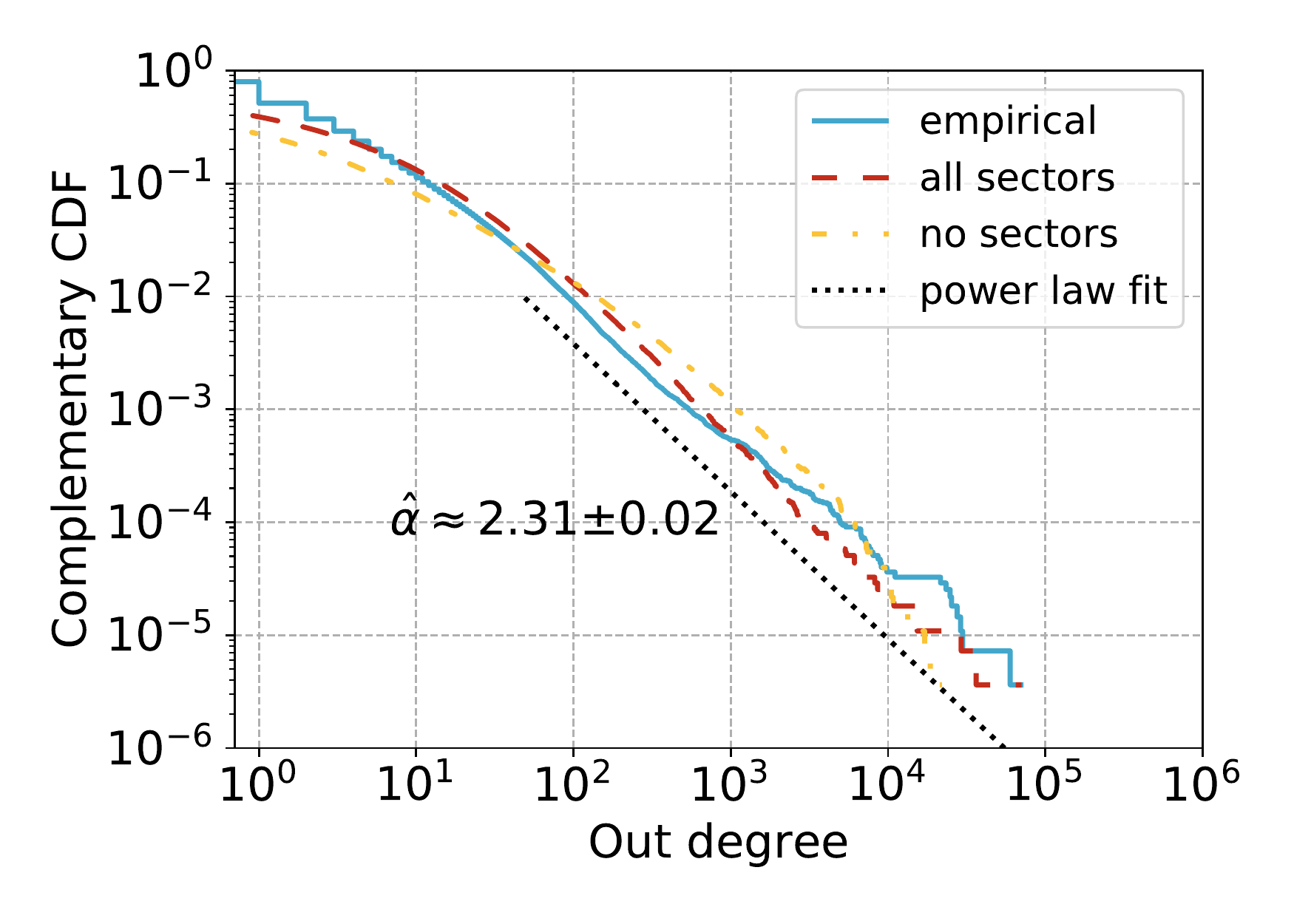}%
    }\vfil
    \subfloat[Institution 1\label{fig:din_cdf_1}]{%
        \includegraphics[width=0.49\textwidth]{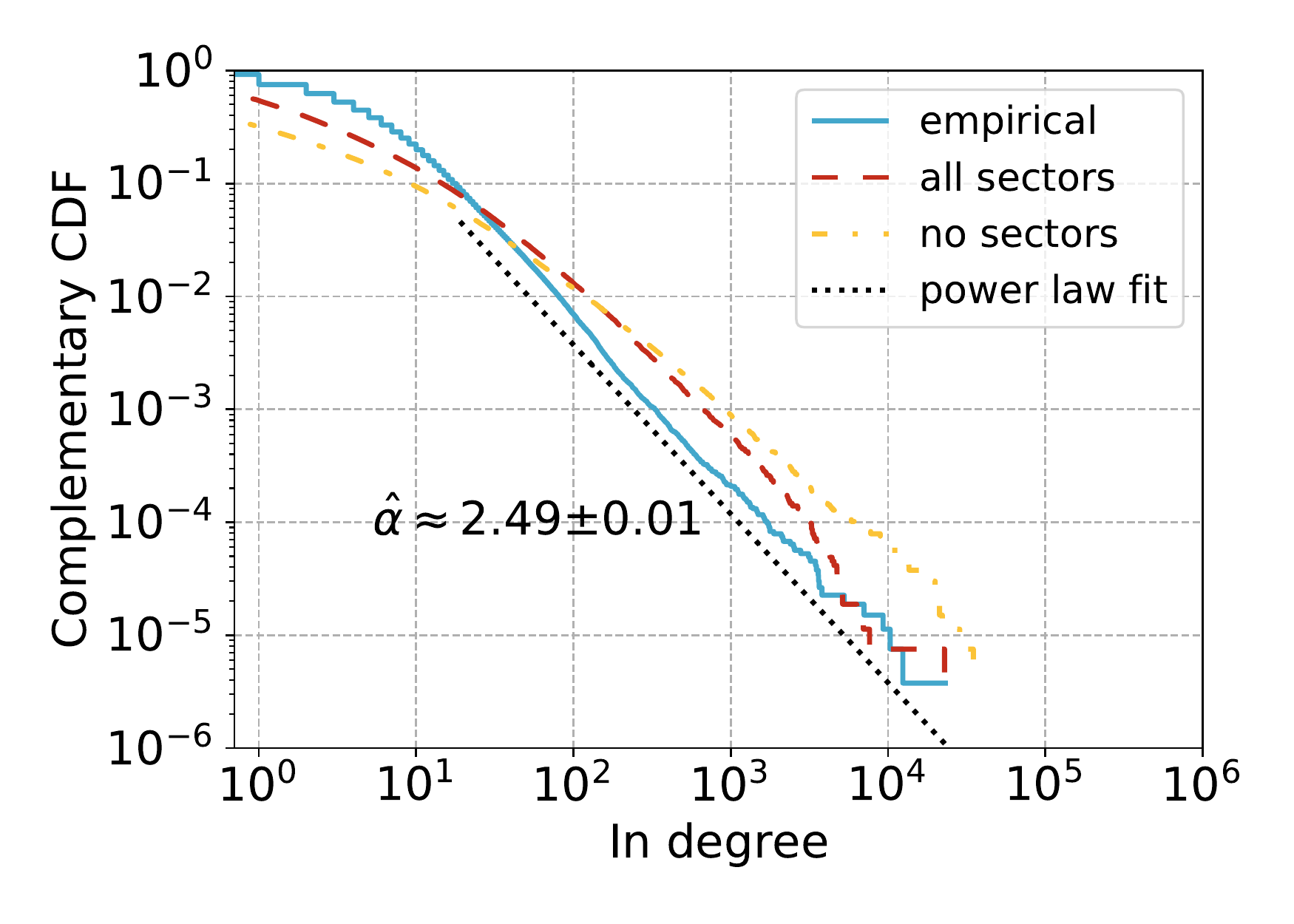}%
    }\hfil
    \subfloat[Institution 2\label{fig:din_cdf_2}]{%
      \includegraphics[width=0.49\textwidth]{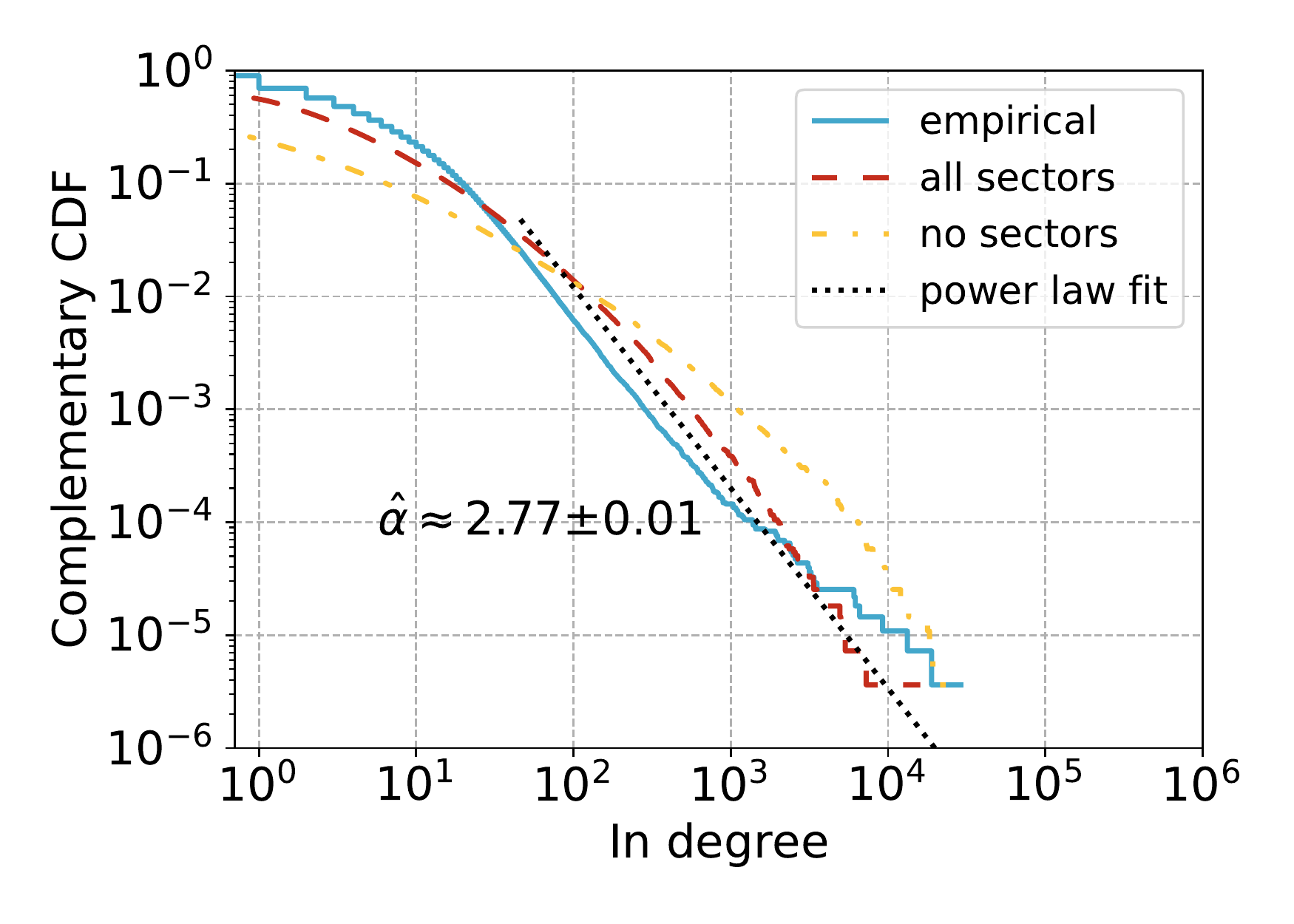}%
    }
    \caption{Complementary cumulative distributions of the out- and in-degree for both networks.}\label{fig:d_cdf}
\end{figure} 

We can further assess the correspondence between model and empirical networks by looking at several structural properties and macroscopic trends.

In figure \ref{fig:d_cdf} we note that both the dcGM (no sectors) and the scGM applied with the level 5 sector definitions (all sectors) accurately reproduce the cumulative distributions of the in- and out-degrees of the empirical networks with only minor differences, e.g. the scGM performs better in reproducing the distribution of in-degrees, especially in the extremes.

We also test which distribution, between exponential, log-normal and power-law, best fits the data using a likelihood ratio test. In all four cases the empirical distribution is fat-tailed with the hypothesis of following an exponential distribution being rejected with a p-value that is less than $10^{-6}$. We find mixed support for power law and log-normal distributions, with a slight preference for the power law: in \ref{fig:dout_cdf_1} power-law (with p-value $0.11$), in \ref{fig:dout_cdf_2} power-law ($0.01$), in \ref{fig:din_cdf_1} log-normal ($0.01$), and in \ref{fig:din_cdf_2} power-law ($0.04$).

\begin{figure}[t!]
    \centering
    \subfloat[Institution 1\label{fig:ksd_params_1}]{%
        \includegraphics[width=0.49\textwidth]{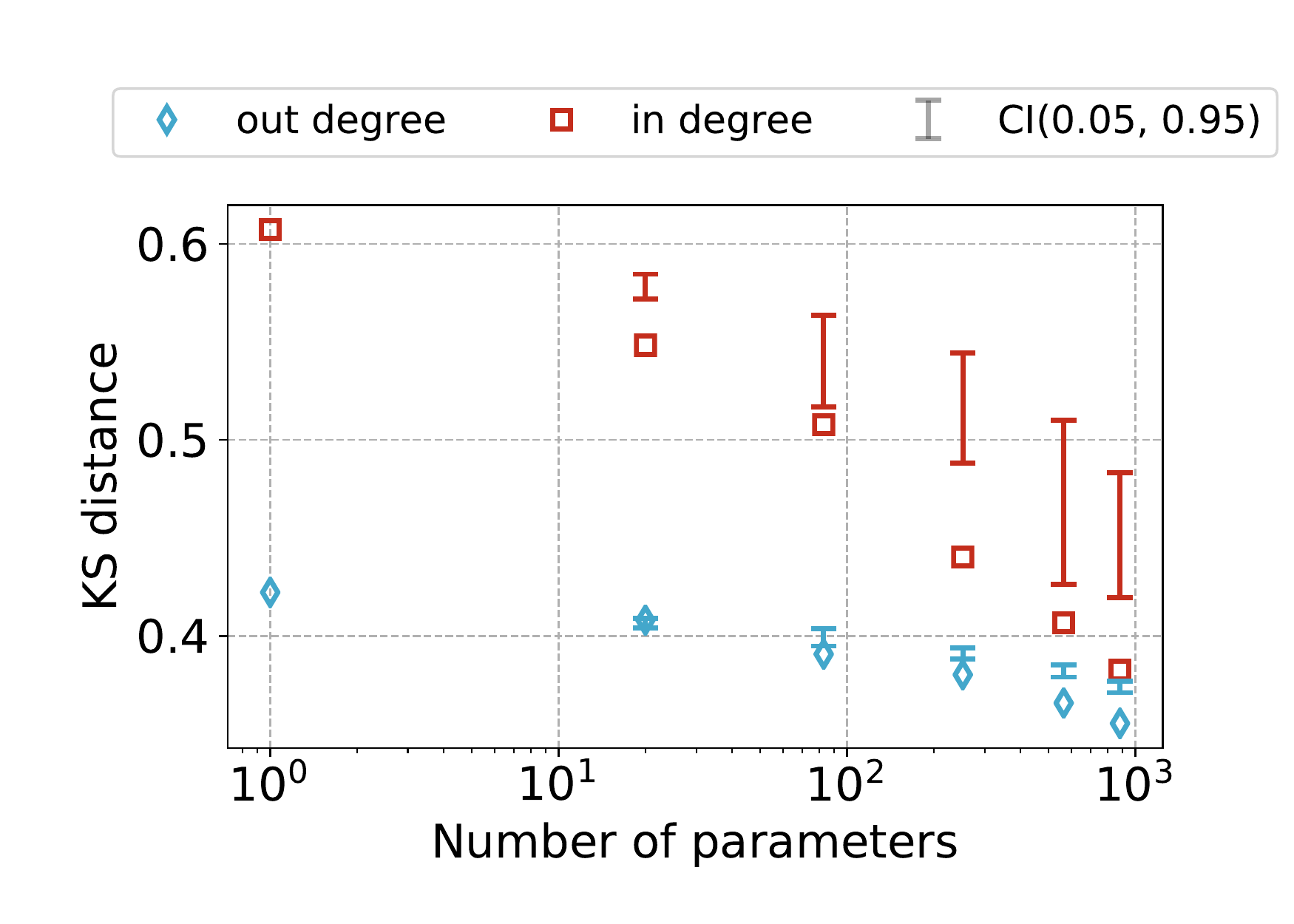}%
    }\hfil
    \subfloat[Institution 2\label{fig:ksd_params_2}]{%
      \includegraphics[width=0.49\textwidth]{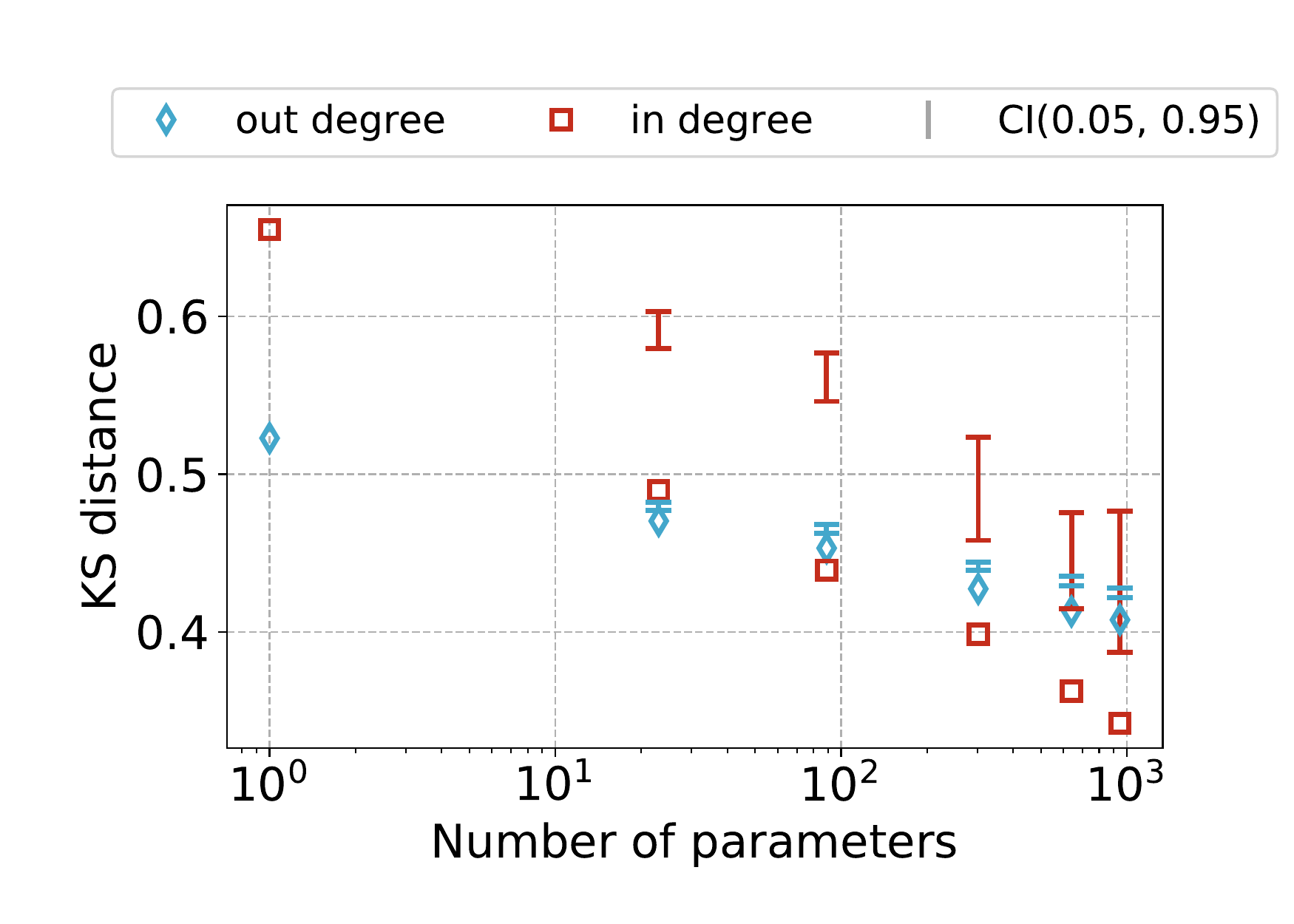}%
    }
    \caption{Kolmogorov Smirnoff distance of the model degree distribution with respect to the empirical one. The confidence interval corresponds to the fitted models where the sector labels had been randomized. Note that in all cases the hypothesis that they belong to the same distribution is rejected.}\label{fig:ksd_params}
\end{figure} 

We, then, quantitatively assess the improvement in the quality of fit of the degree distributions obtained by introducing the sector labels by using the Kolmogorov–Smirnov test. Similarly to what discussed in section \ref{sub:sect_info}, we also perform the test on the models with permuted labels. From the results in figure \ref{fig:ksd_params} we again find a significant improvement in the quality of fit for increasing sector information. We note, as expected, that the effect is larger on the in-degree since it is the in-strength that is being divided by sector.

\begin{figure}[tbp]
    \centering
    \subfloat[dcGM \label{fig:sl_1_dc}]{%
    \includegraphics[width=0.9\textwidth]{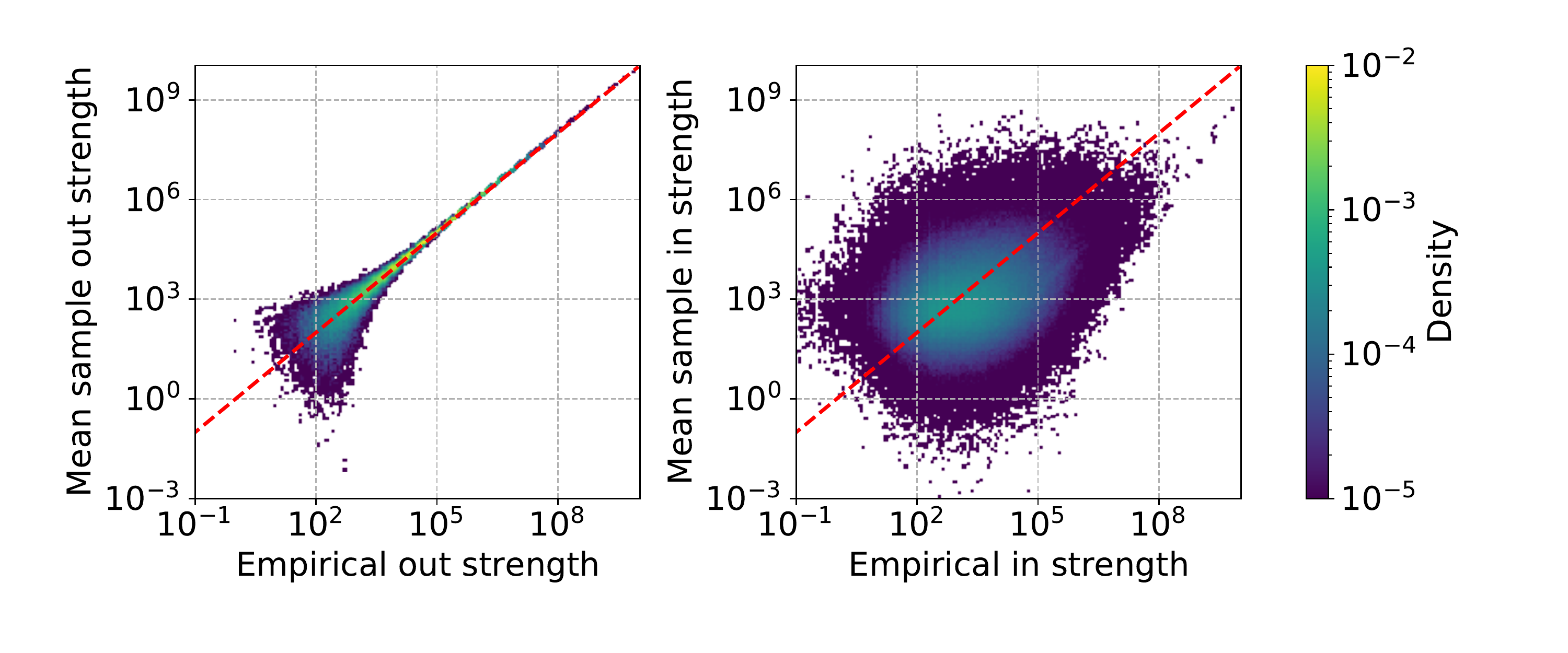}%
    }\vfil
    \subfloat[scGM \label{fig:sl_1_sc}]{%
    \includegraphics[width=0.9\textwidth]{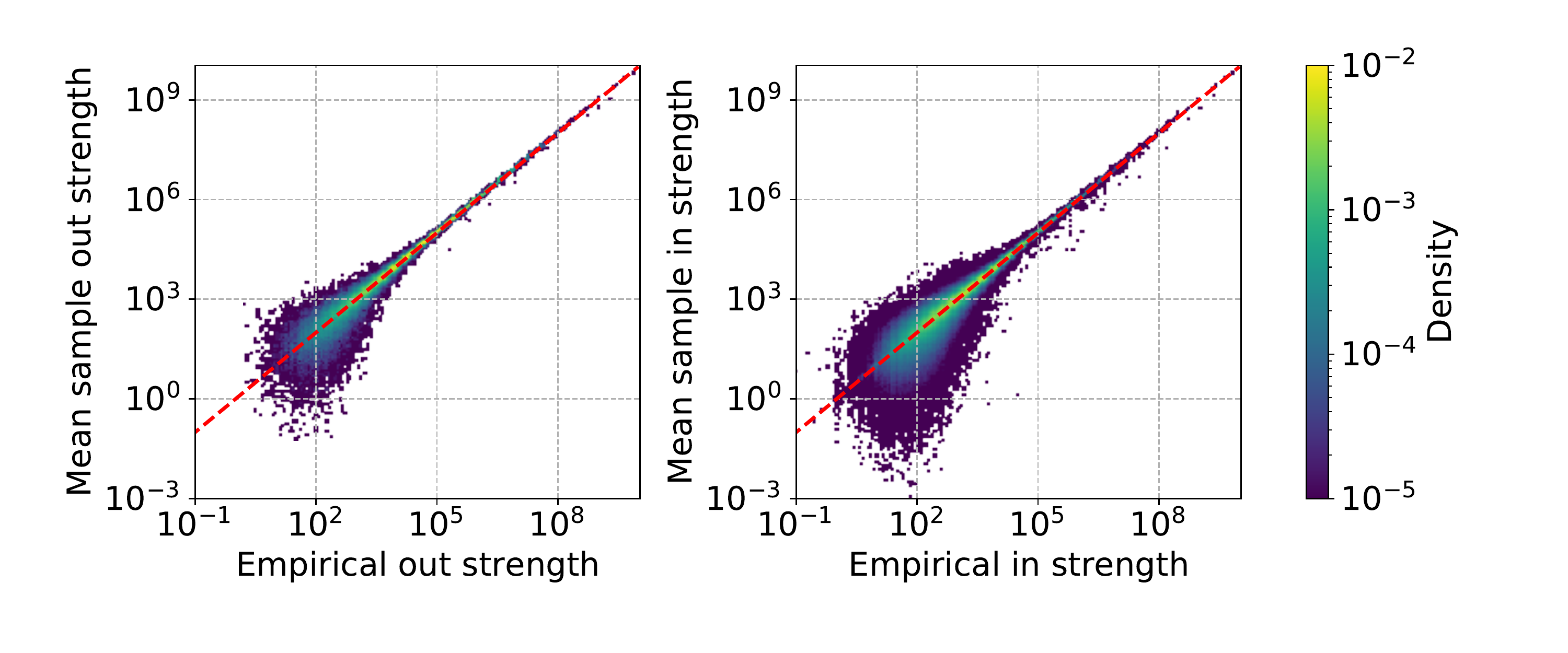}%
    }
    \caption{Empirical versus expected strength by sector for the network of Institution 1.}\label{fig:sl_1}
\end{figure} 

So far, we have shown only properties that are not defined to be sector specific. However, the main difference between the models lies precisely in the ability to preserve sector information. We can visualize this difference by looking at the empirical strength by sector and at the average one, computed over one hundred samples.

The comparison is shown in figure \ref{fig:sl_1}: the plots reveal three important aspects about the way the dcGM and the scGM work. First, we note that there is almost no difference in the accuracy of the out-strength. This is a consequence of the way the scGM is formulated, with the out-strength being always the total rather than the sector-specific one, such that each node will have out-strength by sector equal to zero in all sector except the one they belong to.

\begin{figure}[tbp]
    \centering
    \subfloat[Out-out]{%
        \includegraphics[width=0.49\textwidth]{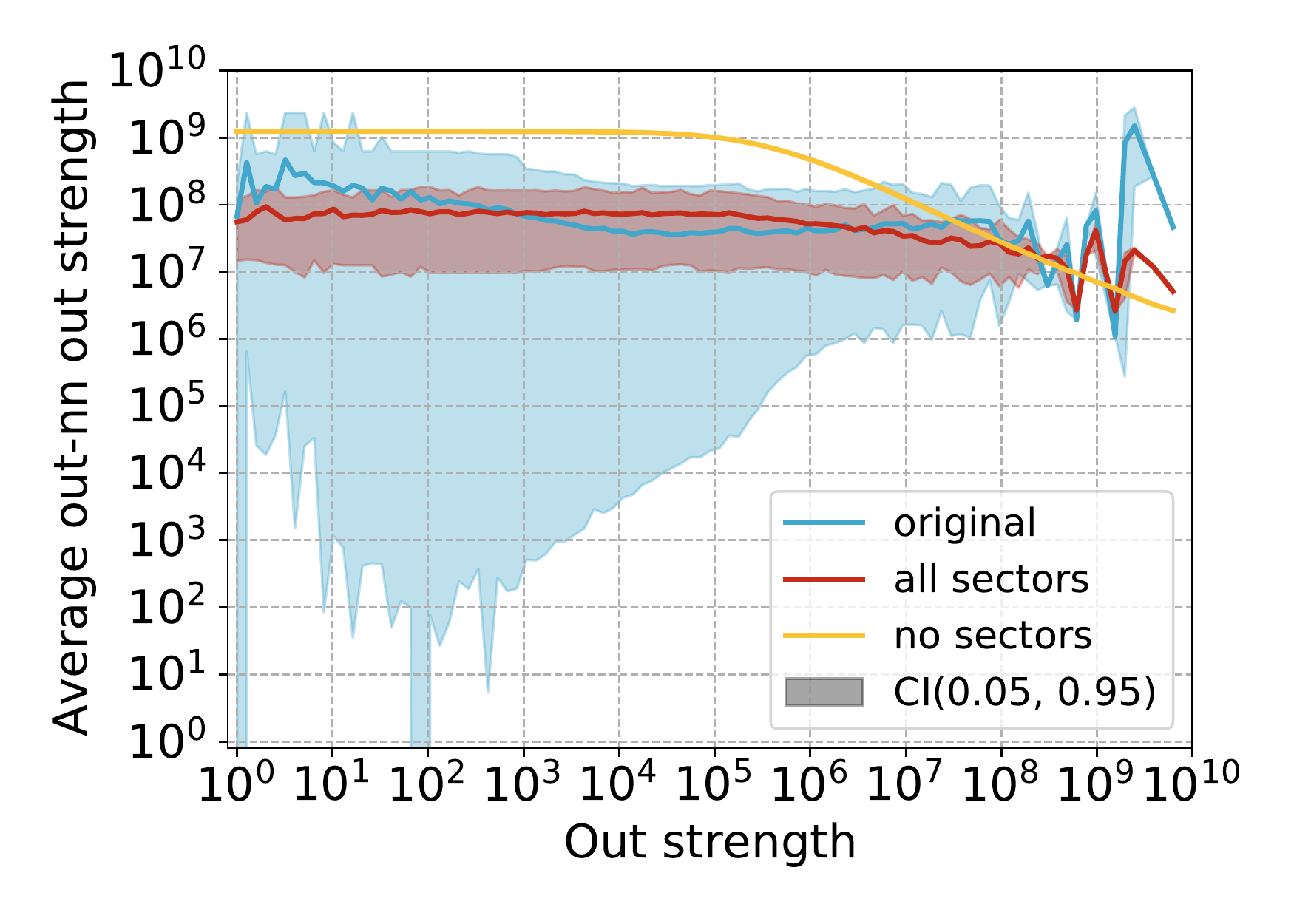}%
    }\hfil
    \subfloat[In-in]{%
      \includegraphics[width=0.49\textwidth]{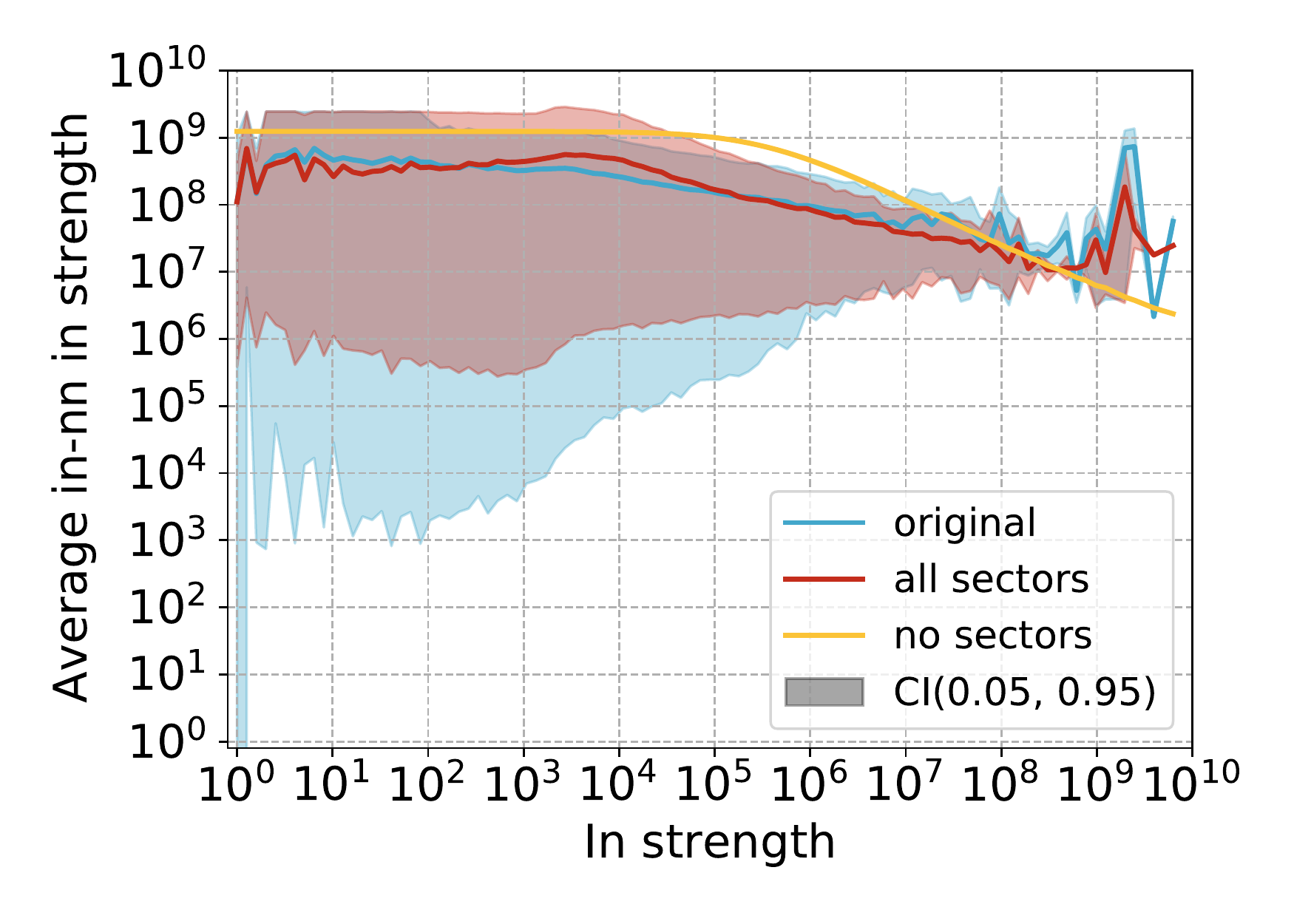}%
    }
    \caption{Average nearest neighbour strength by node strength for Institution 1. The full lines are computed as the average over all firms in the logarithmic binning of the x-axis. The confidence interval is the empirical interval measured over the same bins.}\label{fig:snn_1}
\end{figure} 

\begin{figure}[tp]
    \centering
    \subfloat[Out-out]{%
        \includegraphics[width=0.9\textwidth]{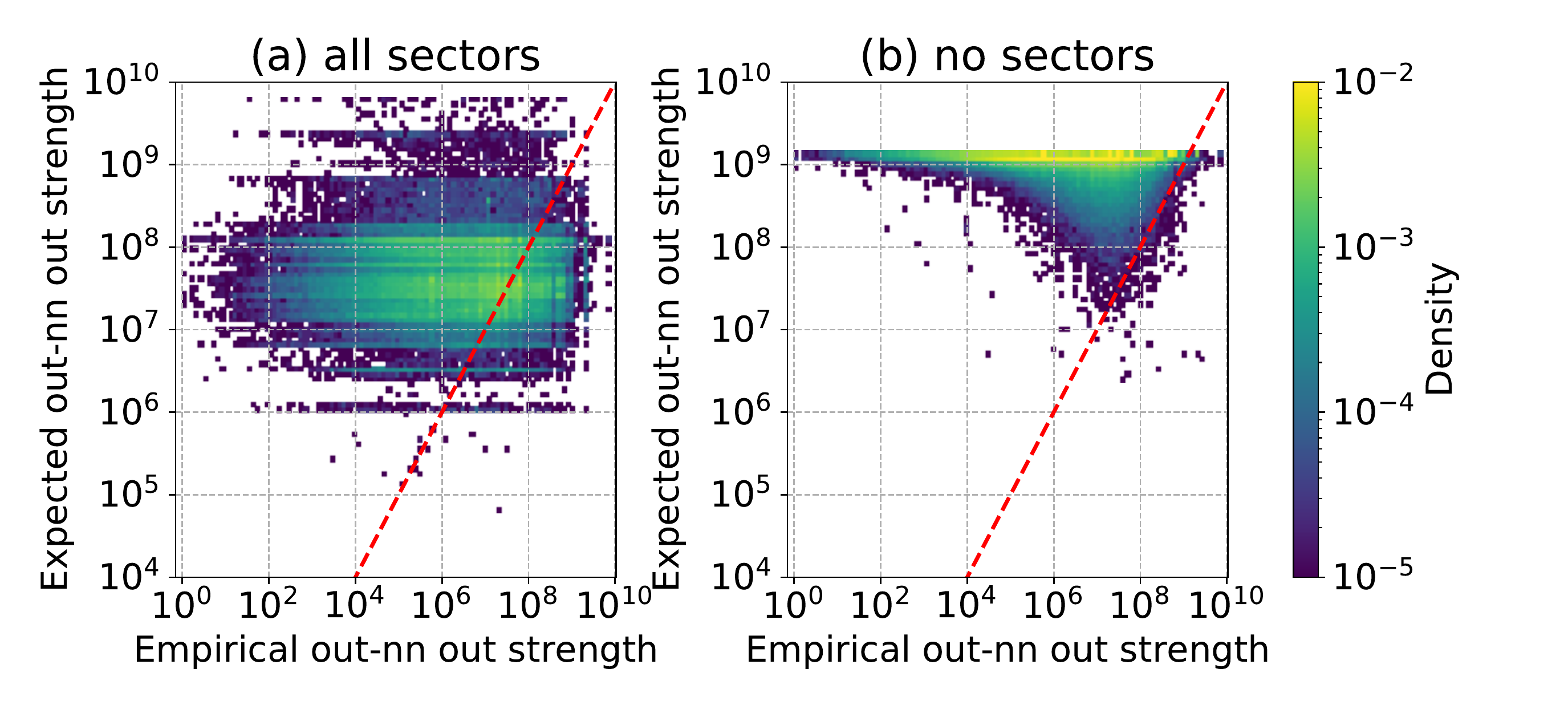}%
    }\vfil
    \subfloat[In-in]{%
      \includegraphics[width=0.9\textwidth]{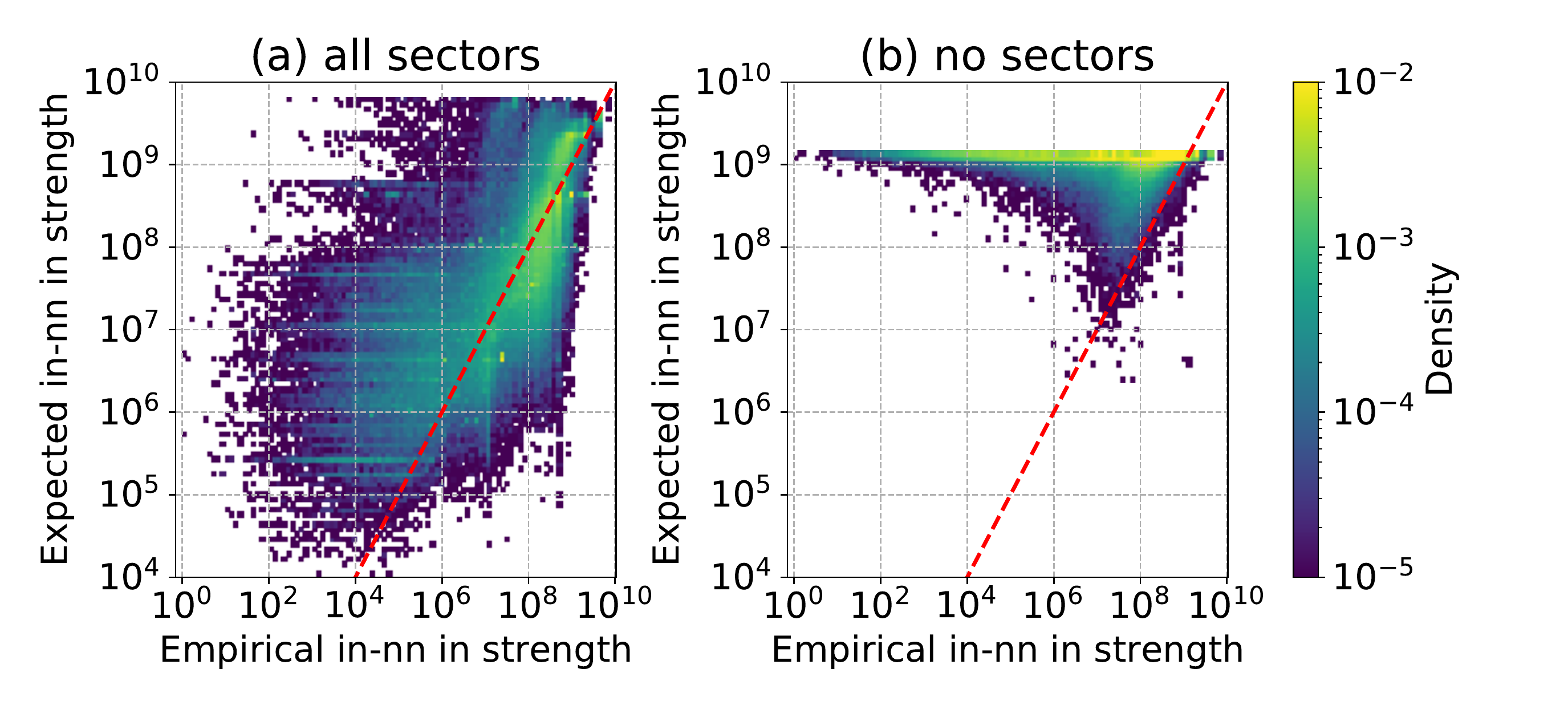}%
    }
    \caption{Per node comparison of the average nearest neighbour strength for Institution 1. The shaded area represents the density of points in the two-dimensional logarithmic binning.}\label{fig:snn_1_node}
\end{figure} 

\begin{figure}[tp]
    \centering
    \includegraphics[width=0.9\textwidth]{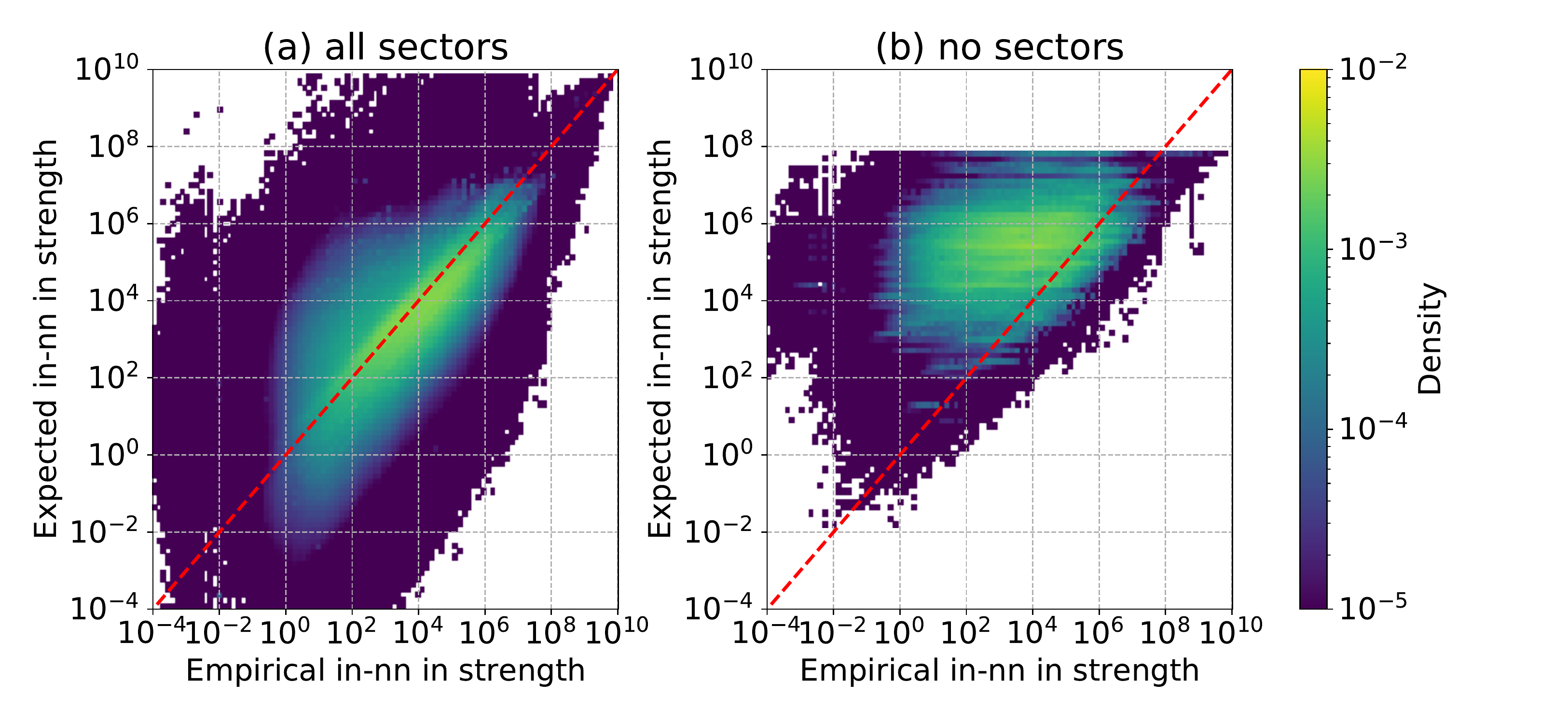}
    \caption{Per node comparison of the average nearest in-neighbour in-strength by sector for Institution 1. The shaded area represents the density of points in the two-dimensional logarithmic binning.}\label{fig:slnn_1_node}
\end{figure}

A second observation concerns the difference in variance between low and high values of strength. This spread is a consequence of limited sampling: indeed the strength is preserved by definition as shown in equations \eqref{eqs_out} and \eqref{eq:sl_in}. However, this is true only on average and not for each specific sampled configuration. The sample variance for these measures is much higher for low values of strength because the probability of observing the link is lower. This means that more often no link will be sampled and that the expected value of the weight given that the link is sampled is further from the expected value as can be deduced from the presence of $p_{ij}$ in the denominator of equation \eqref{eq:exp_sc_weight}. These two mechanism give rise to the higher variance that we observe.

Finally, and most importantly, the way the in-strength is distributed by sector in the dcGM does not resemble the information in the original graph. Note that we are only showing the values when the empirical in-strength is non-zero. The dcGM however almost always assigns a positive value even if empirically it is zero. This is an important point because it tells us a key point in the weakness of the dcGM that motivates the formulation of the scGM: the dcGM is unable to correctly reconstruct the neighbourhood of a node in terms of sectors.

We further explore the quality of reconstruction of each node neighbourhood by looking at the average nearest neighbours strength patterns in the original and reconstructed ensembles. In figure \ref{fig:snn_1} we show the pattern for the out-out and in-in average nearest neighbours strength. We can clearly observe that the scGM (red) follows more closely the empirical trend (blue) than the dcGM (yellow). Furthermore, the narrow confidence interval for the dcGM is due to the fact that the dcGM assigns exactly the same neighbourhood to nodes with the same strength no matter what is the sector they belong to. This is clearly unrealistic and proves that the dcGM is not suitable when applied to production networks.

We can observe this phenomenon also via a direct node-specific comparison of the average nearest neighbours strength shown. Both the right plots in figure \ref{fig:snn_1_node} show that the majority of nodes is assigned a high value of the average nearest neighbours strength because the majority of nodes are likely connected to nodes with highest total strength. By introducing restrictions by sector, the scGM still preserves the same trend, but only within the correct sector. This allows the scGM to predict the neighbours of the nodes and, hence, their properties more accurately. This difference between the dcGM and the scGM is even more evident in figure \ref{fig:slnn_1_node} where we have repeated the same comparison per node but splitting each node strength by its sector components. This allows us to visualize the error in reconstructing the in-strength by sector of the in-neighbours of each node. It is evident that the scGM performs significantly better than the dcGM.

\section{Conclusions}

Production networks are extremely important in understanding the response of an economic system to shocks but the scarcity of relevant data makes this analysis difficult. In this work we proposed a methodology that, based on limited information on firms revenues and expenses by industry origin, can accurately reproduce key structural properties of the original network. We have shown that by introducing this information we can better reproduce macroscopic properties such as the degrees and the average nearest neighbours strength. It is important to remember that being able to estimate these properties is key in any application concerning production networks. In particular capturing the correct statistical behaviour of the average nearest in-neighbour in-strength by sector implies a realistic estimation of sector composition of two-steps in a supply chain. Of course, we cannot expect the scGM to reproduce the neighbourhood exactly with the extremely limited information used to generate the ensemble; nevertheless, it is very encouraging that such a simple methodology can reproduce the average trend so accurately.

The main improvement over previous approaches is given by the ability of the scGM to correctly estimate the true rewiring capabilities of the system. In fact, ignoring the sector specificity of firms leads to an overestimation of the possible interaction between firms which, in practical applications, might result in biased estimates of the resilience of the system. As the scGM is able to correctly encode this information, it is a good candidate to be used for generating realistic probability distributions for inter-firm networks subject to fine production constraints. As such it can also be used to correctly estimate the entropy of the ensemble as a proxy for the rewiring capabilities of the system.

The analytical tractability of this model allows us to find many expected properties over the ensemble without having to generate samples. For all other applications, such as shock propagation exercises, samples can be generated efficiently and without bias. We have shown that this methodology is a considerable improvement with respect to the previous best model, provided that sector information is available at firm level - although this constitutes the main limitation of our approach, it also clearly points out how important this information is in understanding the structure of production networks.

\section{Acknowledgements}

T.S. acknowledges support from the project Artes 4.0 - Advanced Robotics and enabling digital Technologies \& Systems. D.G. acknowledges support from the Dutch Econophysics Foundation (Stichting Econophysics, Leiden, the Netherlands). T.S. and D.G. also acknowledge support from the European Union Horizon 2020 Program under the scheme `INFRAIA-01-2018-2019 - Integrating Activities for Advanced Communities', Grant Agreement n. 871042, `SoBigData++: European Integrated Infrastructure for Social Mining and Big Data Analytics'.

\bibliography{recon}

\begin{thebibliography}{19}%
\makeatletter
\providecommand \@ifxundefined [1]{%
 \@ifx{#1\undefined}
}%
\providecommand \@ifnum [1]{%
 \ifnum #1\expandafter \@firstoftwo
 \else \expandafter \@secondoftwo
 \fi
}%
\providecommand \@ifx [1]{%
 \ifx #1\expandafter \@firstoftwo
 \else \expandafter \@secondoftwo
 \fi
}%
\providecommand \natexlab [1]{#1}%
\providecommand \enquote  [1]{``#1''}%
\providecommand \bibnamefont  [1]{#1}%
\providecommand \bibfnamefont [1]{#1}%
\providecommand \citenamefont [1]{#1}%
\providecommand \href@noop [0]{\@secondoftwo}%
\providecommand \href [0]{\begingroup \@sanitize@url \@href}%
\providecommand \@href[1]{\@@startlink{#1}\@@href}%
\providecommand \@@href[1]{\endgroup#1\@@endlink}%
\providecommand \@sanitize@url [0]{\catcode `\\12\catcode `\$12\catcode
  `\&12\catcode `\#12\catcode `\^12\catcode `\_12\catcode `\%12\relax}%
\providecommand \@@startlink[1]{}%
\providecommand \@@endlink[0]{}%
\providecommand \url  [0]{\begingroup\@sanitize@url \@url }%
\providecommand \@url [1]{\endgroup\@href {#1}{\urlprefix }}%
\providecommand \urlprefix  [0]{URL }%
\providecommand \Eprint [0]{\href }%
\providecommand \doibase [0]{https://doi.org/}%
\providecommand \selectlanguage [0]{\@gobble}%
\providecommand \bibinfo  [0]{\@secondoftwo}%
\providecommand \bibfield  [0]{\@secondoftwo}%
\providecommand \translation [1]{[#1]}%
\providecommand \BibitemOpen [0]{}%
\providecommand \bibitemStop [0]{}%
\providecommand \bibitemNoStop [0]{.\EOS\space}%
\providecommand \EOS [0]{\spacefactor3000\relax}%
\providecommand \BibitemShut  [1]{\csname bibitem#1\endcsname}%
\let\auto@bib@innerbib\@empty
\bibitem [{\citenamefont {Carvalho}\ and\ \citenamefont
  {Tahbaz-Salehi}(2019)}]{carvalho2019production}%
  \BibitemOpen
  \bibfield  {author} {\bibinfo {author} {\bibfnamefont {V.~M.}\ \bibnamefont
  {Carvalho}}\ and\ \bibinfo {author} {\bibfnamefont {A.}~\bibnamefont
  {Tahbaz-Salehi}},\ }\href@noop {} {\bibfield  {journal} {\bibinfo  {journal}
  {Annual Review of Economics}\ }\textbf {\bibinfo {volume} {11}},\ \bibinfo
  {pages} {635} (\bibinfo {year} {2019})}\BibitemShut {NoStop}%
\bibitem [{\citenamefont {Carvalho}\ \emph {et~al.}(2021)\citenamefont
  {Carvalho}, \citenamefont {Nirei}, \citenamefont {Saito},\ and\ \citenamefont
  {Tahbaz-Salehi}}]{carvalho2021jap}%
  \BibitemOpen
  \bibfield  {author} {\bibinfo {author} {\bibfnamefont {V.~M.}\ \bibnamefont
  {Carvalho}}, \bibinfo {author} {\bibfnamefont {M.}~\bibnamefont {Nirei}},
  \bibinfo {author} {\bibfnamefont {Y.~U.}\ \bibnamefont {Saito}},\ and\
  \bibinfo {author} {\bibfnamefont {A.}~\bibnamefont {Tahbaz-Salehi}},\
  }\href@noop {} {\bibfield  {journal} {\bibinfo  {journal} {The Quarterly
  Journal of Economics}\ }\textbf {\bibinfo {volume} {136}},\ \bibinfo {pages}
  {1255} (\bibinfo {year} {2021})}\BibitemShut {NoStop}%
\bibitem [{\citenamefont {Kashiwagi}\ \emph {et~al.}(2021)\citenamefont
  {Kashiwagi}, \citenamefont {Todo},\ and\ \citenamefont
  {Matous}}]{kashiwagi2021propagation}%
  \BibitemOpen
  \bibfield  {author} {\bibinfo {author} {\bibfnamefont {Y.}~\bibnamefont
  {Kashiwagi}}, \bibinfo {author} {\bibfnamefont {Y.}~\bibnamefont {Todo}},\
  and\ \bibinfo {author} {\bibfnamefont {P.}~\bibnamefont {Matous}},\
  }\href@noop {} {\bibfield  {journal} {\bibinfo  {journal} {Review of
  International Economics}\ } (\bibinfo {year} {2021})}\BibitemShut {NoStop}%
\bibitem [{\citenamefont {Acemoglu}\ \emph {et~al.}(2012)\citenamefont
  {Acemoglu}, \citenamefont {Carvalho}, \citenamefont {Ozdaglar},\ and\
  \citenamefont {Tahbaz-Salehi}}]{acemoglu2012net}%
  \BibitemOpen
  \bibfield  {author} {\bibinfo {author} {\bibfnamefont {D.}~\bibnamefont
  {Acemoglu}}, \bibinfo {author} {\bibfnamefont {V.~M.}\ \bibnamefont
  {Carvalho}}, \bibinfo {author} {\bibfnamefont {A.}~\bibnamefont {Ozdaglar}},\
  and\ \bibinfo {author} {\bibfnamefont {A.}~\bibnamefont {Tahbaz-Salehi}},\
  }\href@noop {} {\bibfield  {journal} {\bibinfo  {journal} {Econometrica}\
  }\textbf {\bibinfo {volume} {80}},\ \bibinfo {pages} {1977} (\bibinfo {year}
  {2012})}\BibitemShut {NoStop}%
\bibitem [{\citenamefont {Pichler}\ and\ \citenamefont
  {Farmer}(2021)}]{pichler2021shock}%
  \BibitemOpen
  \bibfield  {author} {\bibinfo {author} {\bibfnamefont {A.}~\bibnamefont
  {Pichler}}\ and\ \bibinfo {author} {\bibfnamefont {J.~D.}\ \bibnamefont
  {Farmer}},\ }\href@noop {} {\bibfield  {journal} {\bibinfo  {journal}
  {Economic Systems Research}\ ,\ \bibinfo {pages} {1}} (\bibinfo {year}
  {2021})}\BibitemShut {NoStop}%
\bibitem [{\citenamefont {Dawid}\ and\ \citenamefont
  {Delli~Gatti}(2018)}]{dawid2018agent}%
  \BibitemOpen
  \bibfield  {author} {\bibinfo {author} {\bibfnamefont {H.}~\bibnamefont
  {Dawid}}\ and\ \bibinfo {author} {\bibfnamefont {D.}~\bibnamefont
  {Delli~Gatti}},\ }\href@noop {} {\bibfield  {journal} {\bibinfo  {journal}
  {Handbook of computational economics}\ }\textbf {\bibinfo {volume} {4}},\
  \bibinfo {pages} {63} (\bibinfo {year} {2018})}\BibitemShut {NoStop}%
\bibitem [{\citenamefont {Bak}\ \emph {et~al.}(1993)\citenamefont {Bak},
  \citenamefont {Chen}, \citenamefont {Scheinkman},\ and\ \citenamefont
  {Woodford}}]{bak1993sect}%
  \BibitemOpen
  \bibfield  {author} {\bibinfo {author} {\bibfnamefont {P.}~\bibnamefont
  {Bak}}, \bibinfo {author} {\bibfnamefont {K.}~\bibnamefont {Chen}}, \bibinfo
  {author} {\bibfnamefont {J.}~\bibnamefont {Scheinkman}},\ and\ \bibinfo
  {author} {\bibfnamefont {M.}~\bibnamefont {Woodford}},\ }\href@noop {}
  {\bibfield  {journal} {\bibinfo  {journal} {Ricerche economiche}\ }\textbf
  {\bibinfo {volume} {47}},\ \bibinfo {pages} {3} (\bibinfo {year}
  {1993})}\BibitemShut {NoStop}%
\bibitem [{\citenamefont {Squartini}\ \emph {et~al.}(2018)\citenamefont
  {Squartini}, \citenamefont {Caldarelli}, \citenamefont {Cimini},
  \citenamefont {Gabrielli},\ and\ \citenamefont
  {Garlaschelli}}]{squartini2018reconstruction}%
  \BibitemOpen
  \bibfield  {author} {\bibinfo {author} {\bibfnamefont {T.}~\bibnamefont
  {Squartini}}, \bibinfo {author} {\bibfnamefont {G.}~\bibnamefont
  {Caldarelli}}, \bibinfo {author} {\bibfnamefont {G.}~\bibnamefont {Cimini}},
  \bibinfo {author} {\bibfnamefont {A.}~\bibnamefont {Gabrielli}},\ and\
  \bibinfo {author} {\bibfnamefont {D.}~\bibnamefont {Garlaschelli}},\
  }\href@noop {} {\bibfield  {journal} {\bibinfo  {journal} {Physics reports}\
  }\textbf {\bibinfo {volume} {757}},\ \bibinfo {pages} {1} (\bibinfo {year}
  {2018})}\BibitemShut {NoStop}%
\bibitem [{\citenamefont {Anand}\ \emph {et~al.}(2018)\citenamefont {Anand},
  \citenamefont {van Lelyveld}, \citenamefont {Banai}, \citenamefont
  {Friedrich}, \citenamefont {Garratt}, \citenamefont {Ha{\l}aj}, \citenamefont
  {Fique}, \citenamefont {Hansen}, \citenamefont {Jaramillo}, \citenamefont
  {Lee} \emph {et~al.}}]{anand2018missing}%
  \BibitemOpen
  \bibfield  {author} {\bibinfo {author} {\bibfnamefont {K.}~\bibnamefont
  {Anand}}, \bibinfo {author} {\bibfnamefont {I.}~\bibnamefont {van Lelyveld}},
  \bibinfo {author} {\bibfnamefont {{\'A}.}~\bibnamefont {Banai}}, \bibinfo
  {author} {\bibfnamefont {S.}~\bibnamefont {Friedrich}}, \bibinfo {author}
  {\bibfnamefont {R.}~\bibnamefont {Garratt}}, \bibinfo {author} {\bibfnamefont
  {G.}~\bibnamefont {Ha{\l}aj}}, \bibinfo {author} {\bibfnamefont
  {J.}~\bibnamefont {Fique}}, \bibinfo {author} {\bibfnamefont
  {I.}~\bibnamefont {Hansen}}, \bibinfo {author} {\bibfnamefont {S.~M.}\
  \bibnamefont {Jaramillo}}, \bibinfo {author} {\bibfnamefont {H.}~\bibnamefont
  {Lee}}, \emph {et~al.},\ }\href@noop {} {\bibfield  {journal} {\bibinfo
  {journal} {Journal of Financial Stability}\ }\textbf {\bibinfo {volume}
  {35}},\ \bibinfo {pages} {107} (\bibinfo {year} {2018})}\BibitemShut
  {NoStop}%
\bibitem [{\citenamefont {Mazzarisi}\ and\ \citenamefont
  {Lillo}(2017)}]{mazzarisi2017methods}%
  \BibitemOpen
  \bibfield  {author} {\bibinfo {author} {\bibfnamefont {P.}~\bibnamefont
  {Mazzarisi}}\ and\ \bibinfo {author} {\bibfnamefont {F.}~\bibnamefont
  {Lillo}},\ }in\ \href@noop {} {\emph {\bibinfo {booktitle} {Econophysics and
  Sociophysics: Recent Progress and Future Directions}}}\ (\bibinfo
  {publisher} {Springer},\ \bibinfo {year} {2017})\ pp.\ \bibinfo {pages}
  {201--215}\BibitemShut {NoStop}%
\bibitem [{\citenamefont {Gandy}\ and\ \citenamefont
  {Veraart}(2019)}]{gandy2019adjustable}%
  \BibitemOpen
  \bibfield  {author} {\bibinfo {author} {\bibfnamefont {A.}~\bibnamefont
  {Gandy}}\ and\ \bibinfo {author} {\bibfnamefont {L.~A.~M.}\ \bibnamefont
  {Veraart}},\ }\href@noop {} {\bibfield  {journal} {\bibinfo  {journal}
  {Journal of Multivariate Analysis}\ }\textbf {\bibinfo {volume} {172}},\
  \bibinfo {pages} {193} (\bibinfo {year} {2019})}\BibitemShut {NoStop}%
\bibitem [{\citenamefont {Ramadiah}\ \emph {et~al.}(2020)\citenamefont
  {Ramadiah}, \citenamefont {Caccioli},\ and\ \citenamefont
  {Fricke}}]{ramadiah2020reconstructing}%
  \BibitemOpen
  \bibfield  {author} {\bibinfo {author} {\bibfnamefont {A.}~\bibnamefont
  {Ramadiah}}, \bibinfo {author} {\bibfnamefont {F.}~\bibnamefont {Caccioli}},\
  and\ \bibinfo {author} {\bibfnamefont {D.}~\bibnamefont {Fricke}},\
  }\href@noop {} {\bibfield  {journal} {\bibinfo  {journal} {Journal of
  Economic Dynamics and Control}\ }\textbf {\bibinfo {volume} {111}},\ \bibinfo
  {pages} {103817} (\bibinfo {year} {2020})}\BibitemShut {NoStop}%
\bibitem [{\citenamefont {Lebacher}\ \emph {et~al.}(2019)\citenamefont
  {Lebacher}, \citenamefont {Cook}, \citenamefont {Klein},\ and\ \citenamefont
  {Kauermann}}]{lebacher2019search}%
  \BibitemOpen
  \bibfield  {author} {\bibinfo {author} {\bibfnamefont {M.}~\bibnamefont
  {Lebacher}}, \bibinfo {author} {\bibfnamefont {S.}~\bibnamefont {Cook}},
  \bibinfo {author} {\bibfnamefont {N.}~\bibnamefont {Klein}},\ and\ \bibinfo
  {author} {\bibfnamefont {G.}~\bibnamefont {Kauermann}},\ }\href@noop {}
  {\bibfield  {journal} {\bibinfo  {journal} {arXiv preprint arXiv:1909.01274}\
  } (\bibinfo {year} {2019})}\BibitemShut {NoStop}%
\bibitem [{\citenamefont {Cimini}\ \emph
  {et~al.}(2015{\natexlab{a}})\citenamefont {Cimini}, \citenamefont
  {Squartini}, \citenamefont {Gabrielli},\ and\ \citenamefont
  {Garlaschelli}}]{cimini2015estimating}%
  \BibitemOpen
  \bibfield  {author} {\bibinfo {author} {\bibfnamefont {G.}~\bibnamefont
  {Cimini}}, \bibinfo {author} {\bibfnamefont {T.}~\bibnamefont {Squartini}},
  \bibinfo {author} {\bibfnamefont {A.}~\bibnamefont {Gabrielli}},\ and\
  \bibinfo {author} {\bibfnamefont {D.}~\bibnamefont {Garlaschelli}},\
  }\href@noop {} {\bibfield  {journal} {\bibinfo  {journal} {Physical Review
  E}\ }\textbf {\bibinfo {volume} {92}},\ \bibinfo {pages} {040802} (\bibinfo
  {year} {2015}{\natexlab{a}})}\BibitemShut {NoStop}%
\bibitem [{\citenamefont {Cimini}\ \emph
  {et~al.}(2015{\natexlab{b}})\citenamefont {Cimini}, \citenamefont
  {Squartini}, \citenamefont {Garlaschelli},\ and\ \citenamefont
  {Gabrielli}}]{cimini2015systemic}%
  \BibitemOpen
  \bibfield  {author} {\bibinfo {author} {\bibfnamefont {G.}~\bibnamefont
  {Cimini}}, \bibinfo {author} {\bibfnamefont {T.}~\bibnamefont {Squartini}},
  \bibinfo {author} {\bibfnamefont {D.}~\bibnamefont {Garlaschelli}},\ and\
  \bibinfo {author} {\bibfnamefont {A.}~\bibnamefont {Gabrielli}},\ }\href@noop
  {} {\bibfield  {journal} {\bibinfo  {journal} {Scientific reports}\ }\textbf
  {\bibinfo {volume} {5}},\ \bibinfo {pages} {1} (\bibinfo {year}
  {2015}{\natexlab{b}})}\BibitemShut {NoStop}%
\bibitem [{\citenamefont {Parisi}\ \emph {et~al.}(2020)\citenamefont {Parisi},
  \citenamefont {Squartini},\ and\ \citenamefont
  {Garlaschelli}}]{parisi2020horse}%
  \BibitemOpen
  \bibfield  {author} {\bibinfo {author} {\bibfnamefont {F.}~\bibnamefont
  {Parisi}}, \bibinfo {author} {\bibfnamefont {T.}~\bibnamefont {Squartini}},\
  and\ \bibinfo {author} {\bibfnamefont {D.}~\bibnamefont {Garlaschelli}},\
  }\href@noop {} {\bibfield  {journal} {\bibinfo  {journal} {New Journal of
  Physics}\ }\textbf {\bibinfo {volume} {22}},\ \bibinfo {pages} {053053}
  (\bibinfo {year} {2020})}\BibitemShut {NoStop}%
\bibitem [{\citenamefont {Jaynes}(1957)}]{jaynes1957information}%
  \BibitemOpen
  \bibfield  {author} {\bibinfo {author} {\bibfnamefont {E.~T.}\ \bibnamefont
  {Jaynes}},\ }\href@noop {} {\bibfield  {journal} {\bibinfo  {journal}
  {Physical review}\ }\textbf {\bibinfo {volume} {106}},\ \bibinfo {pages}
  {620} (\bibinfo {year} {1957})}\BibitemShut {NoStop}%
\bibitem [{\citenamefont {Squartini}\ and\ \citenamefont
  {Garlaschelli}(2011)}]{squartini2011analytical}%
  \BibitemOpen
  \bibfield  {author} {\bibinfo {author} {\bibfnamefont {T.}~\bibnamefont
  {Squartini}}\ and\ \bibinfo {author} {\bibfnamefont {D.}~\bibnamefont
  {Garlaschelli}},\ }\href@noop {} {\bibfield  {journal} {\bibinfo  {journal}
  {New Journal of Physics}\ }\textbf {\bibinfo {volume} {13}},\ \bibinfo
  {pages} {083001} (\bibinfo {year} {2011})}\BibitemShut {NoStop}%
\bibitem [{\citenamefont {Garlaschelli}\ and\ \citenamefont
  {Loffredo}(2008)}]{garlaschelli2008maximum}%
  \BibitemOpen
  \bibfield  {author} {\bibinfo {author} {\bibfnamefont {D.}~\bibnamefont
  {Garlaschelli}}\ and\ \bibinfo {author} {\bibfnamefont {M.~I.}\ \bibnamefont
  {Loffredo}},\ }\href@noop {} {\bibfield  {journal} {\bibinfo  {journal}
  {Physical Review E}\ }\textbf {\bibinfo {volume} {78}},\ \bibinfo {pages}
  {015101} (\bibinfo {year} {2008})}\BibitemShut {NoStop}%
\end{thebibliography}%

\newpage

\appendix

\section{Supplementary material}

\subsection{Entropy and likelihood in Exponential Random Graphs} \label{sub:entropy}
Both the dcGM and the scGM are models that derive from the general family of constrained maximum entropy graph ensembles. We recap here some of the important properties of this group of models.

Maximum entropy graph ensambles comes from the approach proposed by Jaynes to maximise the Shannon entropy subject to constraints \cite{jaynes1957information}. This can be achieved by maximising the functional
\begin{equation} \label{eq:me_lagrangian}
    \mathscr{L}[P] = S[P] - \lambda_0 \left[ \sum_{G \in \mathcal{G}} P(G) - 1 \right] - \sum_{m=1}^M \lambda_m \left[ \sum_{G \in \mathcal{G}} P(G) C_m[G] -\langle C_m \rangle \right]
\end{equation}
where $S[P]$ is the Shannon entropy functional, defined by $S = \sum_{G \in \mathcal{G}} - P(G)\ln P(G)$, and $\left\{C_m[G]\right\}_{m=1}^M$ is the set of constraints encoding known information about the ensemble. Solving the maximisation problem yields the general form of Exponential Random Graphs (ERG) \cite{squartini2018reconstruction}:
\begin{equation}
    P(G|\vec{\lambda}) = \frac{e^{- \sum_{m=1}^M \lambda_m C_m[G]}}{Z(\vec{\lambda})} \quad.
\end{equation}

Here $Z(\vec{\lambda})$ is the partition function defined as $Z(\vec{\lambda}) = \sum_{G \in \mathcal{G}}  e^{- \sum_{m=1}^M \lambda_m C_m[G]}$. Given a real graph $G^*$ we would like to find the set of Lagrange multipliers $\left\{\lambda_m\right\}_{m=1}^M$ that maximise the log-likelihood
\begin{equation}
    \mathcal{L}(G^*|\vec{\lambda}) = \ln P(G^*|\vec{\lambda}) = - \sum_{m=1}^M \lambda_m C_m[G^*] - \ln Z(\vec{\lambda}) \quad.
\end{equation}

It can be shown that the maximum of the log-likelihood is found for the parameter value $\vec{\lambda}^*$ such that the expected value of the constraints matches the value of the original network, i.e. $\langle C_m \rangle = C_m[G^*] = C_m^*$. Notice that the entropy of the ERG model is given by:
\begin{align}
    S(\vec{\lambda}) &= \sum_{G \in \mathcal{G}} - P(G|\vec{\lambda})\ln P(G|\vec{\lambda})\\
    &= - \sum_{G \in \mathcal{G}} P(G|\vec{\lambda}) \left( - \sum_{m=1}^M \lambda_m C_m[G] - \ln Z(\vec{\lambda}) \right)\\
    &= \sum_{m=1}^M \lambda_m \sum_{G \in \mathcal{G}} P(G|\vec{\lambda}) C_m[G] + \ln Z(\vec{\lambda}) \sum_{G \in \mathcal{G}} P(G|\vec{\lambda})\\
    &= \sum_{m=1}^M \lambda_m \langle C_m \rangle + \ln Z(\vec{\lambda}) \quad .\label{anvediquesto}
\end{align}

Given that the maximum likelihood estimation $\vec{\lambda}^*$ of $\vec{\lambda}$ guarantees that $\langle C_m \rangle = C_m^*$, we can write:
\begin{equation}
    S(\vec{\lambda}^*) = \sum_{m=1}^M \lambda^*_m C_m^* + \ln Z(\vec{\lambda}^*) = - \mathcal{L}(G^*|\vec{\lambda}^*) .\label{sobbonitutti}
\end{equation}
We have therefore shown that the entropy of an ERG model is equal to minus the log-likelihood of observing any graph whose constrained quantities match the ensemble expectation. This in practice tells us that given two ERG models, fitted using the maximum likelihood principle, the one with higher likelihood will necessarily have a lower entropy.

It is important to recall here that the first step of the dcGM and scGM is derived from the Directed Configuration Model (DCM) \cite{cimini2015estimating}, which is an ERG model for directed graphs. The DCM is found by constraining the in- and out- degrees $\vec{k}^{in}(G^*),\vec{k}^{out}(G^*)$ of a given graph $G^*$. This results, after maximizing the likelihood, in the probability distribution (detailed derivation can be found in \cite{squartini2011analytical})
\begin{equation}
    P(G|\vec{\alpha}^*, \vec{\beta}^*) = \frac{e^{-\sum_i \alpha^*_i k_i^{\text{out}}(G) + \beta^*_i k_i^{\text{in}}(G) }}{Z(\vec{\alpha}^*, \vec{\beta}^*)} = \prod_{i\neq j} p_{ij}^{a_{ij}}(1 - p_{ij})^{1 - a_{ij}}
\end{equation}
with
\begin{equation} \label{eq:dcm_pij}
    p_{ij} = \frac{e^{-\alpha^*_i -\beta^*_j}}{1 + e^{-\alpha^*_i -\beta^*_j}},
\end{equation}
where the $2N$ Lagrange multipliers $\vec{\alpha}, \vec{\beta}$ have been set to their maximum-likelihood values $\vec{\alpha}^*, \vec{\beta}^*$ ensuring that the expected values of all the in- and out-degrees equal the empirical values $\vec{k}^{in}(G^*),\vec{k}^{out}(G^*)$ \emph{exactly}, i.e.
\begin{eqnarray}
  \langle{k}_i^{in}\rangle =\sum_{i\ne j} \frac{e^{-\alpha^*_j -\beta^*_i}}{1 + e^{-\alpha^*_j -\beta^*_i}}&=&{k}_i^{in}(G^*)\quad\forall i,\label{alimo}\\
\langle{k}_i^{out}\rangle=\sum_{i\ne j} \frac{e^{-\alpha^*_i -\beta^*_j}}{1 + e^{-\alpha^*_i -\beta^*_j}}&=&{k}_i^{out}(G^*)\quad\forall i.\label{taccitu}
\end{eqnarray}
Note that finding the values $\vec{\alpha}^*, \vec{\beta}^*$ requires solving the $2N$ nonlinear coupled equations above.

The dcGM follows from the so-called fitness ansatz, formulated in \cite{cimini2015estimating}, positing that $e^{-\alpha_i^*} \approx \sqrt{z} s_i^{\text{out}}$ and $e^{-\beta_j^*} \approx \sqrt{z} s_j^{\text{in}}$, with $z$ being the only remaining free parameter, which can be used to fit the empirical density of the graph $G^*$. Substituting these values into equation \eqref{eq:dcm_pij} we obtain the dcGM:
\begin{equation}
    p_{ij}\approx p_{ij}^{\text{dcGM}} = \frac{z^* s_i^{\text{out}} s_j^{\text{in}}}{1 + z^* s_i^{\text{out}} s_j^{\text{in}}},
\end{equation}
where $z^*$ is such that the expected link density equals the empirical one, i.e. the expected total number of links $\langle L\rangle$ equals the empirical total number $L(G^*)$ of links of the real graph $G^*$:
\begin{equation}
   \langle L\rangle^{\text{dcGM}}= \sum_{i=1}^N\sum_{j\ne i}\frac{z^* s_i^{\text{out}} s_j^{\text{in}}}{1 + z^* s_i^{\text{out}} s_j^{\text{in}}}=L(G^*).\label{giggetto}
\end{equation}

Note that we can regard the fitness ansatz as restricting the maximum likelihood principle to the parameter $z$ only, since the original $2N$ Lagrange multipliers are assumed to be identifiable with (or approximated by) the empirical node strengths, up to a common parameter $z$. 
Indeed, applying the maximum likelihood principle to $z$ leads precisely to Eq.~\eqref{giggetto} \cite{garlaschelli2008maximum}.

Checking how good the fitness ansatz is for a given network (for which we know both the degrees and the strengths) can be done in two equivalent ways.
Directly, we may explicitly find the values $\vec{\alpha}^*, \vec{\beta}^*$ solving the $2N$ coupled equations \eqref{alimo} and \eqref{taccitu} and then check how good the relationships $e^{-\alpha_i^*} \propto s_i^{\text{out}}$ and $e^{-\beta_j^*} \propto s_j^{\text{in}}$ are.
Alternatively, we may insert the ansatz $e^{-\alpha_i^*} \approx \sqrt{z^*} s_i^{\text{out}}$ and $e^{-\beta_j^*} \approx \sqrt{z^*} s_j^{\text{in}}$ into equations \eqref{alimo} and \eqref{taccitu} and check how well the empirical degrees are approximated by the expected degrees under the dcGM, i.e. how well the following relationships hold:
\begin{eqnarray}
  \langle{k}_i^{in}\rangle^{\text{dcGM}} =\sum_{i\ne j} \frac{z^* s_j^{\text{out}} s_i^{\text{in}}}{1 + z^* s_j^{\text{out}} s_i^{\text{in}}}&\approx&{k}_i^{in}(G^*)\quad\forall i,\label{alimo2}\\
\langle{k}_i^{out}\rangle^{\text{dcGM}}=\sum_{i\ne j} \frac{z^* s_i^{\text{out}} s_j^{\text{in}}}{1 + z^* s_i^{\text{out}} s_j^{\text{in}}}&\approx &{k}_i^{out}(G^*)\quad\forall i.\label{taccitu2}
\end{eqnarray}
Note that the latter procedure does not require solving equations \eqref{alimo} and \eqref{taccitu}, but only the single equation \eqref{giggetto} setting $z^*$.
In our analysis, since we are dealing with a network where $N$ is very large, we adopt the latter test.

It is straightforward to replicate the steps leading to Eq.~\eqref{anvediquesto} in the case of the DCM, irrespective of the values of $\vec{\alpha}$ and $\vec{\beta}$:
\begin{equation}
S(\vec{\alpha},\vec{\beta})= \sum_{i=1}^N (\alpha_i \langle k^{out}_i \rangle +\beta_i \langle k^{in}_i \rangle)  + \ln Z(\vec{\alpha},\vec{\beta}).
\end{equation}
Now, if we insert the values $\vec{\alpha}^*, \vec{\beta}^*$ solving Eqs. \eqref{alimo} and \eqref{taccitu}, we get the exact relationship between entropy and minus log-likelihood as in Eq.~\eqref{sobbonitutti}:
\begin{equation}
    S(\vec{\alpha}^*, \vec{\beta}^*) = \sum_{i=1}^N [\alpha^*_i k^{out}_i(G^*) +\beta^*_i  k^{in}_i(G^*)]  + \ln Z(\vec{\alpha}^*,\vec{\beta}^*) = - \mathcal{L}(G^*|\vec{\alpha}^*, \vec{\beta}^*) .\label{sobbonitutti2}
\end{equation}
If instead we plug the values $\tilde{\alpha}_i=-\ln(\sqrt{z^*}s^{out}_i)$ and $\tilde{\beta}_i=-\ln(\sqrt{z^*}s^{in}_i)$ implied by the fitness ansatz, we obtain 
\begin{equation}
    S(\vec{\tilde{\alpha}}, \vec{\tilde{\beta}}) = \sum_{i=1}^N [\tilde{\alpha}_i \langle k^{out}_i \rangle^{\text{dcGM}} +\tilde{\beta}_i  \langle k^{in}_i \rangle^{\text{dcGM}}]  + \ln Z(\vec{\tilde{\alpha}},\vec{\tilde{\beta}}) = - \mathcal{L}(G^*|\vec{\tilde{\alpha}}, \vec{\tilde{\beta}}),\label{sobbonitutti3}
\end{equation}
where $\mathcal{L}(G^*|\vec{\tilde{\alpha}}, \vec{\tilde{\beta}})$ is exactly the log-likelihood of the dcGM model, while $S(\vec{\tilde{\alpha}}, \vec{\tilde{\beta}})$ is the entropy of a DCM with degrees $\langle \vec{k}^{out} \rangle^{\text{dcGM}}$ and $\langle \vec{k}^{in} \rangle^{\text{dcGM}}$, i.e. the maximum entropy ensemble with degrees given by the fitness ansatz in Eqs.~\eqref{alimo2} and~\eqref{taccitu2}.
So the negative log-likelihood $-\mathcal{L}(G^*|\vec{\tilde{\alpha}}, \vec{\tilde{\beta}})$ is still a measure of the number of alternative network configurations having those degrees as expected values.
If Eqs.~\eqref{alimo2} and~\eqref{taccitu2} are well obeyed by the data, then $-\mathcal{L}(G^*|\vec{\tilde{\alpha}}, \vec{\tilde{\beta}})$ is also a good approximation for the original entropy $S(\vec{\alpha}^*, \vec{\beta}^*)$, i.e. for the number of network configurations having the empirical degrees as expected values.

\subsection{Multiproduct generalization} \label{sub:multi_prod}
In section \ref{sub:scGM} we have discussed that there is a clear relationship between the formulation of the scGM and production functions. This is perhaps clearer if we look at the multiproduct generalization of the model presented. We define $\Gamma$ to be the set of all goods available in our economy and $\alpha \in \Gamma$ to be one such product. Our network is now multi-layered with a generic entry of the directed, weighted adjacency matrix being denoted as $w_{ij}^\alpha$, that is the amount of goods of type $\alpha$ sold by $i$ and bought by $j$. The in- and out- strengths for each node can now be defined per product as follows:

\begin{align}
s_{i, \text{out}}^\alpha &= \sum_{j}w_{ij}^\alpha,\quad\forall\:i; \\
s_{i, \text{in}}^\alpha &= \sum_{j}w_{ji}^\alpha,\quad\forall\:i.
\end{align}

We can now reformulate the two steps for sampling edges and their weights, such that the probability of observing a link between node $i$ and $j$ of product $\alpha$ is given by
\begin{equation}
p_{ij}^\alpha = \frac{zs_{i, \text{out}}^\alpha s_{j, \text{in}}^\alpha}{1+zs_{i, \text{out}}^\alpha s_{j, \text{in}}^\alpha},\quad\forall\:i\neq j ;
\end{equation}
and the weights can be assigned respecting
\begin{equation}
\langle w_{ij}^\alpha | a_{ij}^\alpha = 1 \rangle = \frac{zs_{i, \text{out}}^\alpha s_{j, \text{in}}^\alpha}{w_\alpha^{tot}\cdot p_{ij}^\alpha},\quad\forall\:i\neq j
\end{equation}
where $w_\alpha^{tot} = \sum_{i,j} w_{ij}^\alpha$.

This model might appear different from the scGM given that both the in- and out- strengths are divided per product. However, under the assumption that each firm produces only one good that is the same for each business in a given industry then the generalization above and the model presented in section \ref{sub:scGM} coincide. Looking at this generalization it might also be clearer that input and output strengths are not per se treated differently, rather it is the choice of the proxy used in order to identify goods (in our case the sector classification) that results in only one out-strength being different from zero.

\subsection{Identifying firms} \label{sub:id_firms}
The available data consists of all incoming and outgoing transactions on the accounts of the Dutch clients of the banks. As we are concerned only with inter-firm transactions we exclude from the analysis any account belonging to a private individual. 
For the ING data, it is known from ING accounts whether they belong to private individuals or clients.
For the ABN AMRO dataset this is achieved using the Business Contact Database (BCDB) curated by the Commercial Banking (CB) department. In this database for each firm there is a hierarchical definition of the commercial complex it belongs to. This allows us to both remove private individuals that do not have an assigned commercial complex and to reconstruct the often complex hierarchical relationships that exists between accounts of the same firm. Note that there is a difference between the two approaches: the ING data is more inclusive guaranteeing a maximum coverage of all commercial clients; the ABN approach on the other hand is quite restrictive, any node included will definitely be a firm, however smaller firms might not be included if they have not been correctly labelled as complexes. The commercial complex is a definition that comes from the account manager for the commercial client that together with the client determines which accounts belong to the same firm and which among the available contacts is the main account holder. A similar structured grouping of accounts is achieved in the ING data using the legal (ultimate) parent information available in the transaction data from ING.

Each commercial complex will be a single node in our network and all transactions within the group will be discarded while transactions between groups will be summed irrespective of the specific account on which the transfer was recorded. This step of identifying firm is a delicate but fundamental step. The transaction data is per se already a network as it encodes flows of money from one account to another. However this data cannot be used as is, because it would not be an accurate description of the input-output relationships between firms. This is because we are interested in constructing an input-output network where each node is a functional economic unit such that any incoming money into the account can be interpreted as the remuneration for a product or service of a known type. In order to achieve this, two things must be true: first only transactions that correspond to final consumption or intermediate use of products should be included, and second the accounts must be aggregated such that all expenses and income relating to production are assigned to the same node. You can better understand the importance of the second point by imagining a firm with multiple accounts, one used to pay employees, taxes and receive sales income, and one to pay suppliers. If we considered these accounts as independent nodes we would have broken the functional economic unit and its single balance sheet into two meaningless units and would be considering transfers between accounts as a sale of some product. We could, of course, exaggerate in the other direction, that is by aggregating more than necessary. Consider the case of a financial holding that owns multiple businesses: each business has its own accounts and balance sheet and should be treated as an independent node, if we were to consider all firms owned by the holding as a single node we would be ignoring the fact that these balance sheets operate independently and their production processes are not the same. The commercial complex definition used in our analysis has been developed with a similar objective in mind: that is to be able to analyse the total operating income and expenses of a firm as a whole.

A final step in the identification of the firm is constituted by the correct assignment of an industry label. For the purpose of this analysis we use the Dutch Standaard Bedrijfsindeling (SBI) hierarchical five digit codes for the ABN AMRO data and the North American Industry Classification System (NAICS) hierarchical six digit codes for the ING data. Note that the SBI codes are well aligned with the Statistical Classification of Economic Activities in the European Community (NACE) codes, as they share the same first four digits, and the International Standard Industrial Classification of All Economic Activities (ISIC), as the first two digits are the same.
All Dutch firms are assigned an SBI code when they register at the Chamber of Commerce (Kamer van Koophandel, KvK) however given the complex legal structure that companies may adopt, not always is it clear which of the different codes associated with a commercial complex is the main one. In order to address this internally at ABN AMRO to each commercial complex has been associated a \lq real\rq\ SBI which the account manager believes best captures the real economic activity of the firm. This curated information is particularly useful as often firms have a legal structure with a financial holding at its apex, meaning we would see an over-representation of the financial sector. The reason for using the NAICS codes for ING is that a similar manual overwrite of the sector code is done at ING but only on the NAICS codes. We have chosen to give precedence to this better quality data rather than coherence between the two datasets as the analysis does not rest on the codes being the same for both networks and that they are quite similar classifications. 

\subsection{Filtering transactions} \label{sub:trx_filters}
The bulk of transactions available in the two datasets is composed of SEPA payments. Both however also contain Swift, as well as other international and internal transfers related among others to repayment of loans, credit due to financial investments or deposit of a loan, cash and card payments. As discussed above we want to keep only the payments relating to the sales of the companies, as such we need to eliminate all transactions related to investments and credit, as well as all internal transfers between accounts.
We do so by considering all transactions between firms to be related to an exchange of a product or service and by eliminating all firms from the financial and insurance sector\footnote{SBI codes starting with 64, 65 or 66, and NAICS codes starting with 52.} and the public administration\footnote{SBI codes starting with 84, and NAICS codes starting with 92.}. The remaining transactions will be related to either final consumption or to other credit and debit relationships. For the purpose of this analysis it is however sufficient for us to use only the firm to firm connections and ignore the household consumption.

Another important aspect of the data is the temporal dimension. The available data is extremely granular, however for the purpose of understanding the structure of the relationship between firms we would like to have a more static view. As such we chose to aggregate transactions on a yearly basis, so that a link from node A to node B represents the total flow of goods produced by firm A and bought by firm B in a given year measured in Euros. A year is chosen as it will reduce any effect that seasonality may have giving us a balanced representation of all firms. Indeed, if we chose a monthly or quarterly aggregation those sectors that see most of their activity in a given period of the year would be underrepresented in certain snapshots. 

\subsection{Incomplete information}\label{sub:incompl_info}
It should be noted that unfortunately we cannot claim to have complete coverage of all inter-client transactions. There are two main limitations to our data: mediated transfers and unobserved direct flows. The first case is possibly the more problematic as there exist many payment service providers as well as other systems that mediate between firms. In particular there exist various services offered by financial institutions to pay upfront all credit invoices, or parts of them, taking upon themselves the risk of the customer not paying or paying late. These kind of transactions are precisely what we would like to include in our analysis. We know that these transfers exist in our data and we can in some cases identify one side of the transaction on the accounts, however linking sender and receiver is not easy. Further work could be dedicated to understanding better this issue and how to solve it possibly developing an algorithm to find matching pairs of unconnected flows.
The second issue relates to the heterogeneity in the payment systems available to the clients. Other than SEPA transfers we have batch payments, cash deposits and withdrawals, card payments and credit, international transfers, and many others. In the case of cash it is of course impossible to determine the origin of the payment, however also in many other cases in our datasets it is not possible to clearly identify the counterpart to a transaction.

Furthermore the effect of limiting ourselves to inter-client transactions will introduce a bias in our analysis that will vary firm by firm. The proportion of observed to unobserved links will be determined by how connected the client is to firms that are not part of the bank's network and to businesses abroad. Unfortunately we are currently not able to determine if those accounts belong to a private individual or a firm, nor group the accounts as discussed in section \ref{sub:id_firms}. We are also not able to correctly assign an industrial classification code to those firms. How this will affect our analysis is not easy to determine. The quality of our reconstruction depends fundamentally on the functional relation there is between fitness and probability of a link existing. In our case, where the fitness is the strength of the node, we can see two possibilities for a given firm: we observe a very high strength but a low degree because few counterparties to this firm are in the network, but the size of the transaction that is observed is in line with the real size of the firm; we observe a very low strength because the only transactions that are in our network are of low value, and hence the strength is not representative of the real probability of observing a link. The second case in particular could result in fully disconnected nodes in our reconstructed samples and very low expected degree.

\subsection{Descriptive statistics}
The network constructed with the methodology so far described is composed of one giant connected component containing 99.66 \% of nodes for the ABN data and 99.998\% for the ING data. Removing the financial and government sectors does not affect significantly the number of components with the largest component still retaining 99.54\% of all nodes for the ABN data and 99.01\% for the ING data. In both cases the other nodes are mostly in dyadic relations with the maximum component size of 27 nodes. In the table below we have summarized the sector compositions of the two datasets. Note that some of the discrepancies are due to the difficulty in precisely translating between the two classifications standards. 

\begin{table}[htb]
\scriptsize 
\caption{\label{tab:firms_per_area}%
Percentages of number of firms per highest level of industrial classification\footnote{We use here the definitions of the NACE system and adapt the NAICS to these areas.}
}
\begin{ruledtabular}
\begin{tabular}{lp{0.7\linewidth}dd}
\textrm{Symbol}&
\textrm{Area}&
\multicolumn{1}{c}{\textrm{ABN}}&
\multicolumn{1}{c}{\textrm{ING}}\\
\colrule
A & AGRICULTURE, FORESTRY AND FISHING & 3.09 & 1.10\\
B & MINING AND QUARRYING & 0.07 & 0.07\\
C & MANUFACTURING& 4.33 & 4.59\\
D & ELECTRICITY, GAS, STEAM AND AIR CONDITIONING SUPPLY & 0.09 & 0.19\\
E & WATER SUPPLY; SEWERAGE, WASTE MANAGEMENT AND REMEDIATION ACTIVITIES & 0.15 & 0.02\\
F & CONSTRUCTION & 10.84 & 4.53\\
G & WHOLESALE AND RETAIL TRADE; REPAIR OF MOTOR VEHICLES AND MOTORCYCLES & 15.88 & 15.24\\
H & TRANSPORTATION AND STORAGE & 3.07 & 2.94\\
I & ACCOMMODATION AND FOOD SERVICE ACTIVITIES & 3.94 & 3.84\\
J & INFORMATION AND COMMUNICATION & 4.96 & 1.50\\
K & FINANCIAL AND INSURANCE ACTIVITIES & 5.50 & 7.99\\
L & REAL ESTATE ACTIVITIES & 2.43 & 11.49\\
M & PROFESSIONAL, SCIENTIFIC AND TECHNICAL ACTIVITIES & 16.96 & 12.71\\
N & ADMINISTRATIVE AND SUPPORT SERVICE ACTIVITIES & 4.95 & 3.80\\
O & PUBLIC ADMINISTRATION AND DEFENCE; COMPULSORY SOCIAL SECURITY & 0.00 & 0.00\\
P & EDUCATION & 2.81 & 1.63\\
Q & HUMAN HEALTH AND SOCIAL WORK ACTIVITIES & 8.11 & 5.29\\
R & ARTS, ENTERTAINMENT AND RECREATION & 3.96 & 6.23\\
S & OTHER SERVICE ACTIVITIES & 8.78 & 16.86\\
T & ACTIVITIES OF HOUSEHOLDS AS EMPLOYERS; UNDIFFERENTIATED GOODS- AND SERVICES-PRODUCING ACTIVITIES OF HOUSEHOLDS FOR OWN USE & 0.01 & n/a\\
U & ACTIVITIES OF EXTRATERRITORIAL ORGANISATIONS AND BODIES & 0.05 & n/a
\end{tabular}
\end{ruledtabular}
\end{table}

\newpage

\subsection{Other measures and complementary plots for Institution 2}
We include for completeness other measures computed on the networks and fitted models.


\begin{figure}[tbp]
    \centering
    \subfloat[Institution 1\label{fig:sout_cdf_1}]{%
        \includegraphics[width=0.49\textwidth]{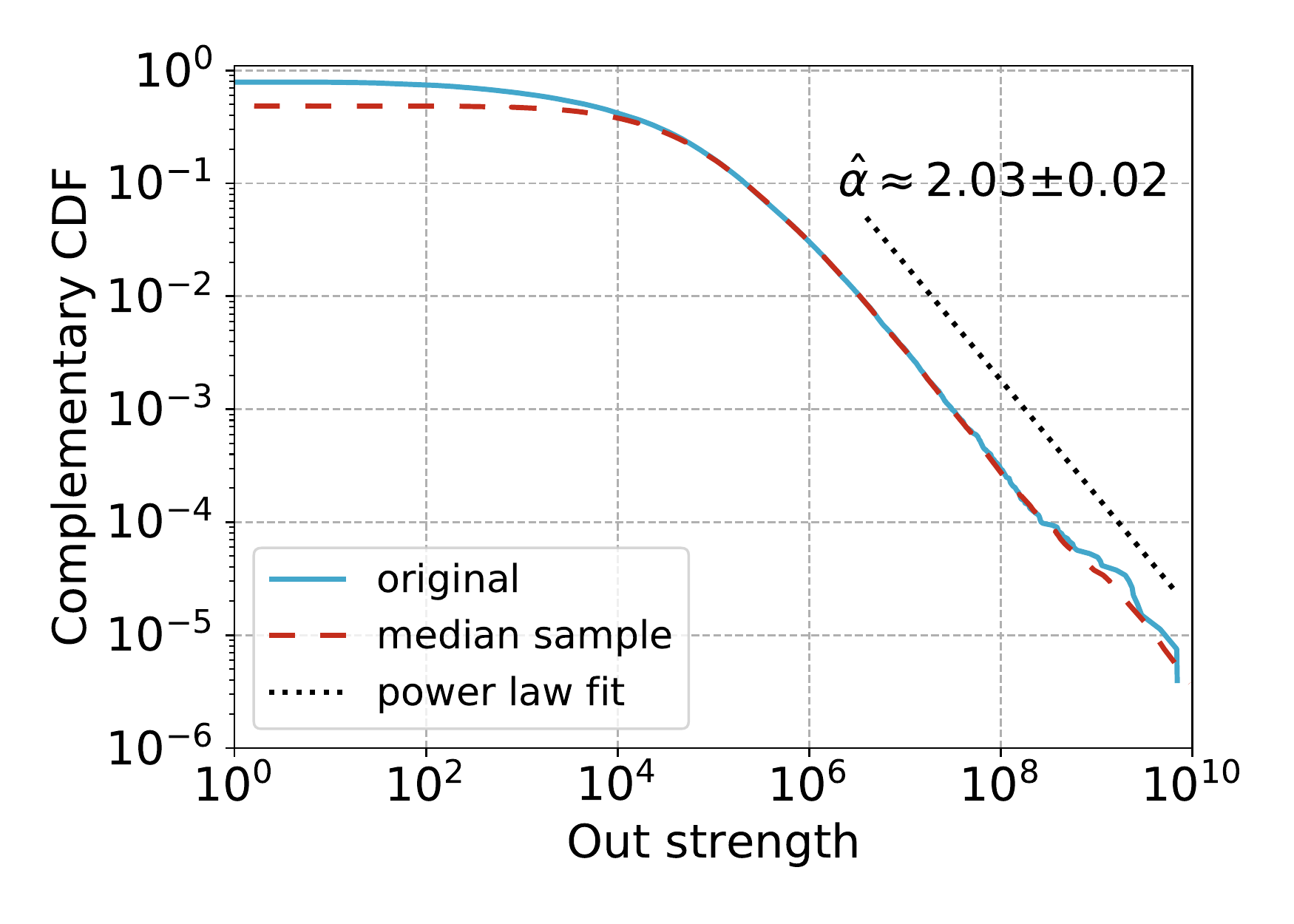}%
    }\hfil
    \subfloat[Institution 2\label{fig:sout_cdf_2}]{%
      \includegraphics[width=0.49\textwidth]{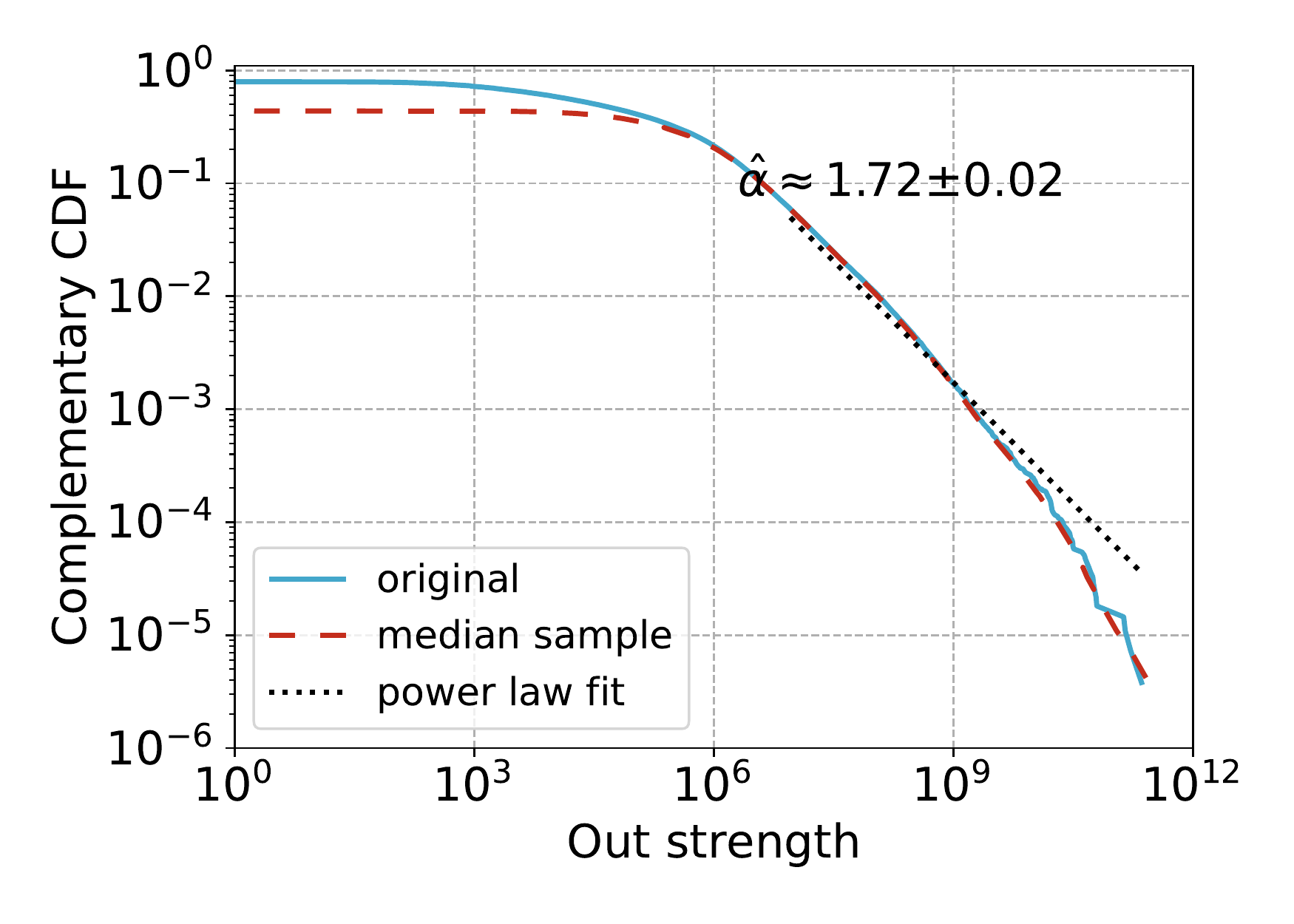}%
    }\vfil
    \subfloat[Institution 1\label{fig:sin_cdf_1}]{%
        \includegraphics[width=0.49\textwidth]{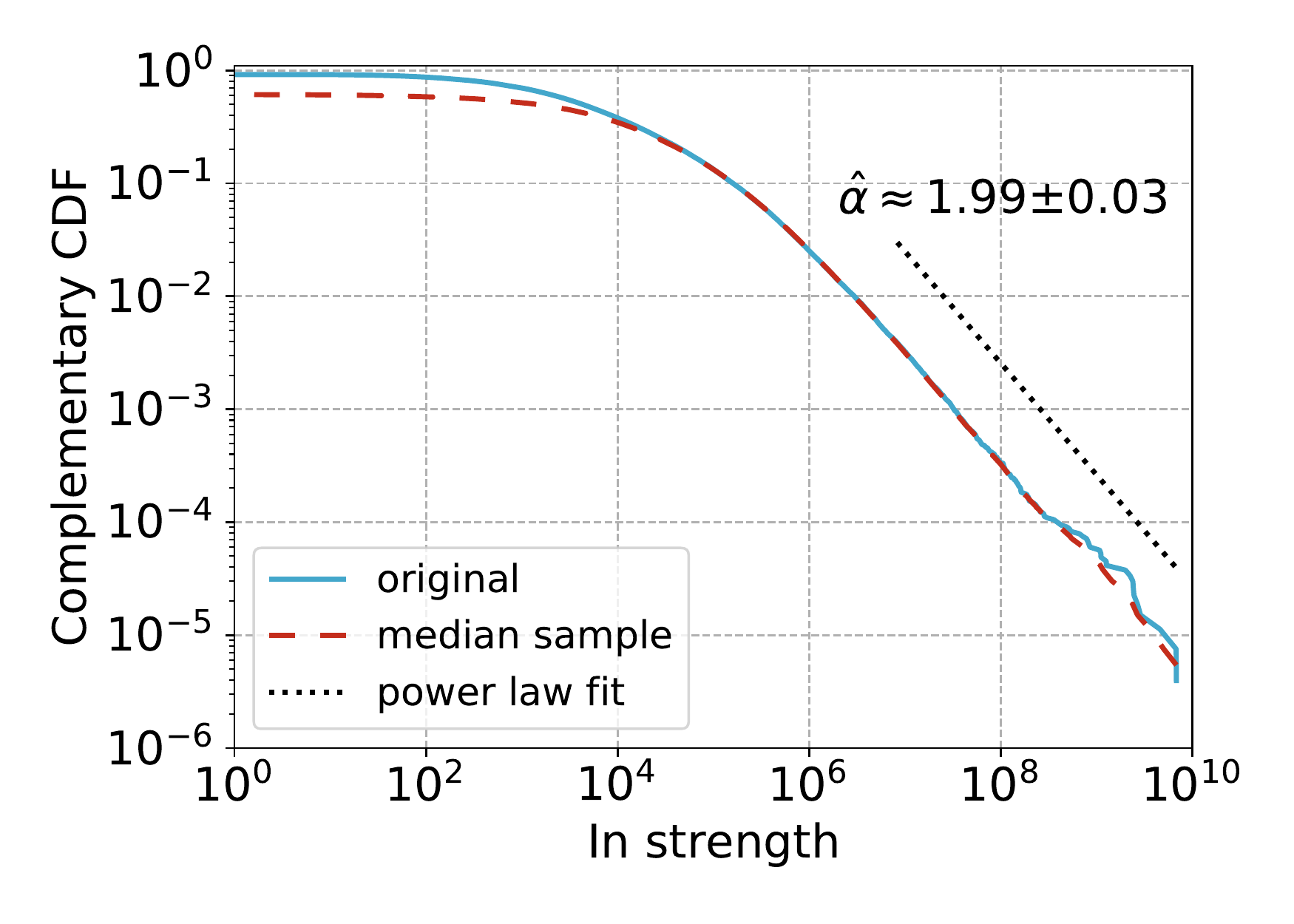}%
    }\hfil
    \subfloat[Institution 2\label{fig:sin_cdf_2}]{%
      \includegraphics[width=0.49\textwidth]{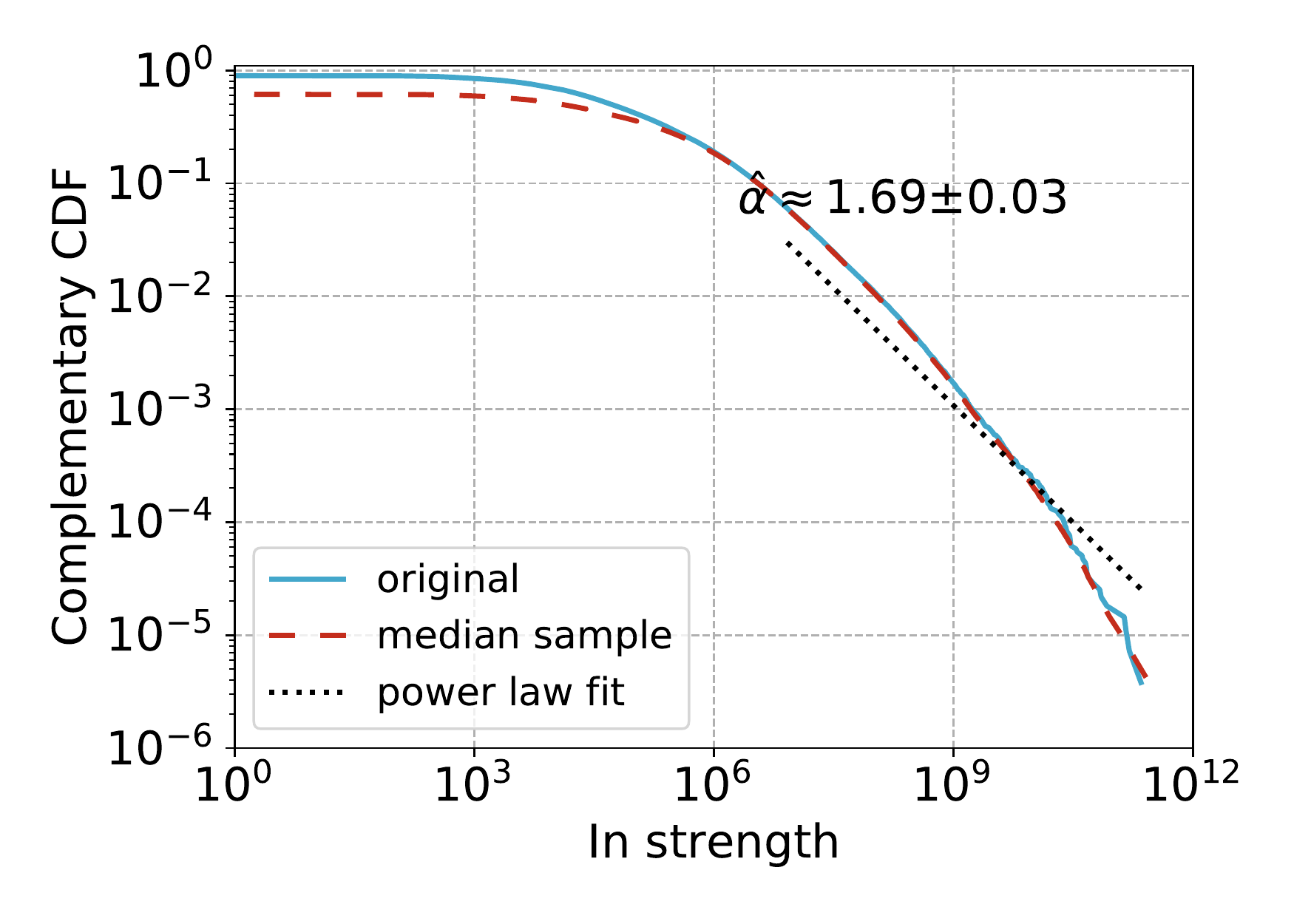}%
    }
    \caption{Complementary cumulative distribution of the out- and in-strength for both networks. We find that the distributions of institution 1 are almost as likely to follow a power law or a log-normal distribution but with slight preference for power law (p-value 0.92 for \ref{fig:sout_cdf_1} and 0.66 for \ref{fig:sin_cdf_1}). On the contrary for institution 2 we find a statistically significant support for log-normality (p-value $2.2\cdot 10^{-5}$ for \ref{fig:sout_cdf_2} and $2.4\cdot 10^{-8}$ for \ref{fig:sin_cdf_2}).}\label{fig:s_cdf}
\end{figure} 

\begin{figure}[tbp]
    \centering
    \subfloat[scGM \label{fig:dl_1_sc}]{%
        \includegraphics[width=0.9\textwidth]{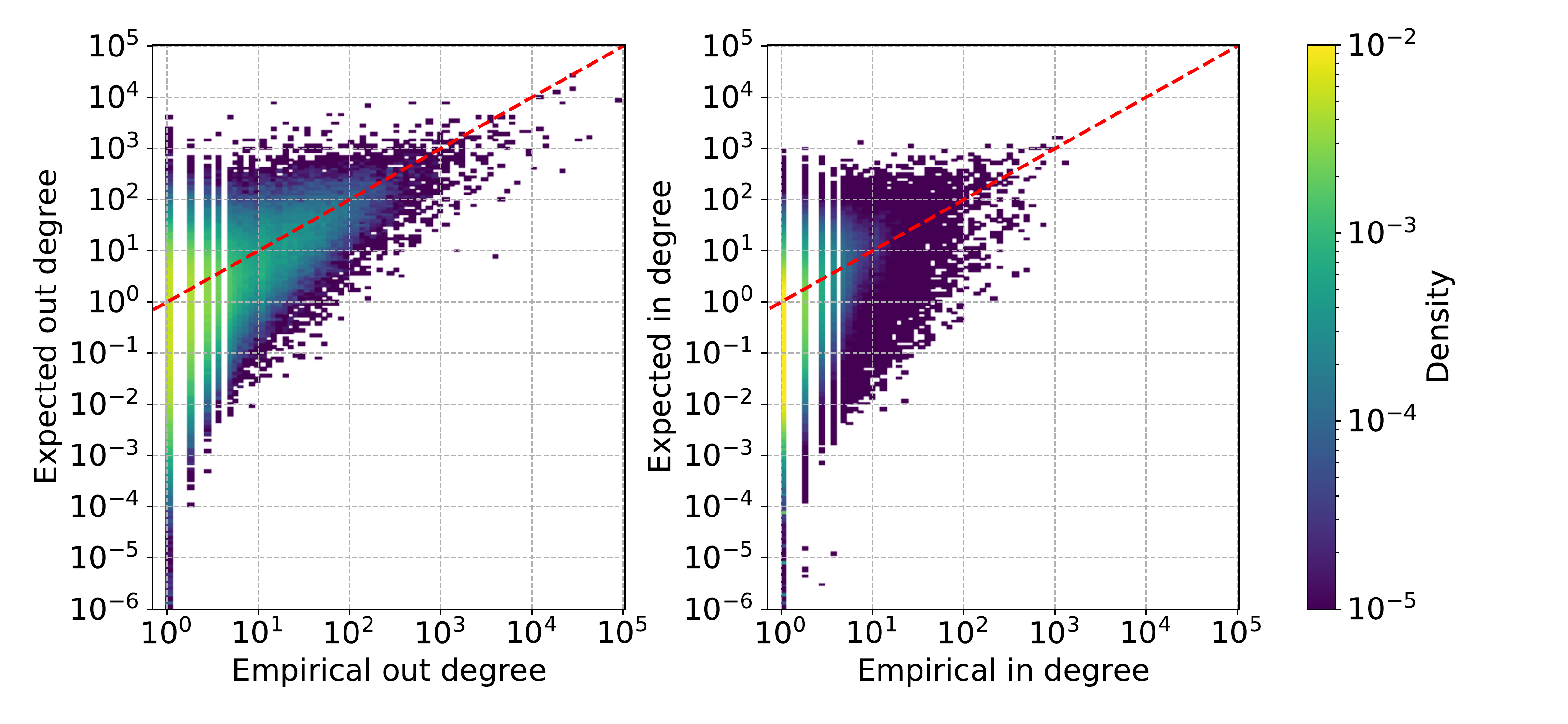}%
    }\vfil
    \subfloat[dcGM \label{fig:dl_1_dc}]{%
      \includegraphics[width=0.9\textwidth]{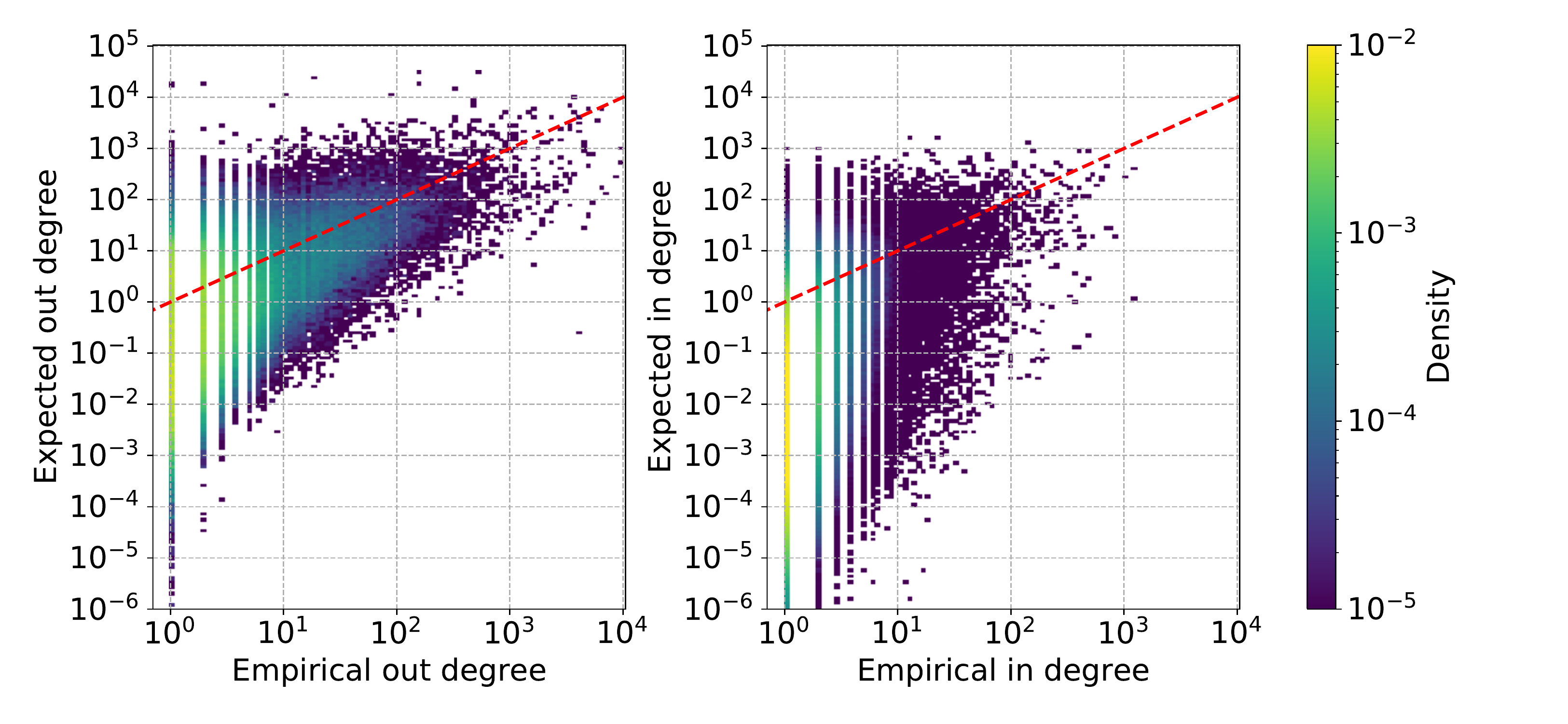}%
    }
    \caption{Empirical versus expected degree by sector for the network of Institution 1.}\label{fig:dl_1}
\end{figure} 

\begin{figure}[tbp]
    \centering
    \subfloat[scGM \label{fig:dl_2_sc}]{%
        \includegraphics[width=0.9\textwidth]{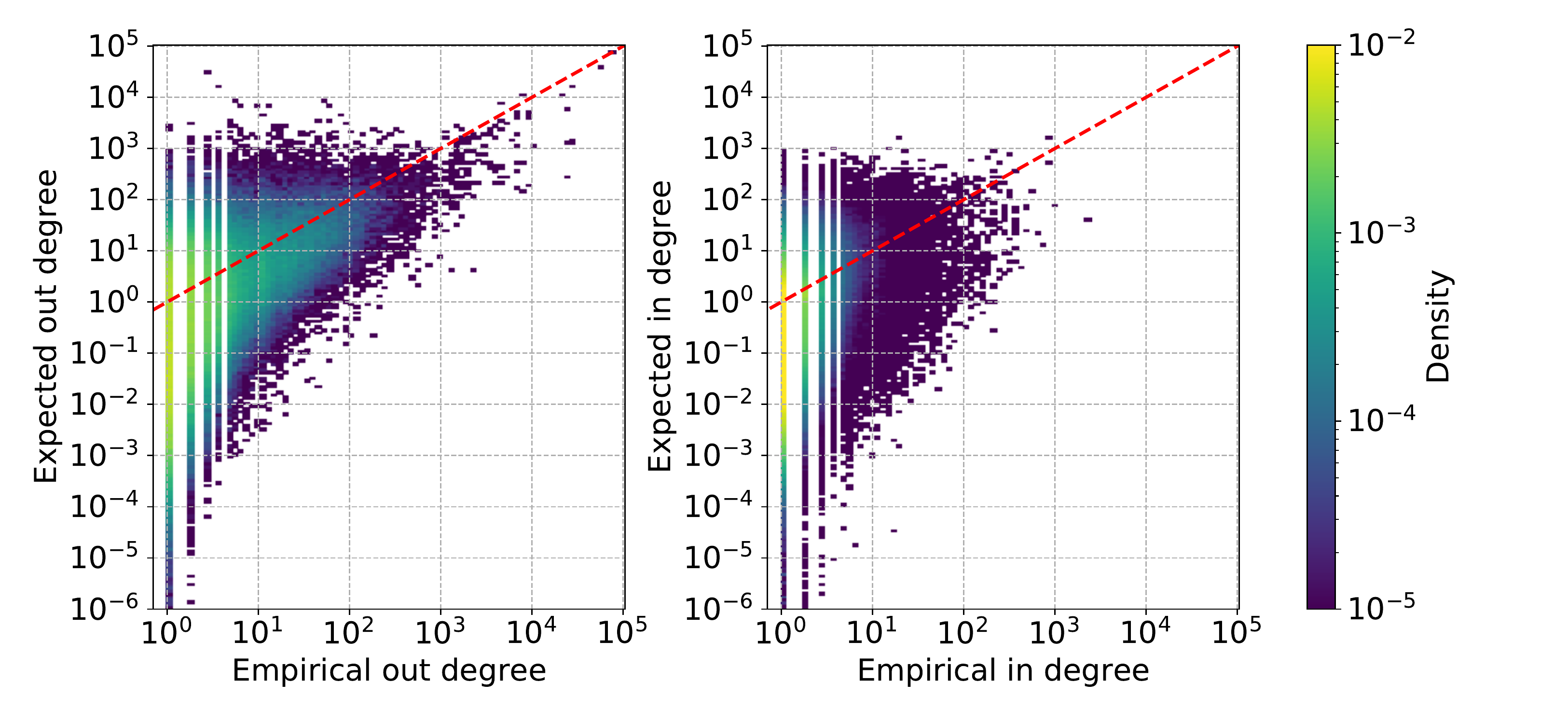}%
    }\vfil
    \subfloat[dcGM \label{fig:dl_2_dc}]{%
      \includegraphics[width=0.9\textwidth]{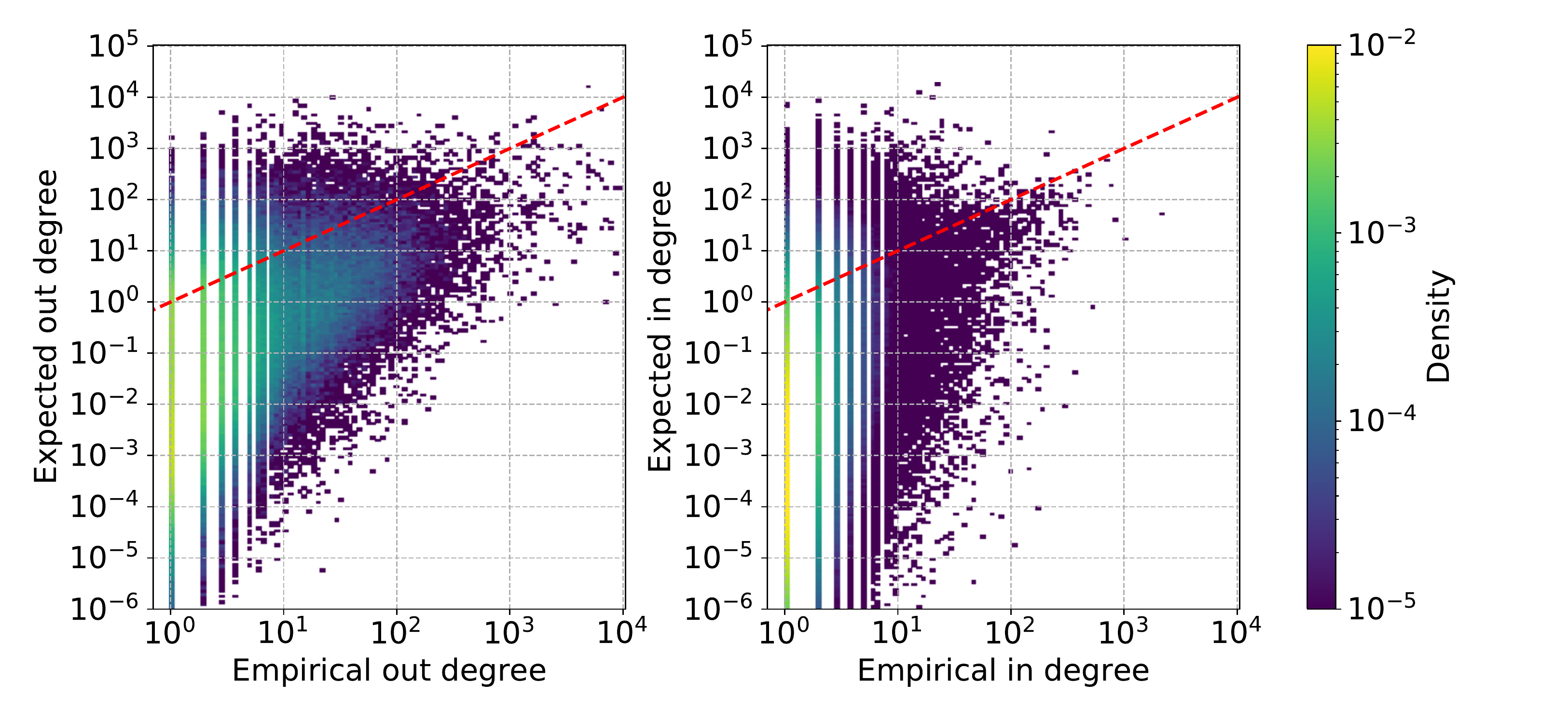}%
    }
    \caption{Empirical versus expected degree by sector for the network of Institution 2.}\label{fig:dl_2}
\end{figure} 


\begin{figure}[p]
    \centering
    \subfloat[All sectors \label{fig:ansatz_0_2}]{%
        \includegraphics[width=0.9\textwidth]{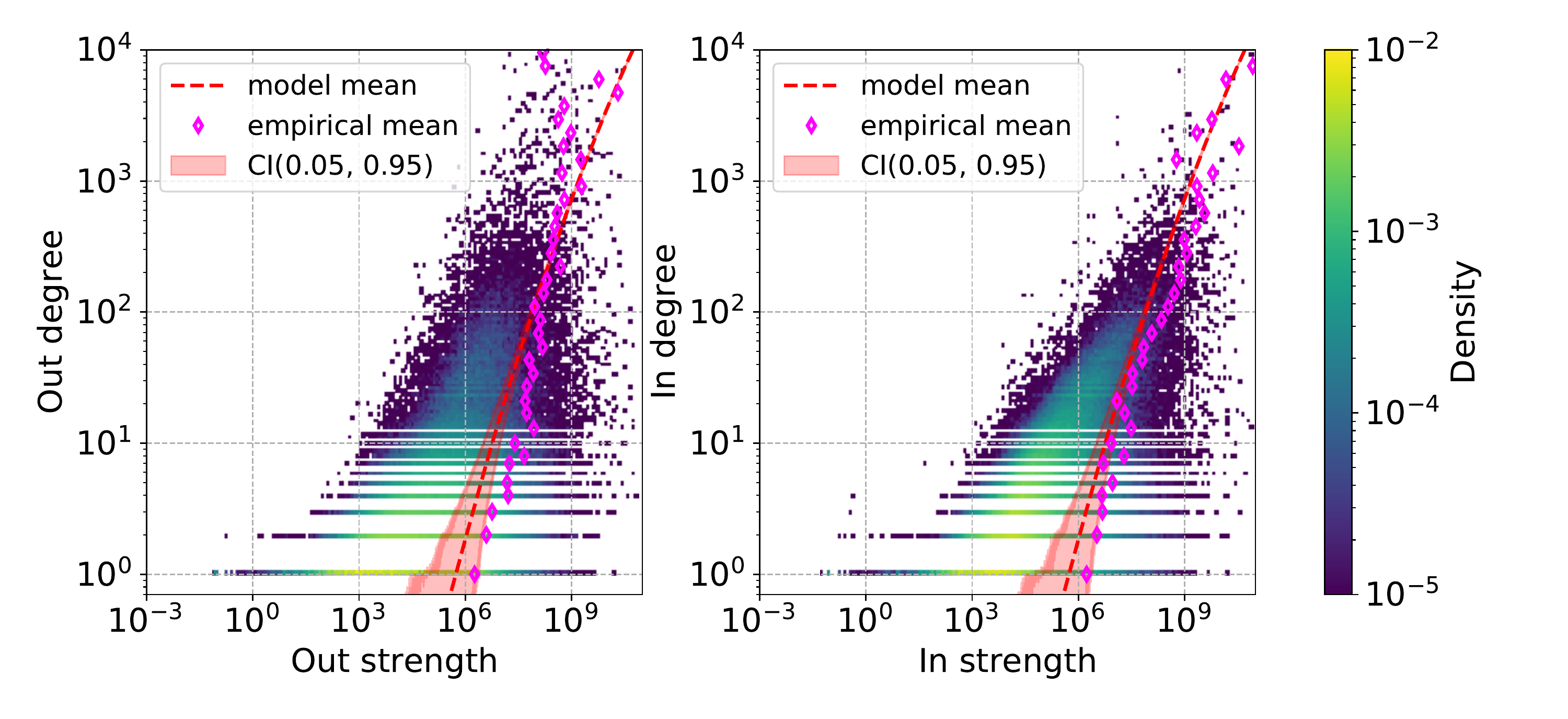}
    }\vfil
    \subfloat[C - Manufacturing\label{fig:ansatz_1_2}]{%
        \includegraphics[width=0.9\textwidth]{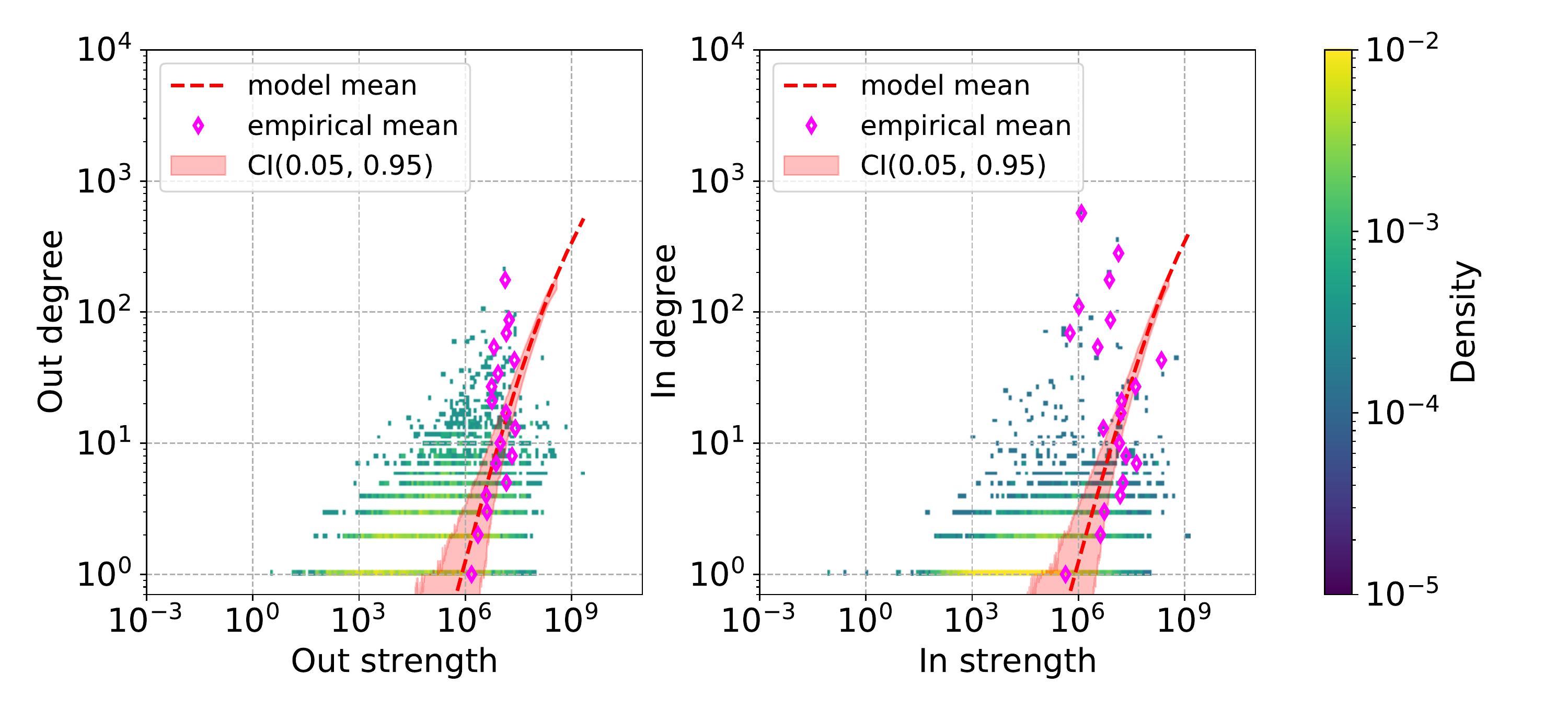}%
    }
    \caption{Degree as a function of strength for the specified sector. Both in the left and right plots we represent degree and strength by sector only for those firms that have non-zero values. This means firms in the sector for the left and firms with a relation to the sector for the right. Coloured area represents the density of points in the empirical network of Institution 2. In red is the fitted model mean and the confidence interval computed from one hundred samples taken from the ensemble and smoothed using a moving average.}\label{fig:ansatzA_2}
\end{figure} 

\begin{figure}[hp]
    \centering
    \subfloat[33 - Repair and installation of machinery and equipment\label{fig:ansatz_2_2}]{%
        \includegraphics[width=0.9\textwidth]{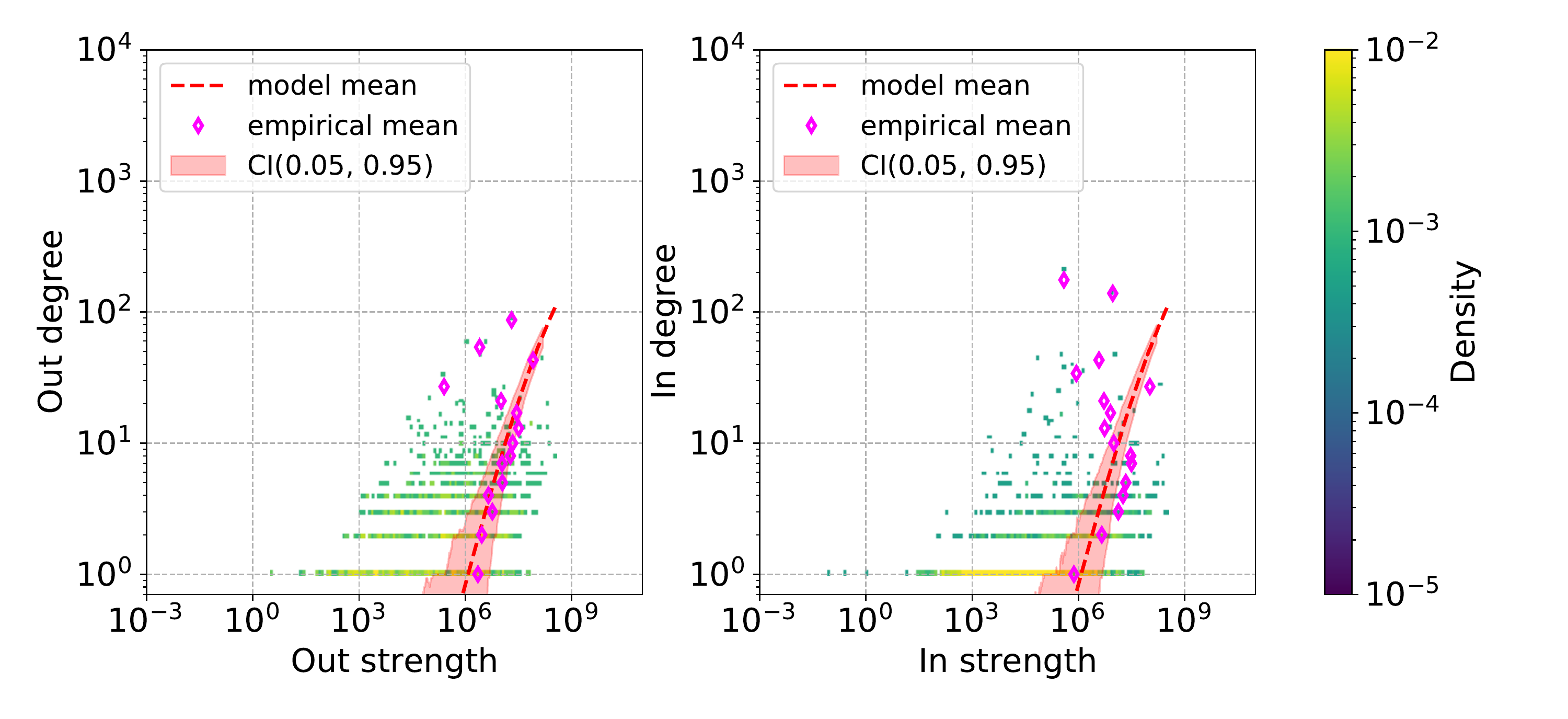}%
    }\vfil
    \subfloat[33121 - Repair and maintenance of machinery for general use and machine parts \label{fig:ansatz_5_2}]{%
        \includegraphics[width=0.9\textwidth]{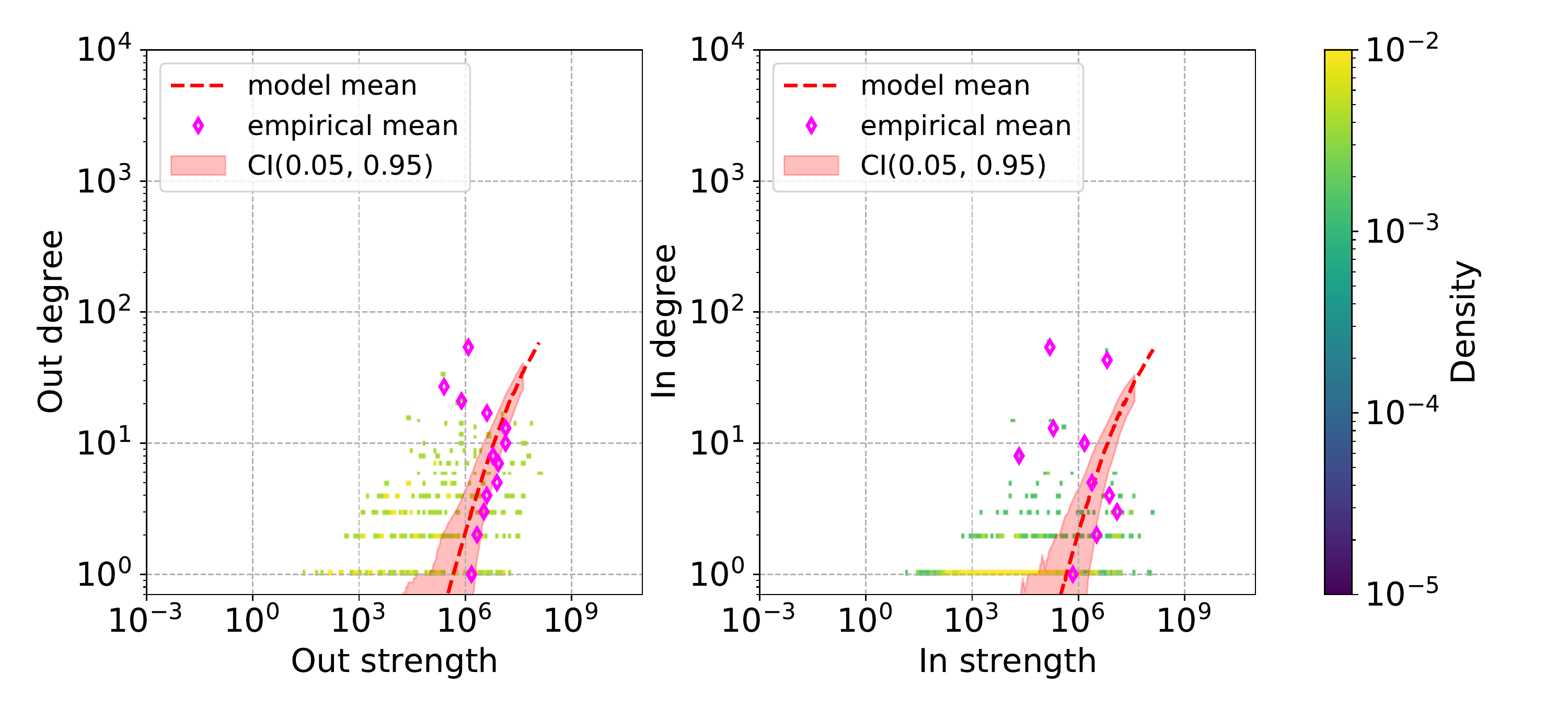}%
    }
    \caption{Degree as a function of strength for the specified sector. Both in the left and right plots we represent degree and strength by sector only for those firms that have non-zero values. This means firms in the sector for the left and firms with a relation to the sector for the right. Coloured area represents the density of points in the empirical network of Institution 2. In red is the fitted model mean and the confidence interval computed from one hundred samples taken from the ensemble and smoothed using a moving average.}\label{fig:ansatzB_2}
\end{figure} 

\begin{figure}[tbp]
    \centering
    \subfloat[scGM \label{fig:sl_2_sc}]{%
        \includegraphics[width=0.9\textwidth]{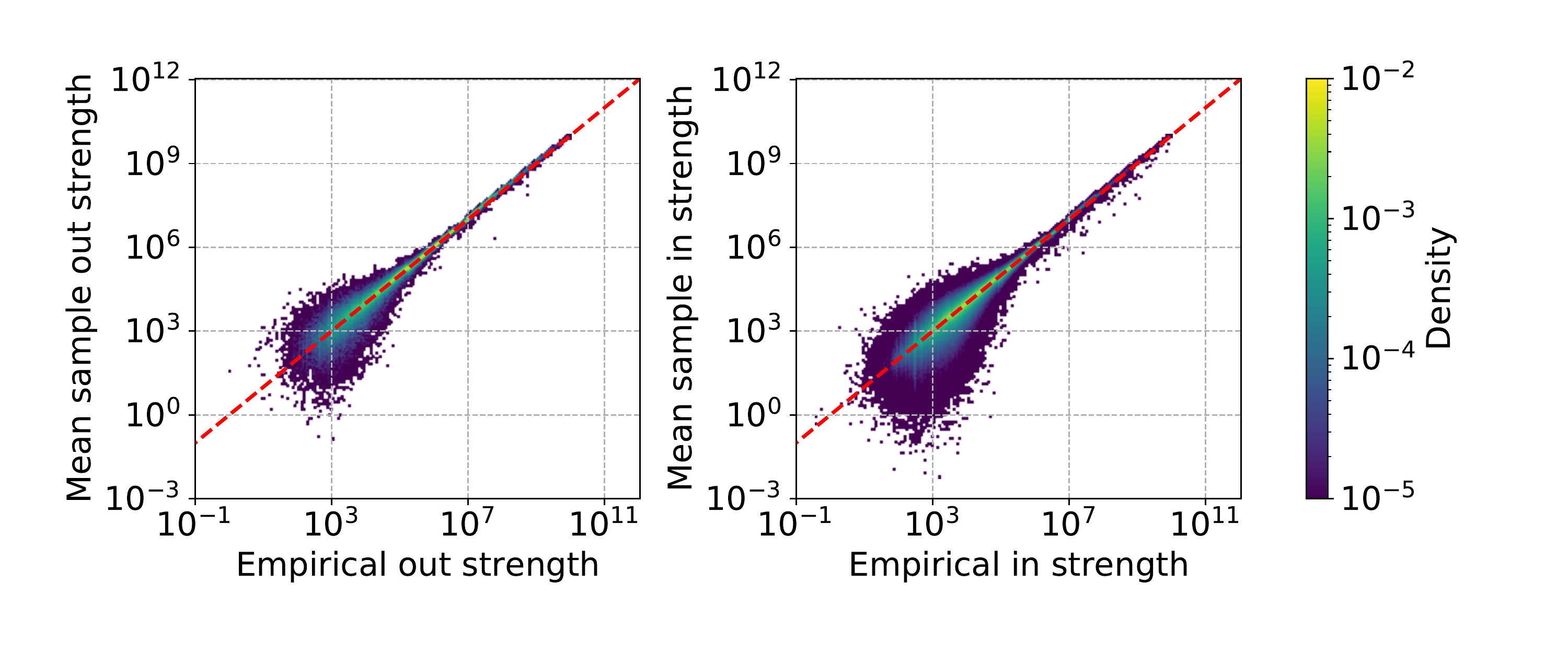}%
    }\vfil
    \subfloat[dcGM \label{fig:sl_2_dc}]{%
      \includegraphics[width=0.9\textwidth]{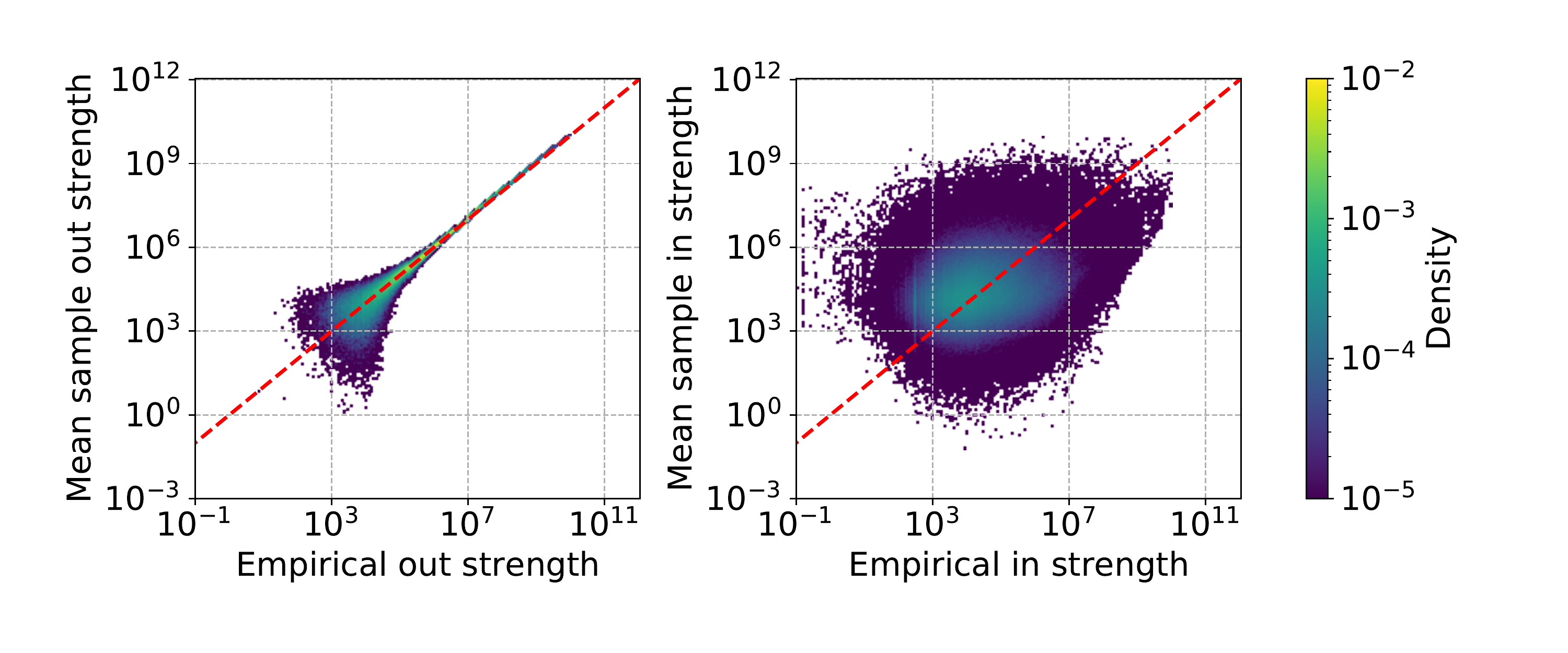}%
    }
    \caption{Empirical versus expected strength by sector for the network of Institution 2.}\label{fig:sl_2}
\end{figure} 

\begin{figure}[tbp]
    \centering
    \subfloat[Out-out]{%
        \includegraphics[width=0.49\textwidth]{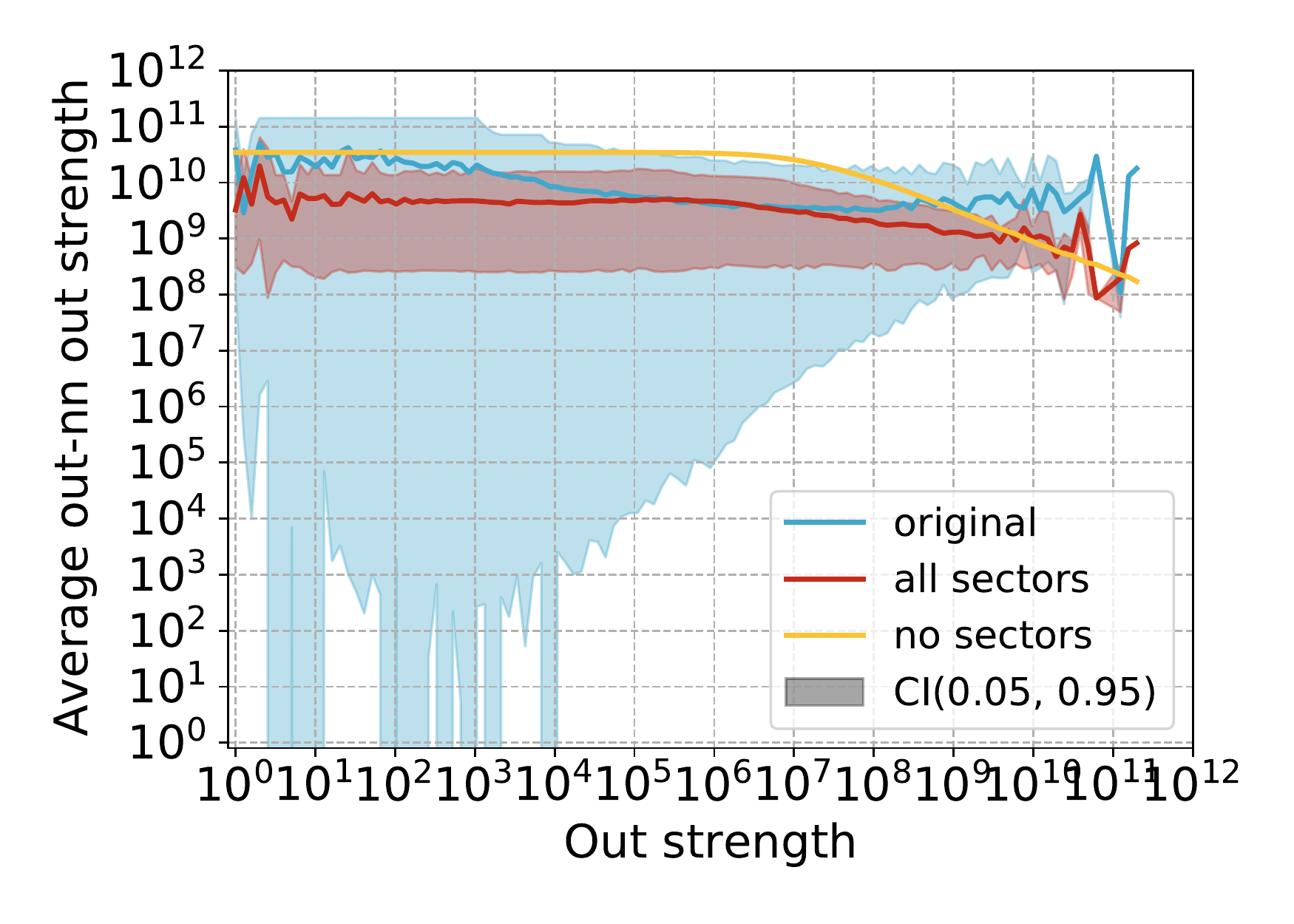}%
    }\hfil
    \subfloat[In-in]{%
      \includegraphics[width=0.49\textwidth]{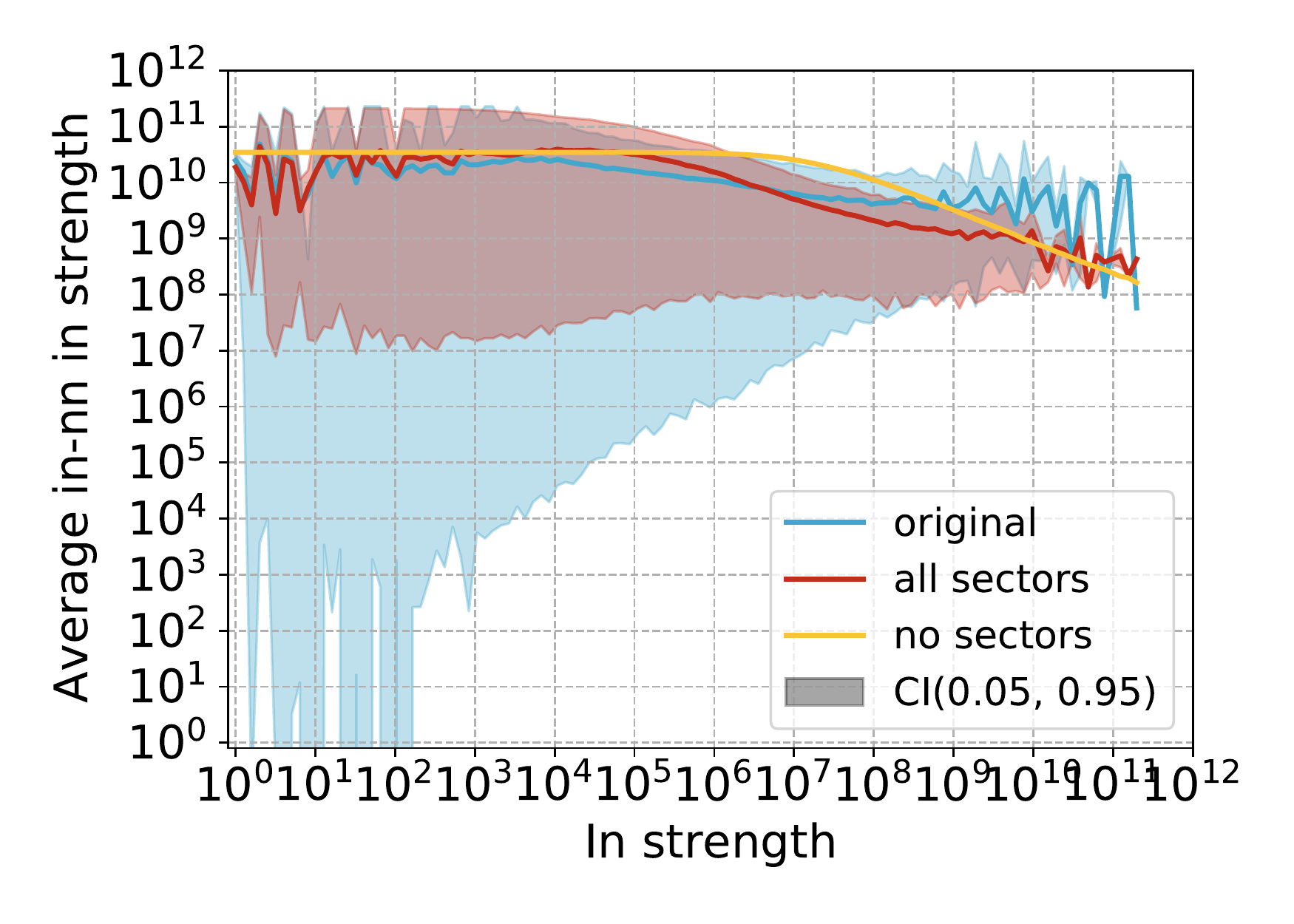}%
    }
    \caption{Average nearest neighbour strength by node strength for Institution 2. The full lines are computed as the average over all firms in the logarithmic binning of the x-axis. The confidence interval is the empirical interval measured over the same bins.}\label{fig:snn_2}
\end{figure} 

\begin{figure}[tbp]
    \centering
    \subfloat[Out-out]{%
        \includegraphics[width=0.9\textwidth]{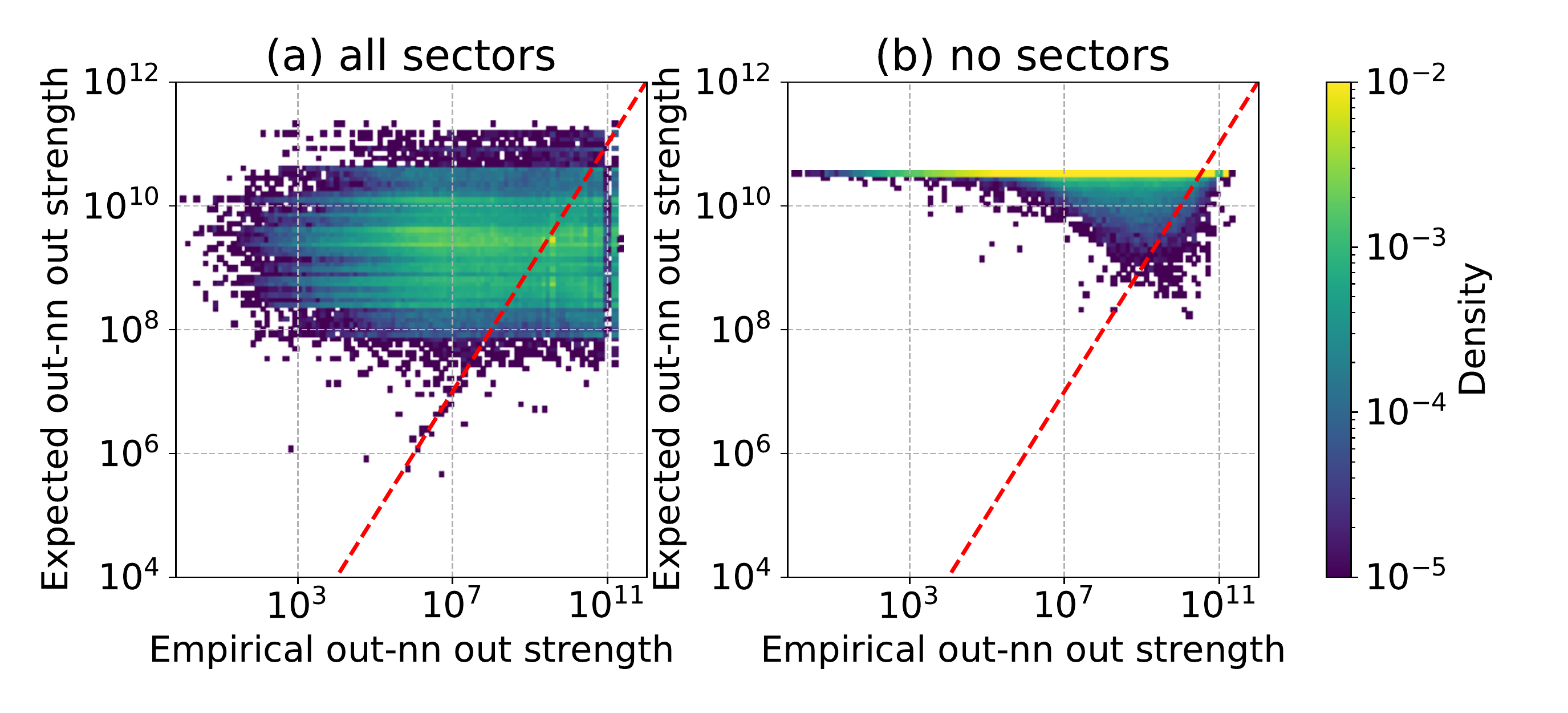}%
    }\vfil
    \subfloat[In-in]{%
      \includegraphics[width=0.9\textwidth]{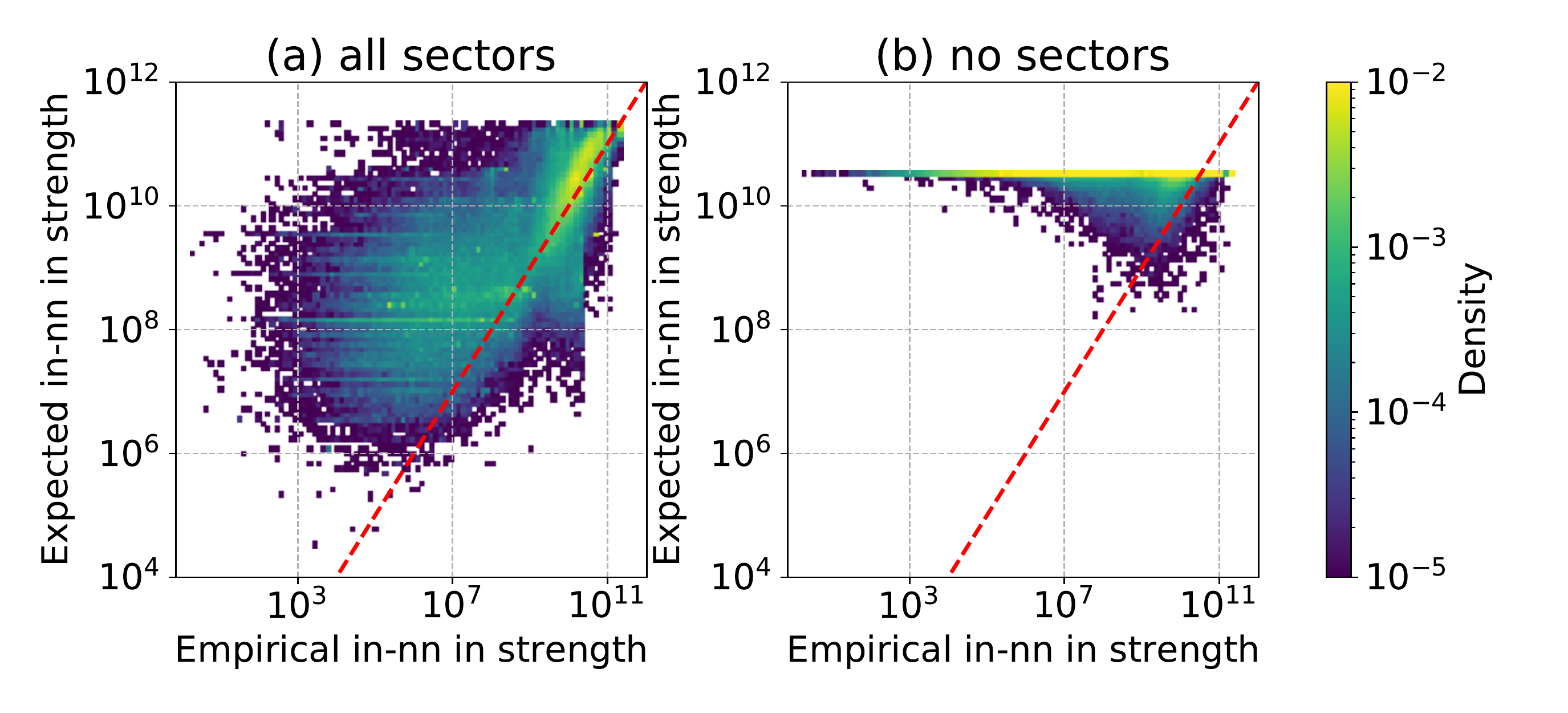}%
    }
    \caption{Per node comparison of the average nearest neighbour strength for Institution 2. The shaded area represents the density of points in the two-dimensional logarithmic binning.}\label{fig:snn_2_node}
\end{figure} 

\begin{figure}[tbp]
    \centering
    \includegraphics[width=0.9\textwidth]{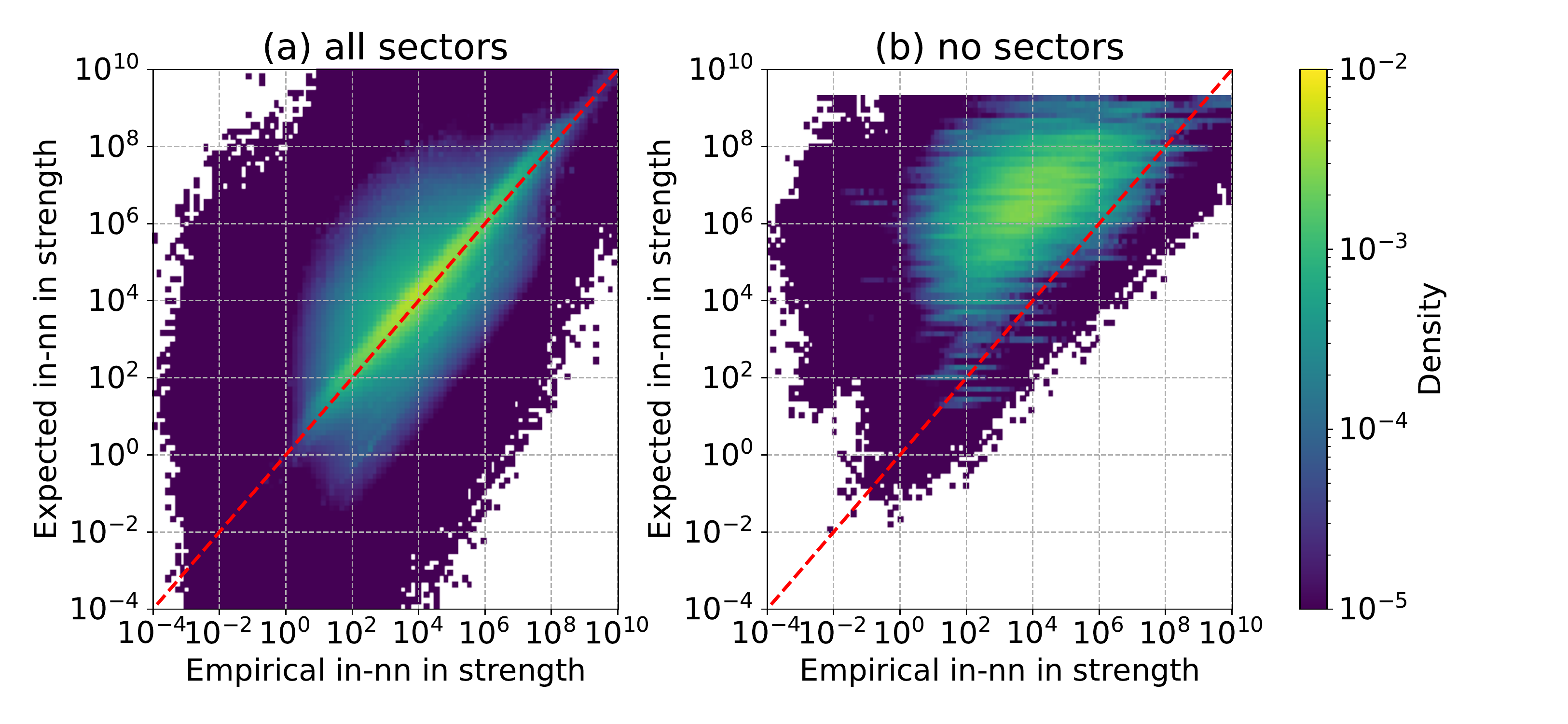}
    \caption{Per node comparison of the average nearest in-neighbour in-strength by sector for Institution 2. The shaded area represents the density of points in the two-dimensional logarithmic binning.}\label{fig:slnn_2_node}
\end{figure} 

\end{document}